\documentclass[showpacs,aps,prd,floatfix,amsmath,amssymb,nofootinbib,superscriptaddress]{revtex4-2}
\usepackage{graphicx}
\usepackage{epstopdf}
\usepackage{amssymb}
\usepackage{bbm}
\usepackage{hyperref}
\usepackage{slashed}
\usepackage{float}
\usepackage{subfigure}
\usepackage{dcolumn}
\usepackage{xcolor}
\usepackage{hyperref}

\usepackage{multirow}
\begin{document}

\title{A new evaluation of the  $HZZ$ coupling: direct bounds on anomalous contributions and  $CP$-violating effects via  a new asymmetry}
\author{A. I. Hern\'andez-Ju\'arez}
\affiliation{Facultad de Ciencias F\'isico Matem\'aticas, Benem\'erita Universidad Aut\'onoma de Puebla, Apartado Postal 1152, Puebla, Puebla, M\'exico}
%\email{alaban7\_3@hotmail.com}
\author{A. Fern\'andez-T\'ellez}
\affiliation{Facultad de Ciencias F\'isico Matem\'aticas, Benem\'erita Universidad Aut\'onoma de Puebla, Apartado Postal 1152, Puebla, Puebla, M\'exico}\author{G. Tavares-Velasco}
\affiliation{Facultad de Ciencias F\'isico Matem\'aticas, Benem\'erita Universidad Aut\'onoma de Puebla, Apartado Postal 1152, Puebla, Puebla, M\'exico}
\date{\today}

\date{\today}

\begin{abstract}
The standard model (SM) one-loop contributions to the most general $H^*Z^*Z^*$ coupling are obtained via the background field method in terms of  Passarino-Veltman scalar functions, from which the contributions to the $H^*ZZ$ and $HZZ^*$ couplings are obtained in terms of two $CP$-conserving $h_{1,2}^V$  and one $CP$-violating $h_3^V$ form factors ($V=H, Z$). The current CMS  constraints on the $HZZ$ coupling ratios are then used to obtain bounds on the real and absorptive parts of the anomalous  $HZZ$ couplings. The former  are up to two orders of magnitude tighter than previous ones, whereas the latter are the first one of this kind.  The effects of the absorptive parts of the $HZZ$ anomalous couplings, which have been overlooked in the past, are analyzed via the partial decay width  $\Gamma_{H^\ast\rightarrow ZZ}$,  and a significant deviation from the SM tree-level contribution is observed at low energies, though it becomes negligible at high energies.  We also explore the possibility that polarized $Z$ gauge bosons are used for the study of non-SM $HZZ$ contributions via a  new left-right asymmetry $\mathcal{A}_{LR}$, which is sensitive to $CP$-violating complex form factors and can be as large as the unity at most, though in a more conservative scenario it is  four to five orders of magnitude larger than the SM prediction arising up to the three-loop level. 
The partial decay widths  $\Gamma_{H^\ast\rightarrow Z_LZ_L}$ and $\Gamma_{H^\ast\rightarrow Z_RZ_R}$ are also studied in several scenarios and it is observed that the deviations from the SM  can be large at high energies and  increases as the energy increases. Thus, the use of polarized $Z$ gauge bosons could give hints of $CP$ violation.  The Mathematica code for our analytical results and the numerical evaluation  is available in our GitLab site.

\end{abstract}

%\pacs{12.60.Cn, 13.35.DX}

\date{\today}

\maketitle

\section{Introduction}

The observation of the Higgs boson  at the CERN LHC \cite{CMS:2012qbp,ATLAS:2012yve} was  clear evidence that the mechanism of electroweak gauge symmetry  breaking is realized in nature as conjectured by the Glashow-Weinberg-Salam Standard Model (SM) of elementary particles. Up to now, the data collected at the LHC has confirmed that the properties of the Higgs particle are consistent with the SM, though  some of its couplings are yet to be measured, such as those to light fermions and its self-couplings as well. It is expected that the  LHC Run 3 can also explore hints of some anomalous Higgs couplings. Along these lines, the CMS collaboration has reported for the first time data on the off-shell $H^*ZZ$ coupling  via off-shell Higgs boson production (O-SHBP) $pp\rightarrow H^*\rightarrow ZZ$ \cite{CMS:2022ley}, which requires that the  four-momentum of the off-shell Higgs boson is above the threshold $\|q\|=2m_Z$. According to the SM, 10\% of  $V$-pair production events at the LHC are due to the $H^*VV$ coupling \cite{Kauer:2012hd}, which is statistically large enough   to allow measurement. Moreover, it has been found that the  $VV$ invariant mass kinematic distribution is sensitive to  the off-shell $H^*ZZ$ contribution, whereas the ratio of off-shell to on-shell production rates can be used to determine the Higgs decay width $\Gamma_H$ \cite{Caola:2013yja,Campbell:2013una}. 

The phenomenological and experimental implications  of the $H^\ast ZZ$ coupling at the LCH  and  future colliders  have been explored in the past and also very recently \cite{Bolognesi:2012mm,Anderson:2013afp,Sahin:2019wew,Gritsan:2020pib,Goncalves:2020vyn,Azatov:2022kbs,Sharma:2022epc,Aguilar-Saavedra:2022wam,Nguyen:2022ubf}.  Furthermore, the off-shell $HZZ^\ast$ coupling has   several phenomenological  implications, which have been studied  through  $HZ$ production  \cite{Kniehl:1991hk,Hagiwara:1993sw,Hagiwara:2000tk,Kniehl:2001jy,Biswal:2005fh,Choudhury:2006xe,Godbole:2007cn,Dutta:2008bh,Rindani:2009pb,Cakir:2013bxa,Rao:2019hsp,Kumar:2019bmk,Rao:2020hel,Chen:2020gae,Bizon:2021rww,Rao:2022olq,Bittar:2022wgb,Gritsan:2022php,Chen:2022mre}. Thus, the  $HZZ$ coupling is worth studying, thereby requiring a highly precise determination of all its lowest-order contributions. In particular, the search for any anomalous contribution to the $H^\ast ZZ$ coupling  is expected to play an important role in the LHC future program.

Off-shell couplings have been of great interest in recent years \cite{Haisch:2021hvy,Englert:2014ffa,Hernandez-Juarez:2021xhy} as they can develop an imaginary (absorptive) part because of the optical theorem, thereby giving rise to interesting effects in physical processes. Such class of effects have already been studied by some of us, for instance, in the off-shell chromomagnetic and chromoeletric dipole moments of quarks \cite{Hernandez-Juarez:2020drn,Hernandez-Juarez:2020gxp} and also in the trilinear neutral gauge boson couplings \cite{Hernandez-Juarez:2021mhi,Hernandez-Juarez:2022kjx}. The latter require that at least one of the three gauge bosons is off-shell to be non-vanishing \cite{Gounaris:2000tb,Choudhury:2000bw}. On the other hand, the phenomenological implications of the absorptive part of the off-shell Higgs boson coupling $H^*ZZ$ have been  brought to attention by many authors   \cite{Hagiwara:2000tk,Biswal:2005fh,Godbole:2007cn,Rao:2020hel}, nonetheless, to our knowledge, a precise determination of the SM contribution has not been reported yet, which may stem from the fact that there has been some controversy on whether or not off-shell couplings represent valid observable quantities, but we will dwell on this issue below.  The study of the absorptive part of the $H^\ast ZZ$ and $HZZ^*$ couplings may be relevant at the LHC and future colliders as it can explain slight deviations on some physical observables, such as in the case of the absorptive part of the $ttg$ coupling, which has direct effects on top pair production at the LHC \cite{Hernandez-Juarez:2021xhy}. 
 
The one-loop corrections to the $HZZ$ coupling  were calculated long ago in the SM \cite{Fleischer:1980ub,Kniehl:1990mq} and also recently \cite{Phan:2022amy}, whereas new physics  contributions have  been reported in several SM extensions, such as the two-Higgs doublet model \cite{Kanemura:2017wtm}, the minimal Higgs triplet model (HTM) \cite{Aoki:2012jj}, the Higgs singlet model (HSM) \cite{Kanemura:2017wtm}, and the minimal supersymmetric standard model (MSSM) \cite{Englert:2014ffa}. Nevertheless, some of those results were reported in a somewhat old-fashioned notation, which could lead to some confusion if a numerical calculation is required to cross check results. Therefore, a new evaluation of the SM one-loop contributions to the off-shell $H^*ZZ$ coupling with a more up-to-date notation for the analytical results is in order. The purpose of this work is to present such an evaluation. Afterwards, we will use our analytical results and the current LHC data to obtain limits on the real and absorptive parts of the anomalous couplings of the $HZZ$ vertex. These new bounds  can be relevant because of the O-SHBP results. Furthermore, their implications could be observed in  $Z$ boson pair production.

The rest of this work is organized as follows. In Section \ref{theofram} we analyze the most general effective Lagrangian for the $HZZ$ coupling up to dimension-six operators, along with a short discussion on the most popular parametrizations used to study the effects of this coupling at particle colliders, with some lengthy formulas used in the derivation of such parametrizations being presented in  Appendix \ref{ParRel}. Section \ref{AnRe} is devoted to the calculation of the SM one-loop contributions to the off-shell $H^*ZZ$ and $HZZ^*$ couplings via the background field method, which in the Feynman-t'Hooft gauge yields identical results to those obtained via the Pinch technique. Both methods are known to lead to gauge-independent and ultraviolet finite off-shell Green's functions.  The analytical results are presented in Appendix \ref{PassVel} in terms of  Passarino-Veltman scalar functions.  In Section \ref{NA} we present the numerical analysis of the behavior of the $H^*ZZ$ and $HZZ^*$ couplings as functions of the off-shell boson transfer momentum, whereas in Sec. \ref{BOTAC} we  obtain bounds on the anomalous contributions. The results  are used in Sec. \ref{WidthSec} to study  $Z$ pair production off an off-shell Higgs boson at the LHC, for which we introduce a new asymmetry  defined for polarized $Z$ gauge bosons. Finally, in Section \ref{Conc} we present the conclusions and outlook.

\section{Theoretical framework}\label{theofram}

Several  parametrizations have been introduced in the literature  to  study the phenomenology of the $HZZ$ coupling. It is thus worth  presenting an overview of the most popular of such parametrizations and the corresponding limits from experimental data with the aim that our analysis can be straightforwardly compared with similar studies if required. 

\subsection{Hagiwara Basis}
\begin{figure}[H]
\begin{center}
\includegraphics[width=10cm]{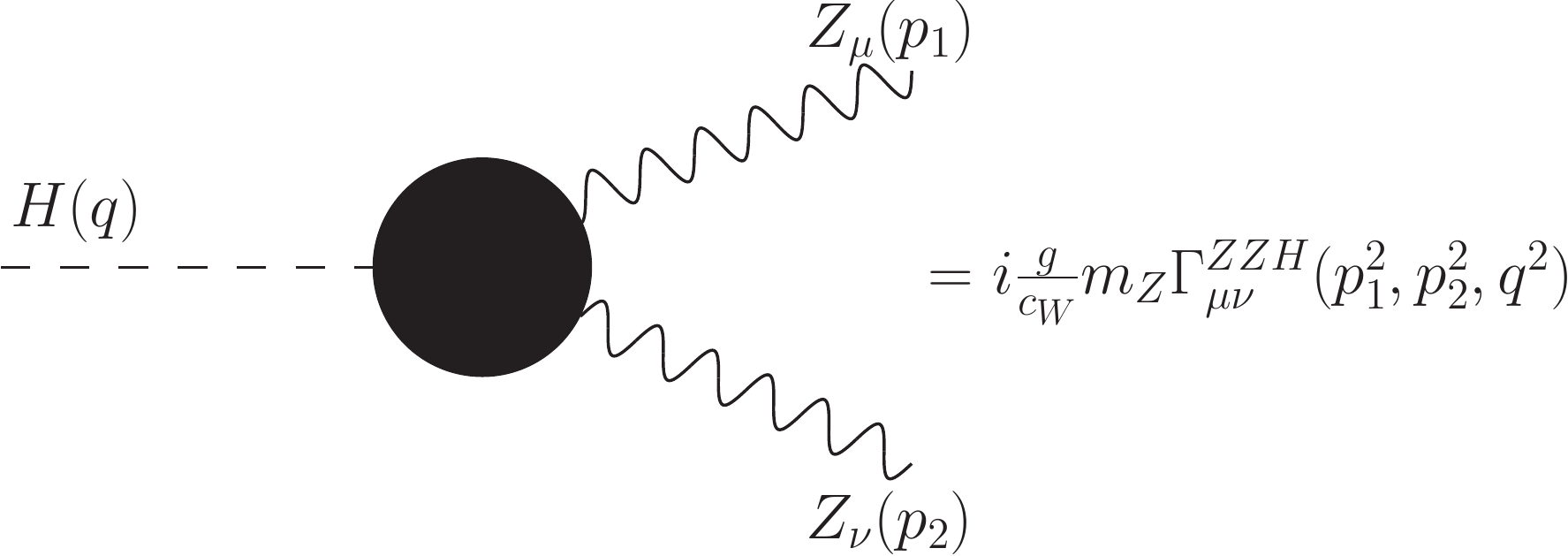}
\caption{Nomenclature for the $HZZ$ coupling and the $\Gamma^{ZZH}_{\mu\nu}$ vertex function.} \label{ZHHvertex}
\end{center}
\end{figure}

First of all, the effective Lagrangian up to dimension-six operators  for the $HZZ$ interaction can be written in  the so-called Hagiwara basis \cite{Hagiwara:1993sw,Hagiwara:2000tk,Dutta:2008bh} as  follows
\begin{align}
\label{Lag}
\mathcal{L}^{HZZ}=&\frac{g }{c_W}m_Z \Big[\frac{(1-a_Z)}{2}  H Z_\mu Z^\mu+\frac{1}{2m^2_Z} \Big\{b_Z HZ_{\mu\nu}Z^{\mu\nu}+c_Z \Big(\big(\partial_\mu H\big)Z_\nu- \big(\partial_\nu H\big)Z_\mu \Big)Z^{\mu\nu}+ \widetilde{b}_Z H Z_{\mu\nu}\widetilde{Z}^{\mu\nu}\Big\}\Big],
\end{align}
where $Z_{\mu\nu}=\partial_\mu Z_\nu -\partial_\nu Z_\mu$ and $\widetilde{Z}_{\mu\nu}=\epsilon _{\mu\nu\alpha\beta}Z^{\alpha\beta}/2$. In the SM, the  $a_Z$,  $b_Z$, $c_Z$ and $\widetilde{b}_Z$ form factors vanish at the tree level, but they can receive contributions at higher orders of perturbation theory.
While the $CP$ conserving form factors $b_Z$ and $c_Z$ are induced at the one-loop level in the SM \cite{Kniehl:1990mq},  the $CP$ violating form factor $\widetilde{b}_Z$  arises up to the three-loop level \cite{Bolognesi:2012mm} and its magnitude has been estimated to be of the order of $10^{-11}$ \cite{Soni:1993jc}. Anomalous contributions to the $b_Z$, $c_Z$ and $\widetilde{b}_Z$ form factors  can also arise from new physics theories at the one-loop-level or higher orders of perturbation theory. In Fig. \ref{ZHHvertex},  we introduce the notation used throughout the rest of this work for the vertex function $\Gamma^{ZZH}_{\mu\nu}$. 

Although we are  mainly interested in processes where the Higgs boson is off-shell, namely, those induced by the $H^\ast ZZ$ coupling, for completeness we will also study the $HZZ^\ast$ coupling, which has also been widely studied in the literature as it has several  implications at particle colliders \cite{Kniehl:1991hk,Hagiwara:1993sw,Hagiwara:2000tk,Kniehl:2001jy,Biswal:2005fh,Choudhury:2006xe,Godbole:2007cn,Dutta:2008bh,Rindani:2009pb,Cakir:2013bxa,Rao:2019hsp,Kumar:2019bmk,Rao:2020hel,Chen:2020gae,Bizon:2021rww,Rao:2022olq,Bittar:2022wgb,Gritsan:2022php}. Furthermore, we will consider that the $Z$ gauge bosons are coupled to conserved currents, thus in   Eq. \eqref{Lag}, we drop the longitudinal component of the $Z$ gauge bosons by setting $\partial_\mu Z^\mu=0$.

We now denote the off-shell boson  by $V$ and consider the kinematics  $H\rightarrow ZZ$ ($Z\rightarrow HZ$) for $V=H\,(Z)$ along with the notation of Fig. \ref{ZHHvertex} to obtain the corresponding vertex function:
\begin{align}
\label{vertex}
\Gamma_{\mu\nu}^{ZZH}(p_1^2,p^2_2,q^2)=h^V_1(q^2,p_1^2,p^2_2) g_{\mu\nu}+\frac{h_2^V(q^2,p_1^2,p^2_2)}{m_Z^2} p_{1\nu}p_{2\mu}+\frac{h_3^V(q^2,p_1^2,p^2_2) }{m_Z^2}\epsilon_{\mu\nu\alpha\beta}p_{1}^\alpha p_{2}^\beta,
\end{align}
where  the dependence of the $h_i^V$ form factors have been written out explicitly. The relations between the form factors  $h_i^V$ and the parameters of Lagrangian \eqref{Lag} for $V=H\,(Z)$ are
\begin{align}
 h_1^V(q^2,p_1^2,p^2_2)&=1+ a_Z-   b_Z \frac{q^2-p_1^2-p_2^2}{m_Z^2}+ c_Z \frac{q^2}{m_Z^2}, \label{h11}\\
h_2^V(q^2,p_1^2,p^2_2)&=\pm 2\big(b_Z-c_Z\big), \label{h22}\\
h_3^V(q^2,p_1^2,p^2_2)&= \pm 2 \widetilde{b}_Z, \label{h33}
\end{align}
where  $p_1^2=m_Z^2$ ($q^2=m_H^2$) for $V=H\,(Z)$, and  $p_2^2=m_Z^2$ in either case.

Note that the structure of Eq. \eqref{vertex} is the same for one, two, and three off-shell particles as long as both $Z$ gauge bosons are coupled to conserved currents. It is also worth noticing that the basis used in the Lagrangian \eqref{Lag} is not unique. In fact,
the use of the equations of motion leads to the following alternative form for the Lagrangian of the $HZZ$ coupling
\begin{align}
\label{Lag2}
\mathcal{L}^{HZZ}=&\frac{g }{c_W}m_Z \left[\frac{(1-a_Z)}{2}  H Z_\mu Z^\mu+\frac{1}{2m^2_Z} \Big\{\hat{b}_Z HZ_{\mu\nu}Z^{\mu\nu}+\hat{c}_Z HZ_\mu\partial_\nu Z^{\mu\nu}+ \widetilde{b}_Z H Z_{\mu\nu}\widetilde{Z}^{\mu\nu}\Big\}\right],
\end{align}
where the $\hat{b}_Z$ and $\hat{c}_Z$ form factors  obey
$\hat{b}_Z=b_Z-c_Z$ and $\hat{c}_Z=2 c_Z$.
In this notation, the form factors $h_i^V$ of Eq. \eqref{vertex} are now given by
\begin{align}
& h_1^V(q^2,p_1^2,p^2_2)=1+ a_Z-   \hat{b}_Z \frac{q^2-p_1^2-p_2^2}{m_Z^2}+ \frac{\hat{c}_Z}{2} \frac{p_1^2+p_2^2}{m_Z^2}, \label{H11}\\
&h_2^V(q^2,p_1^2,p^2_2)=\pm 2 \hat{b}_Z, \label{H22}\\
&h_3^V(q^2,p_1^2,p^2_2)= \pm 2 \widetilde{b}_Z. \label{H33}
\end{align}
The main difference between the basis of Lagrangians \eqref{Lag} and \eqref{Lag2} is that in the latter the form factor $h_2^V$ is given in terms of only one parameter. Although one can calculate the contributions to  each $h_i^V$ in a specific theory,  it is not possible to extract the partial contribution from  $b_Z$ and $c_Z$. Furthermore, it turns out that both parametrizations are used indistinctly  in the literature, which may be confusing for a cross-check of the numerical results. Thus, to avoid such a situation, we present our results in terms of the parametrization of Lagrangian \eqref{Lag2}, which is more suited for our work since we can calculate explicitly the $h_2^V$ form factor, out of which the exact SM contribution to the $\hat{b}_Z$ coefficient can be extracted. 

Direct bounds on the $HZZ$ anomalous couplings have been obtained using polarization observables of the $Z$ gauge boson produced via the $Z^\ast\rightarrow HZ$ coupling at the LHC at $\sqrt{s}=14$ TeV \cite{Rao:2020hel}. In the basis of Lagrangian \eqref{Lag2} such bounds read as
\begin{equation}
\big|{\rm Re}\big[\hat{b}_Z\big]\big|\leqslant3.5\times 10^{-4}\text{,} \quad \big|{\rm Im}\big[\hat{b}_Z\big]\big|\leqslant7.94\times 10^{-3},
\end{equation}
\begin{equation}
\big|{\rm Re}\big[\widetilde{b}_Z\big]\big|\leqslant4.76\times 10^{-3}, \quad \big|{\rm Im}\big[\widetilde{b}_Z\big]\big|\leqslant6.64\times 10^{-3}.
\end{equation}
Additional limits on the anomalous $HZZ$ couplings have been obtained  from the analysis of several processes at $e^+e^-$  \cite{Hagiwara:2000tk,Dutta:2008bh}, $ep$ \cite{Cakir:2013bxa} and $\gamma e$ colliders \cite{Sahin:2019wew}.

For completeness we now discuss two alternative parametrizations used in the literature for the study of the phenomenology  of the $HZZ$ coupling and their mappings with the above parametrizations.

\subsection{The LHC framework}
In the analyses of the CMS collaboration \cite{Gao:2010qx},   the scattering amplitude of the processes mediated by the $H^*ZZ$ coupling is parametrized as follows  in the notation  of Fig. \ref{ZHHvertex}:

\begin{equation}
\label{vertex2}
A(H\to ZZ)\sim \frac{1}{v}\Big[ a_1^{ZZ}+\frac{\kappa_1^{ZZ} p_1^2+\kappa_2^{ZZ} p_2^2}{\big(\Lambda^{ZZ}_1\big)^2}\Big]m_Z^2\epsilon^\ast_1\epsilon^\ast_2+\frac{a_2^{ZZ}}{v}f^{\ast(1)}_{\mu\nu}f^{\ast(2)\mu\nu}+\frac{a_3^{ZZ}}{v}f^{\ast(1)}_{\mu\nu}\widetilde{f}^{\ast(2)\mu\nu},
\end{equation}
where   $f^{(i)\mu\nu}=\epsilon^\mu_i p_i^\nu-\epsilon_i^\nu p_i^\mu$ and $\widetilde{f}^{(i)}_{\mu\nu}=\epsilon_{\mu\nu\rho\sigma}f^{(i)\rho\sigma}$  stand for the field and dual field strength tensors of the $Z$ gauge bosons in momentum space.
In the SM $a_1^{ZZ}=2$ at the tree level. The relations between the coefficients of Eq. \eqref{vertex2} and the form factors of the basis discussed above are derived in Appendix \ref{ParRel}, where we also show that one $\kappa_i$ form factor is redundant.

Indirect bounds on the $a_i^{ZZ}$ couplings have been obtained by the CMS collaboration \cite{CMS:2017len,CMS:2019ekd,CMS:2021nnc,CMS:2022ley} through  effective fractional cross sections $f_{ai}$, which minimize the uncertainties and are independent of the coupling parametrization. The effective  cross-section ratios are defined as
\begin{equation}
\label{fafi}
f_{ai}^{ZZ}=\frac{|a_i^{ZZ} |^2 \alpha_{ii}^{(2e2\mu)}}{\sum_j |a_j^{ZZ} | ^2 \alpha_{jj}^{(2e2\mu)}} \text{sign}\left(\frac{a^{ZZ}_i}{a_1^{ZZ}}\right),
\end{equation}
where the coefficients $\alpha_{ii}^{(2e2\mu)}$  stand for the cross-sections of $2e+2\mu$ production  via $H\to Z\gamma^\ast/\gamma^\ast \gamma^\ast\rightarrow 2e2\mu$ when $a_i^{ZZ}=1$. The corresponding numerical values, which can be obtained through Monte-Carlo simulation and are normalized with respect to the coefficient $\alpha_{11}^{(2e2\mu)}$, are shown in Table \ref{TabAlphaii}. In the CMS analysis, the $a_i^{ZZ}$  couplings are considered real, thus their relative phase is 0 or $\pi$. The current bounds  \cite{CMS:2022ley} are shown in Table \ref{TabBound1}.

\begin{table}[H]
  \centering 
  \caption{Anomalous couplings $a_i^{ZZ}$, cross sections ratios $f_{ai}^{ZZ}$ and coefficients $\alpha_{ii}/ \alpha_{11}$ considered in this work.  We use the relationship $a_{i}^{ZZ}=a_{i}^{WW}$ and the value $\Lambda_1=m_Z$ for $\kappa_1^{ZZ}$. The negative sign arises from the convention adopted in \cite{CMS:2014nkk} for Eq. \eqref{fafi} .} \label{TabAlphaii}
  \begin{tabular}{ p{1.9cm} c  p{1.6cm}    }
\hline \hline
% after \\ : \hline or \cline{col1-col2} \cline{col3-col4} ...
  $a_i^{ZZ}$ & $f_{ai}^{ZZ}$ &$\alpha_{ii}/\alpha_{11}$\\
  \hline
  $a_3^{ZZ}$&$f_{a3}^{ZZ}$&0.153\\
  $a_2^{ZZ}$&$f_{a2}^{ZZ}$&0.361\\
-$k_1^{ZZ}$&$f_{\Lambda1}^{ZZ}$&1.016\\
  \hline\hline
\end{tabular}
\end{table}

\begin{table}[H]
  \centering 
  \caption{Allowed intervals at 95\% C.L. for the coupling parameters $f_{ai}$ obtained by the CMS collaboration \cite{CMS:2022ley} through a combined analysis of off-shell and on-shell events. Two scenarios are considered: $\Gamma_H=\Gamma_H^{SM}$=4.1 MeV and $\Gamma_H$ unconstrained. The sign of the relative phase between $a_i^{ZZ}$ and $a_1^{ZZ}$ is absorbed into the definition of $f_{ai}$. \cite{CMS:2021nnc}.}\label{TabBound1}
  \begin{tabular}{ p{1.9cm} c  p{1.6cm}    }
\hline \hline
% after \\ : \hline or \cline{col1-col2} \cline{col3-col4} ...
  Parameter in units $\times10^{-5}$ & Scenario & Observed  at 95\% CL\\
  \hline
  \multirow{2}{4em}{$f_{a2}^{ZZ}$}&$\Gamma_H=\Gamma_H^{SM}$&$\big[-32\text{,}514\big]$\\
  & $\Gamma_H$ unconstrained & $\big[-38\text{,}503\big]$\\
\multirow{2}{4em}{$f_{a3}^{ZZ}$}  &$\Gamma_H=\Gamma_H^{SM}$ & $\big[-46\text{,}107\big]$ \\
&$\Gamma_H$ unconstrained &$\big[-46\text{,}110\big]$\\
\multirow{2}{4em}{$f_{\Lambda_1}^{ZZ}$}&$\Gamma_H=\Gamma_H^{SM}$&$\big[-11\text{,}46\big]$\\
&$\Gamma_H$ unconstrained &$\big[-10\text{,}47\big]$\\
  \hline\hline
\end{tabular}
\end{table}

The coupling fractions are also useful to study the limits on the anomalous couplings. They can be obtained from the cross sections ratios of Eq. \eqref{fafi} as
\begin{equation}
\label{aiaj}
\frac{a_i^{ZZ}}{a_j^{ZZ}}=\sqrt{\frac{|f_{ai}^{ZZ}|\alpha_{jj}^{2e2\mu}}{|f_{aj}^{ZZ}|\alpha_{ii}^{2e2\mu}}}\text{sign}\Big(f_{ai}^{ZZ}f_{aj}^{ZZ}\Big).
\end{equation}

\subsection{The standard model effective field theory framework}
A recent approach  well-suited for the analysis of anomalous couplings is offered by 
the standard model effective field theory (SMEFT). The  effective Lagrangian up to dimension six includes 2499   effective operators $\mathcal{O}_i$  \cite{Alonso:2013hga}  
\begin{equation}
\mathcal{L}_{\text{SMEFT}}=\mathcal{L}_\text{SM}+\sum_i^{2499}C_i \mathcal{O}_i,
\end{equation}
where  the Wilson coefficients $C_i$ along with the SM parameters constitute the parameter space of the SMEFT, whereas the operators $\mathcal{O}_i$ can be described via the Warsaw basis \cite{Grzadkowski:2010es}, the SILH basis \cite{Contino:2013kra} or the Higgs boson basis \cite{LHCHiggsCrossSectionWorkingGroup:2016ypw}. These bases are equivalent and their Wilson coefficients can be mapped onto each other. Although the Higgs boson basis is well suited to study the Higgs boson interactions at the LHC,  this is not always true: for instance, in this basis the parameter space for diboson production at the LHC  is larger than the ones of the other bases. 

In the Higgs boson basis \cite{Azatov:2022kbs}, the lowest dimension effective Lagrangian for the $HZZ$ interaction  can be written, after a redefinition of the couplings,  as follows
\begin{align}
\label{LagSMEFT}
  \mathcal{L}^{HZZ}_{\text{SMEFT}}  &=\frac{H}{\upsilon} \Big[\big(1+\delta c_z\big)\frac{\big(g_L^2+g_Y^2\big)v^2}{4} Z_\mu Z^\mu +c_{zz}\frac{g_L^2+g_Y^2}{4}Z_{\mu\nu}Z^{\mu\nu}+c_{z\Box} g_L^2 Z_\mu\partial_\nu Z^{\mu\nu}+\widetilde{c}_{zz}\frac{g_L^2+g_Y^2}{4}Z_{\mu\nu}\widetilde{Z}^{\mu\nu} \Big],
\end{align}
where $\upsilon$ is the vacuum expectation value (VEV) of the Higgs field, whereas  $g_L\equiv g$ and $g_Y\equiv g'$ stand for the $SU(2)_L\times U_Y(1)$ coupling constants. Again, the relations between the form factors of Lagrangians \eqref{LagSMEFT} and \eqref{Lag2} are presented in Appendix \ref{ParRel}.

Some remarks  concerning the $\delta c_z$, $c_{zz}$, $c_{z\Box}$, and $\widetilde{c}_{zz}$ couplings are in order here. Although these couplings are assumed to be real in the SMEFT \cite{Dedes:2017zog}, we will consider a more general scenario where they are complex. Our assumption is motivated by the fact that in some variations of the SMEFT  \cite{Brehmer:2017lrt, Degrande:2021zpv} complex Wilson coefficients in the Higgs-gauge sector have been considered, which leads to absorptive parts of the Higgs boson couplings to gauge boson pairs and induce new physics scenarios through $CP$-violating effects.  Such absorptive parts can receive contributions from dimension-8 SMEFT operators, which in some cases could be of the same order of magnitude as the contributions to the real parts arising from dimension-6 operators: for instance, in the case of diboson production via an off-shell Higgs boson at the LHC \cite{Degrande:2023iob}. Along these lines, complex couplings have also been considered in other EFT analyses, such as in the study of $b\rightarrow s\ell\ell$ transitions  \cite{DiLuzio:2019jyq,Biswas:2020uaq} and in a possible solution to the apparent $R_K$ and $R_{K^{\ast 0}}$ anomalies \cite{Alda:2018mfy}. The assumption of complex Higgs boson couplings is of special interest since they can lead to new sources of $CP$ violation in the Higgs sector. Thus, we assume a naive SMEFT with complex Wilson coefficients in the Higgs-gauge sector, which preserves the structure given by the Higgs boson basis but leads to complex   $HZZ$ couplings.

For easy reference, we present in Table \ref{mappings} the mapping between all the four parametrizations discussed above.

\begin{table}[!hbt]
\caption{Mappings between form factors for the most popular parametrizations of the $HZZ$ vertex. In our analysis we work in the Basis of Eq. \eqref{Lag2}.\label{mappings}}
\begin{tabular}{cccc}
\hline\hline
Basis of Eq. \eqref{Lag2}&Hagiwara basis& LHC framework&SMEFT\\
\hline\hline
$a_Z$&$a_Z$&$a_1^{ZZ}=2(1+a_Z)$&$\delta c_z=a_Z$\\
$\hat b_Z$&$b_Z=\hat b_Z+\dfrac{\hat c_Z}{2}$&$a_2^{ZZ}=-2\hat b_Z$&$c_{zz}=\dfrac{4}{g^2+g'^2}\hat{b}_Z$\\
$\hat c_Z$&$c_Z=\dfrac{\hat c_Z}{2}$&$\kappa_1^{ZZ}=\hat c_Z$&$c_{z\Box}= \dfrac{1}{g_L^2} \hat{c}_Z$\\
$\widetilde b_Z$&$\widetilde b_Z$&$a_3^{ZZ}=-2\widetilde b_Z$&$\widetilde{c}_{zz}=\dfrac{4}{g^2+g'^2}\widetilde{b}_Z$\\
\hline\hline 
\end{tabular}
\end{table}

\section{Analytical results}\label{AnRe}

We now turn to present our analytical results for the SM one-loop contributions to the $HZZ$ coupling and consider the most general case with off-shell particles, which is known to yield ill-behaved Green's functions that can be gauge-dependent and ultraviolet divergent. 
This issue can be tackled by two approaches: the first one is the so-called pinch technique (PT) \cite{Cornwall:1981zr}, which is a diagrammatic approach that allows one to systematically  obtain gauge-independent and well-behaved off-shell Green's functions by inserting the off-shell vertex into a physical process and judiciously  combining self-energy, vertex, and box diagrams in order to remove the gauge-dependent terms. This approach may turn into a cumbersome task due to the large number of  Feynman diagrams involved in the calculation and the use of a general $R_\xi$ gauge. Nevertheless, it was shown that, at least up to one-loop level, the well-behaved Green's functions obtained via the PT are identical to those obtained via  the background field method (BFM) provided that the Feynman-t'Hooft gauge is used in the calculation \cite{Papavassiliou:1994yi,Hashimoto:1994ct}.  This provides a more economic method to obtain well-behaved off-shell Green's functions as the calculation involves less Feynman diagrams and the corresponding amplitudes can be more easily  manipulated in the Feynman-t'Hooft gauge. We thus find it more convenient the use  of the BFM  to obtain gauge-independent form factors for the $HZZ$ coupling.

\begin{figure}[!hbt]
\begin{center}
\includegraphics[width=6cm]{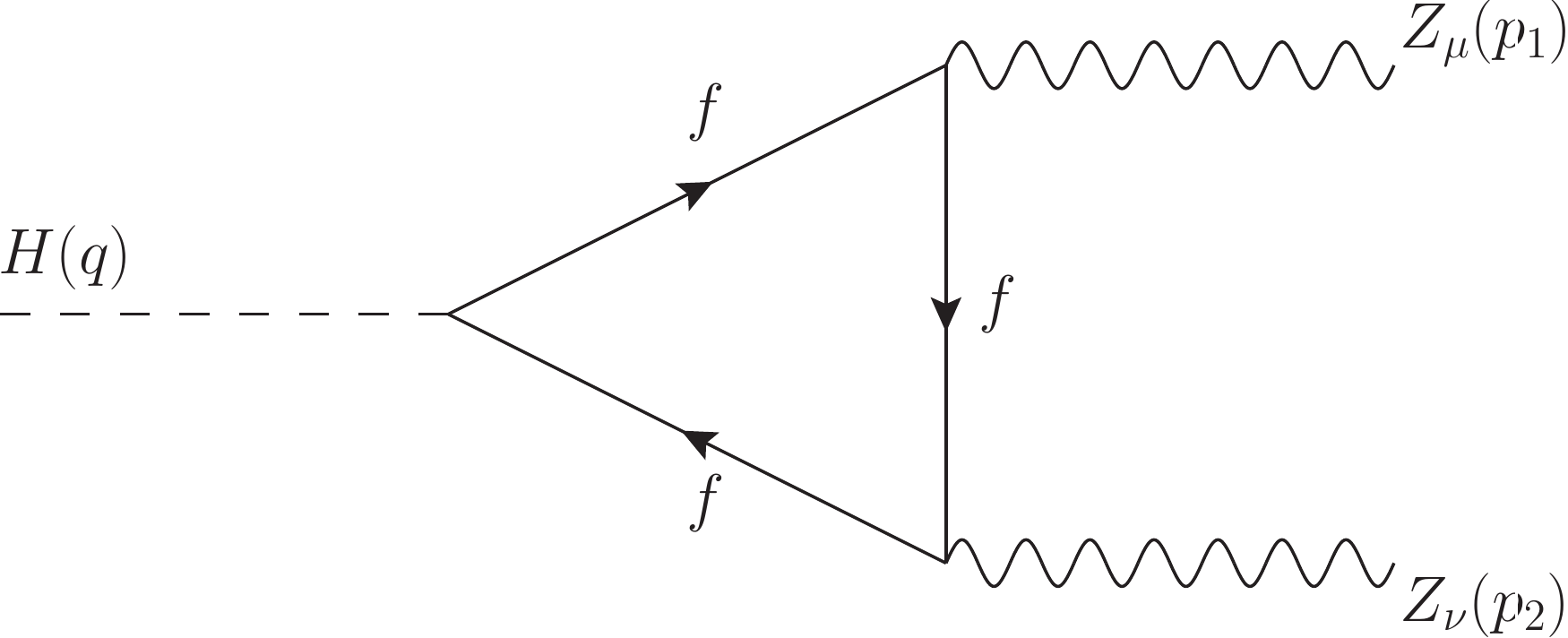}
\caption{Feynman diagram for the  fermion  contribution to the $HZZ$ coupling at the one-loop level.} \label{Fc}
\end{center}
\end{figure}

\begin{figure}[!hbt]
\begin{center}
\subfigure[]{\includegraphics[width=5cm]{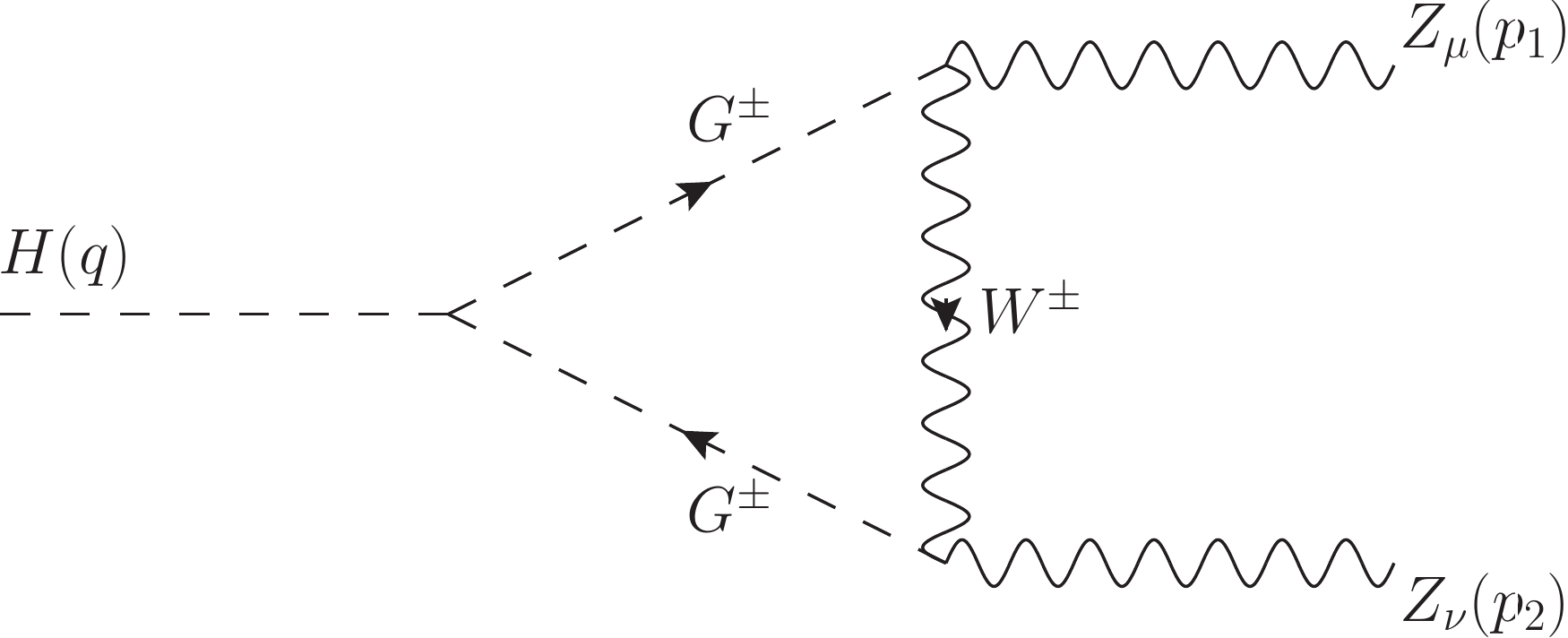}}
\subfigure[]{\includegraphics[width=5cm]{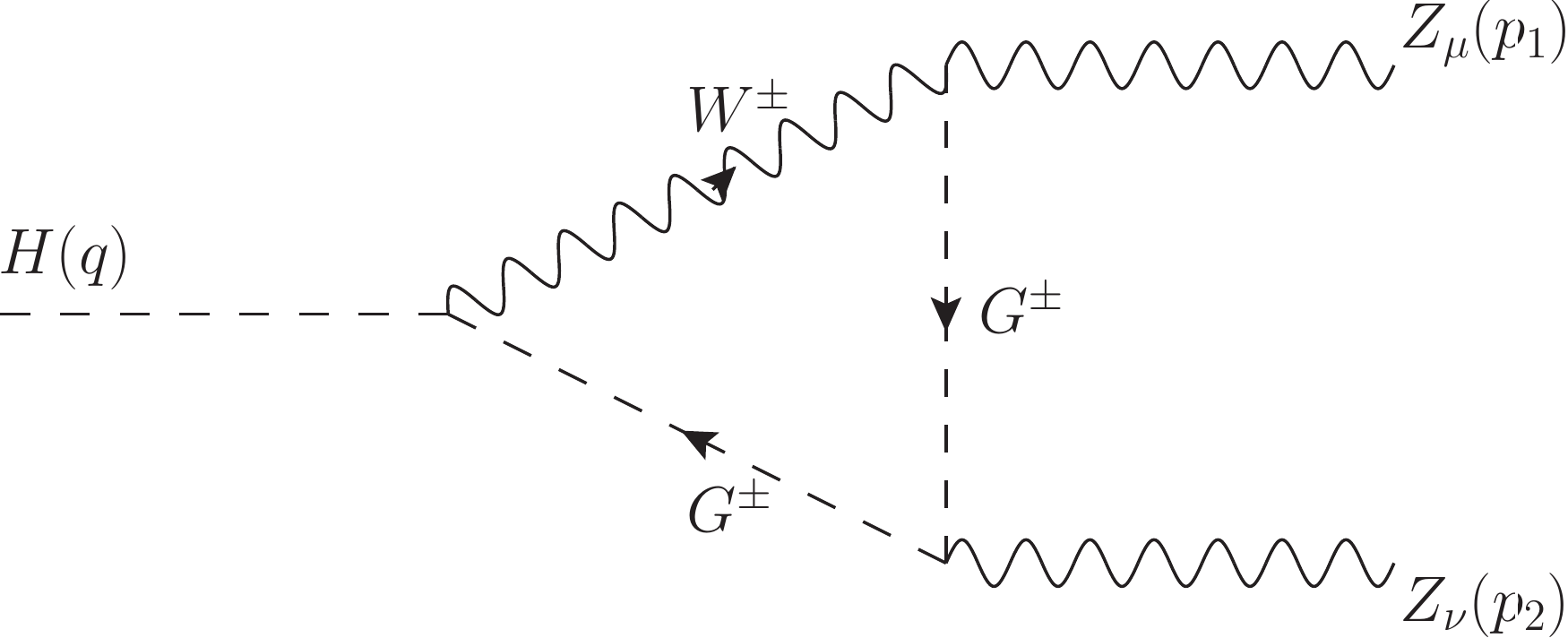}}
\subfigure[]{\includegraphics[width=5cm]{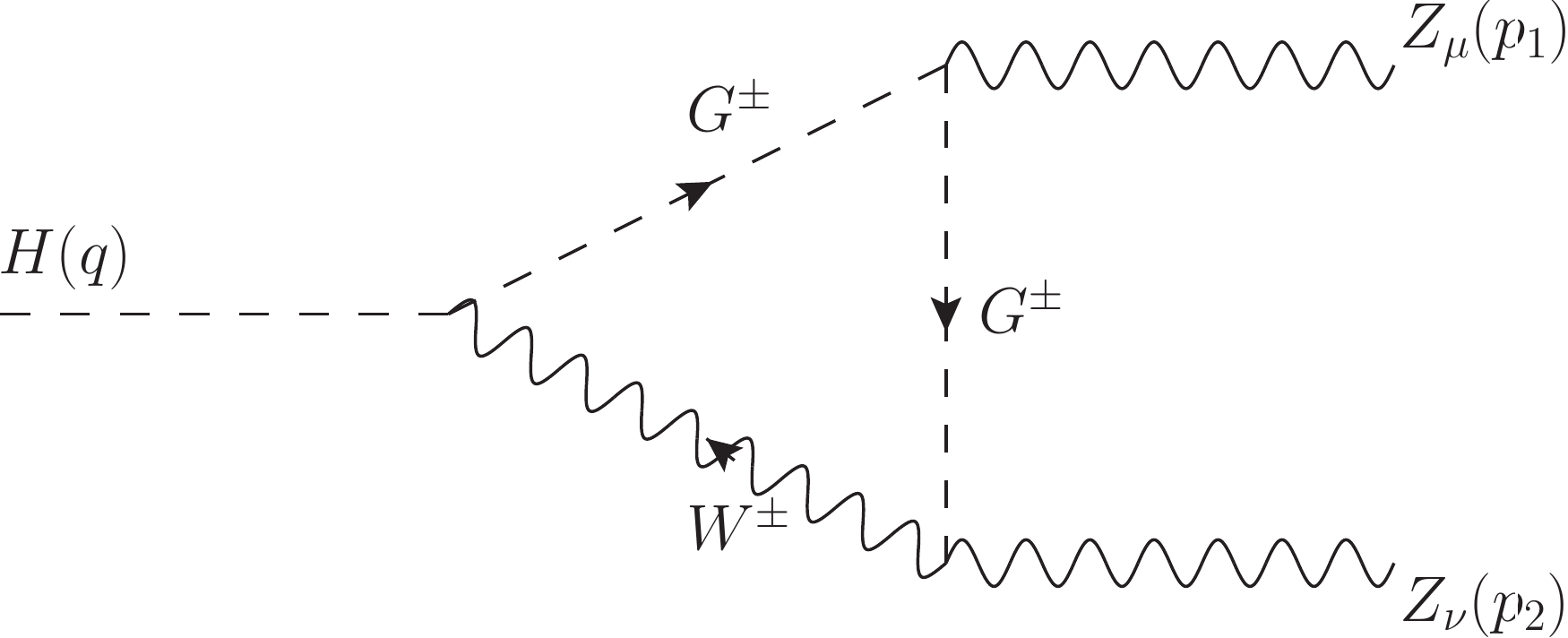}}
\subfigure[]{\includegraphics[width=5cm]{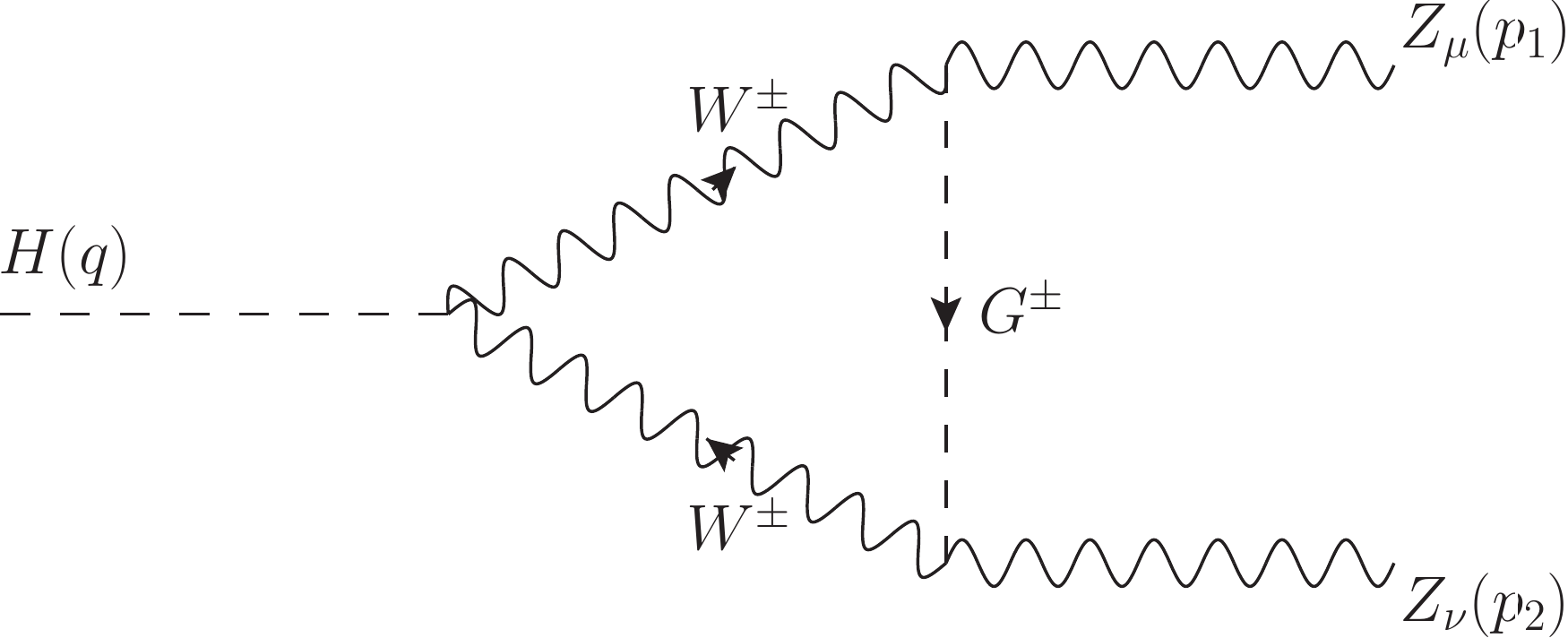}}
\subfigure[]{\includegraphics[width=5cm]{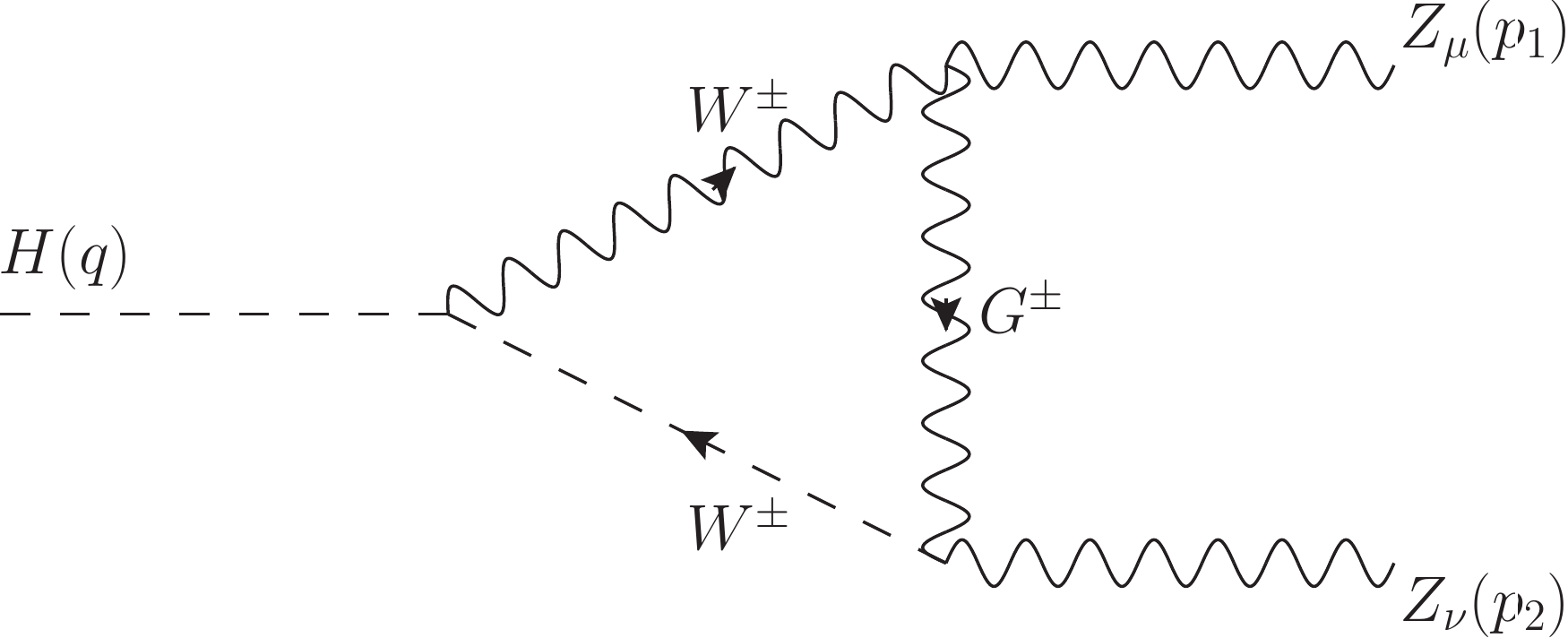}}
\subfigure[]{\includegraphics[width=5cm]{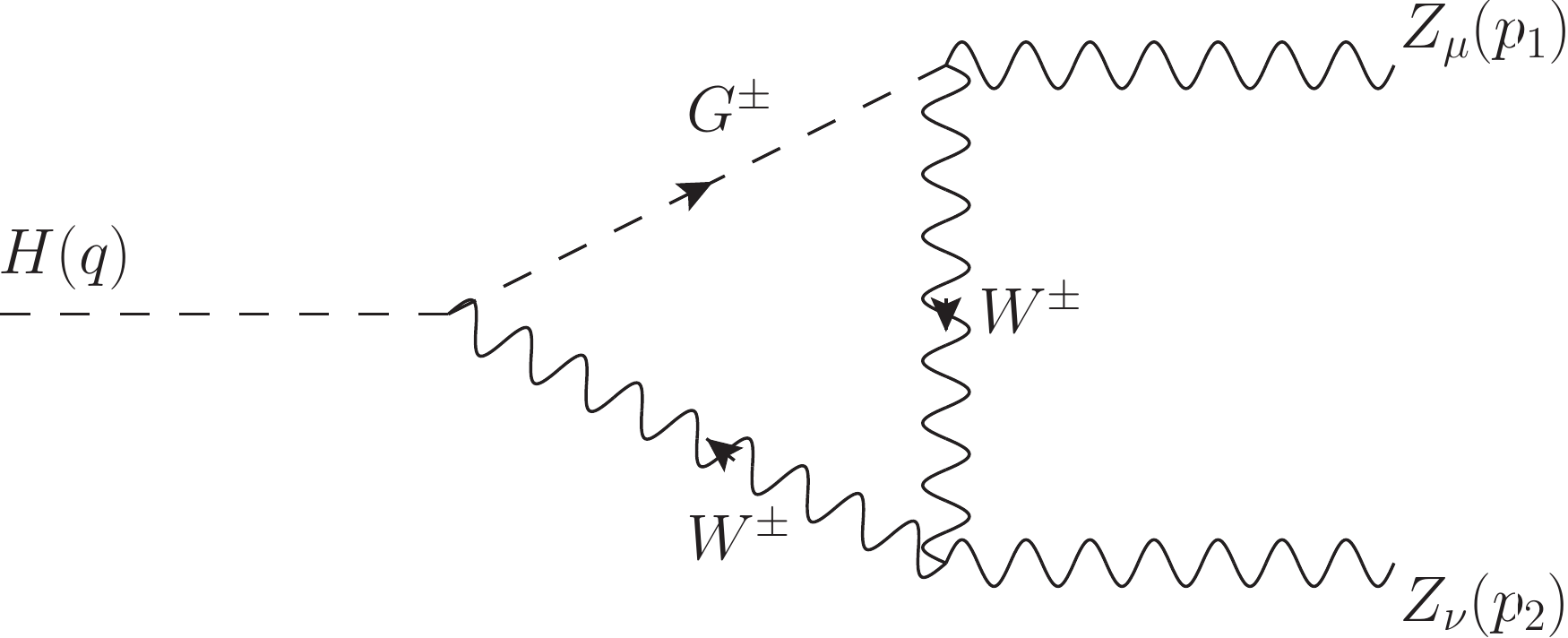}}
\subfigure[]{\includegraphics[width=5cm]{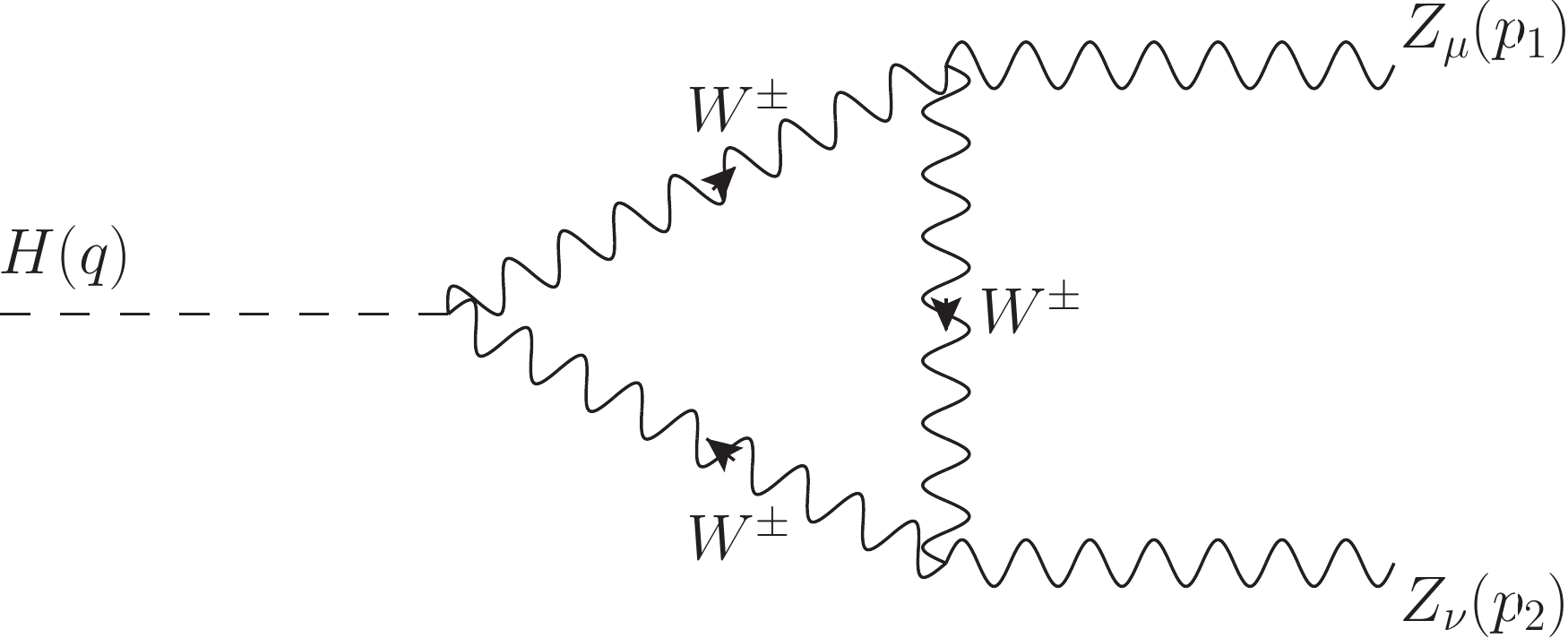}}
\subfigure[]{\includegraphics[width=5cm]{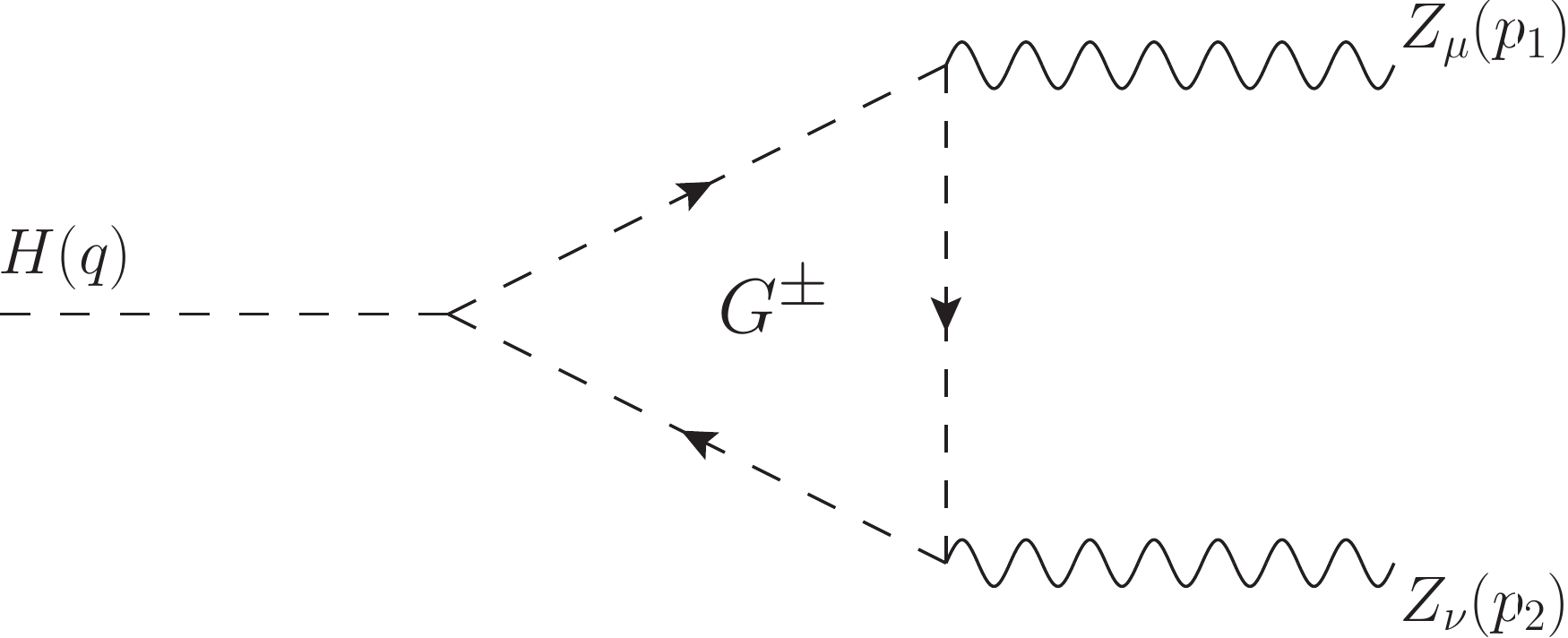}}
\subfigure[]{\includegraphics[width=5cm]{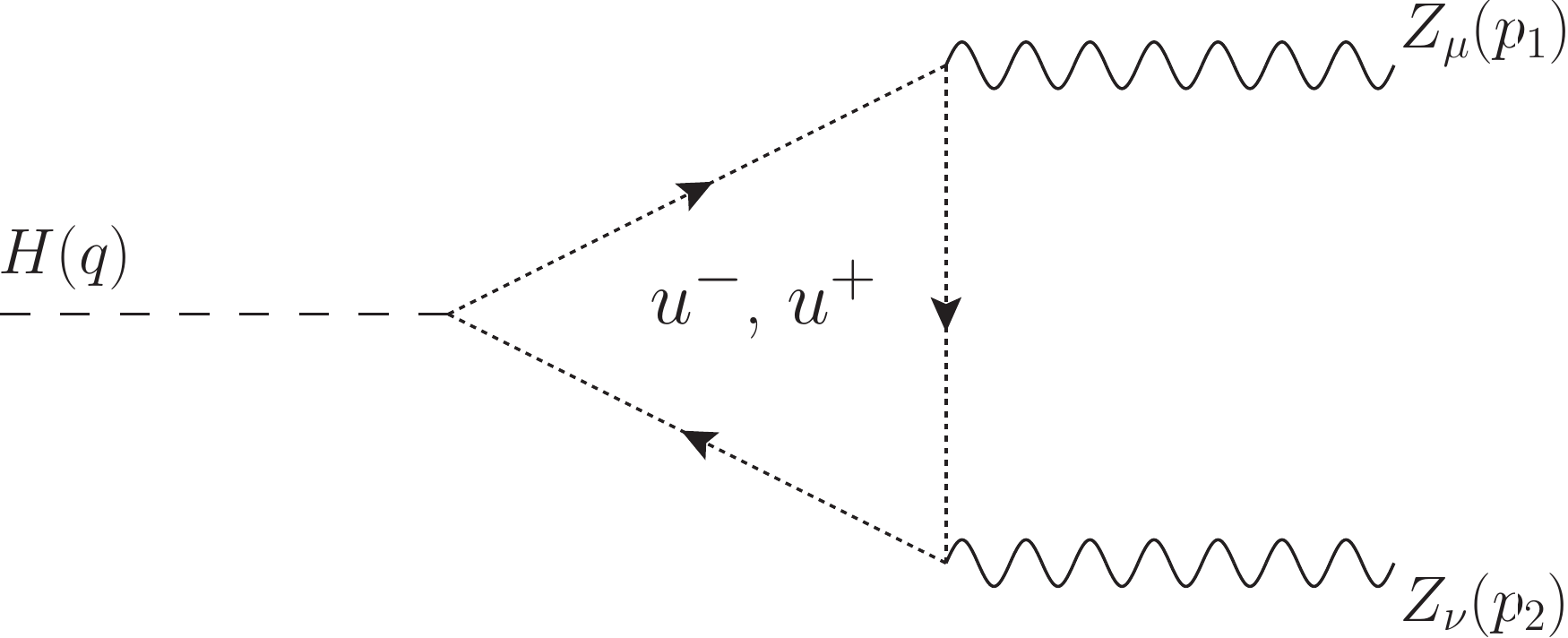}}
\subfigure[]{\includegraphics[width=5cm]{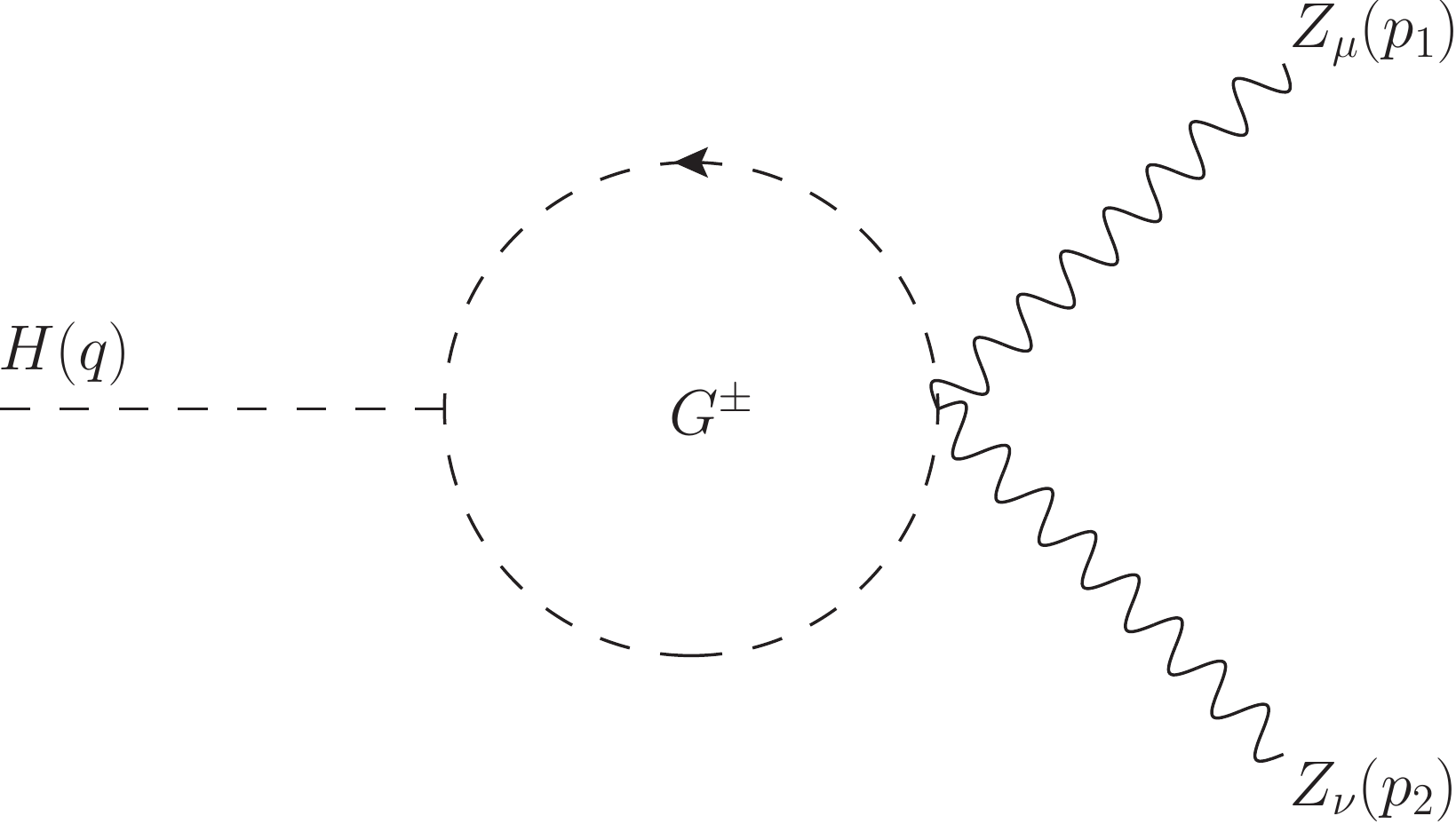}}
\subfigure[]{\includegraphics[width=5cm]{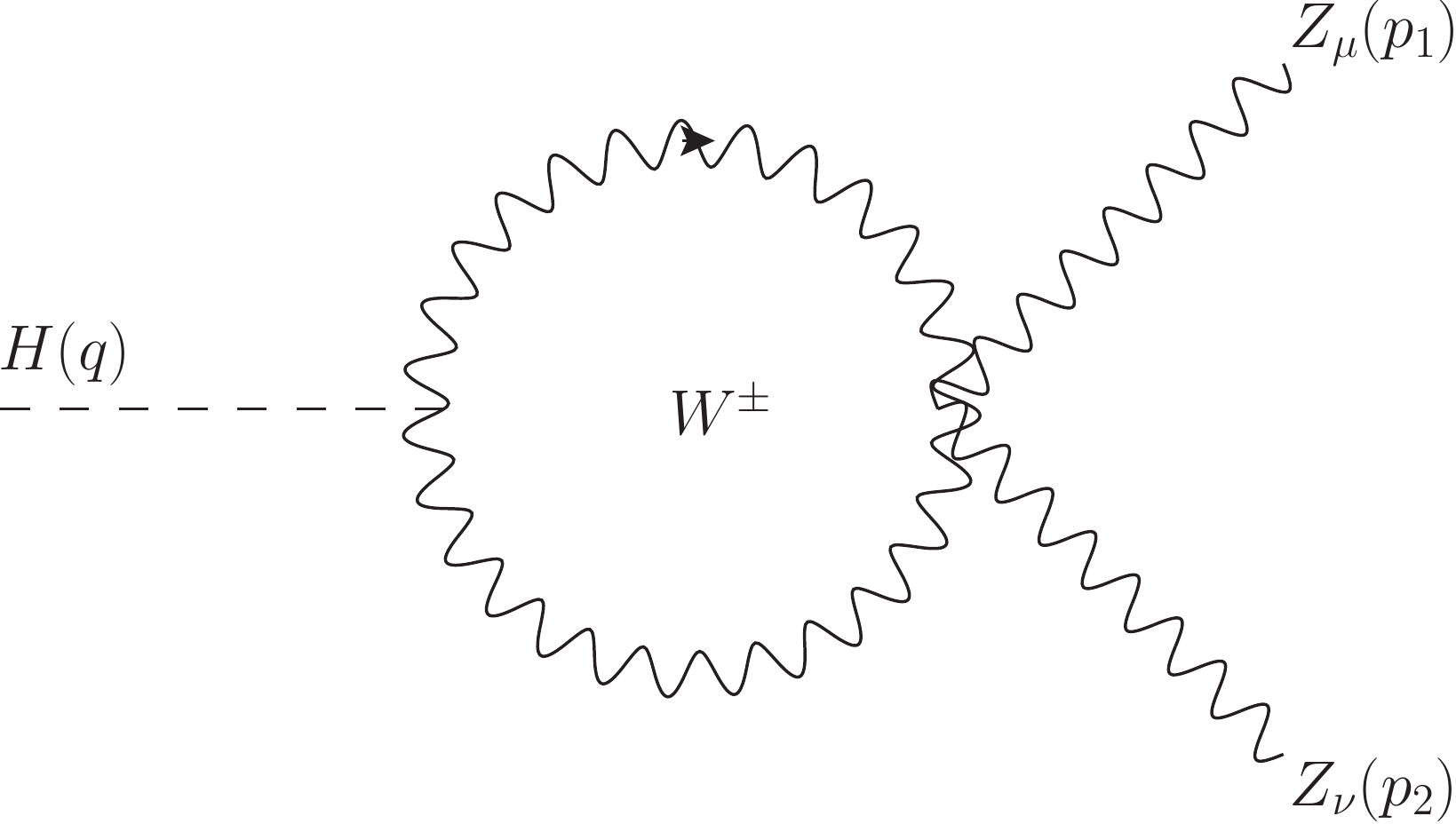}}
\subfigure[]{\includegraphics[width=5cm]{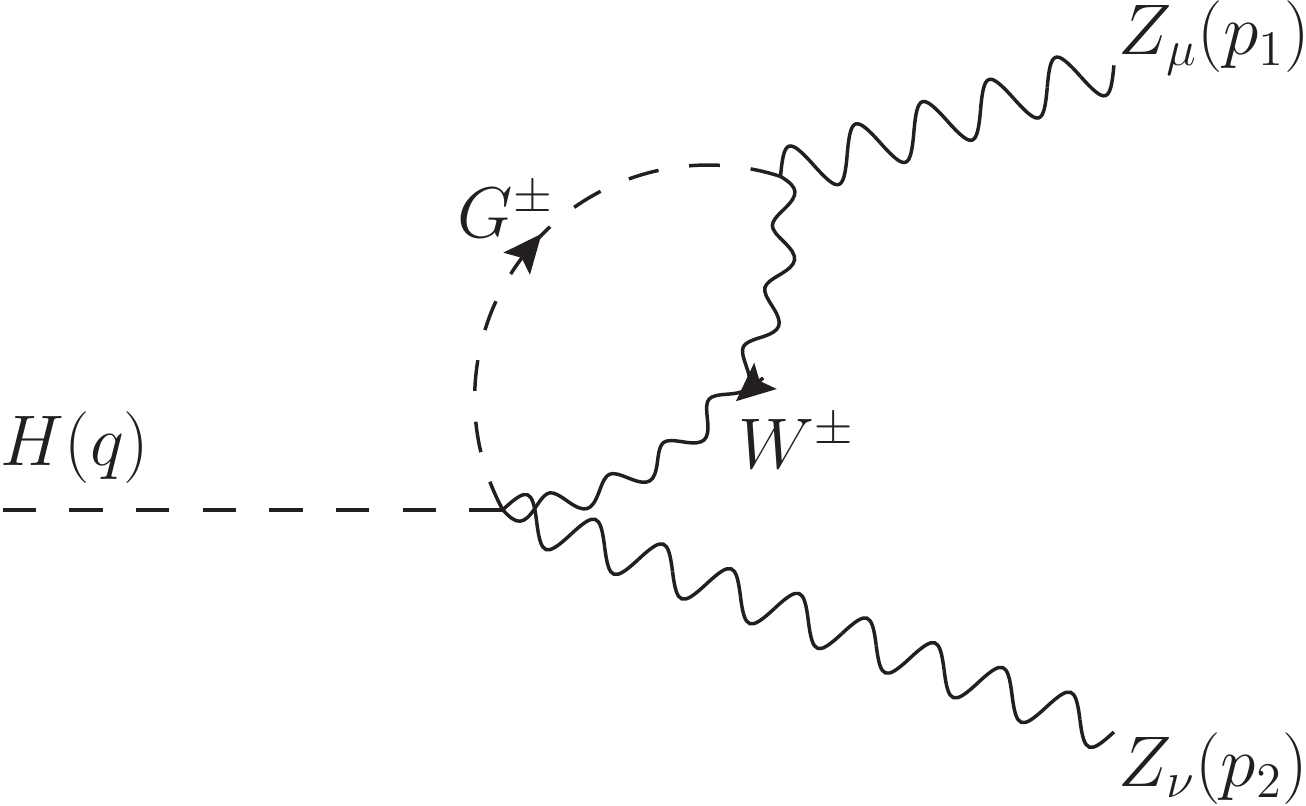}}
\subfigure[]{\includegraphics[width=5cm]{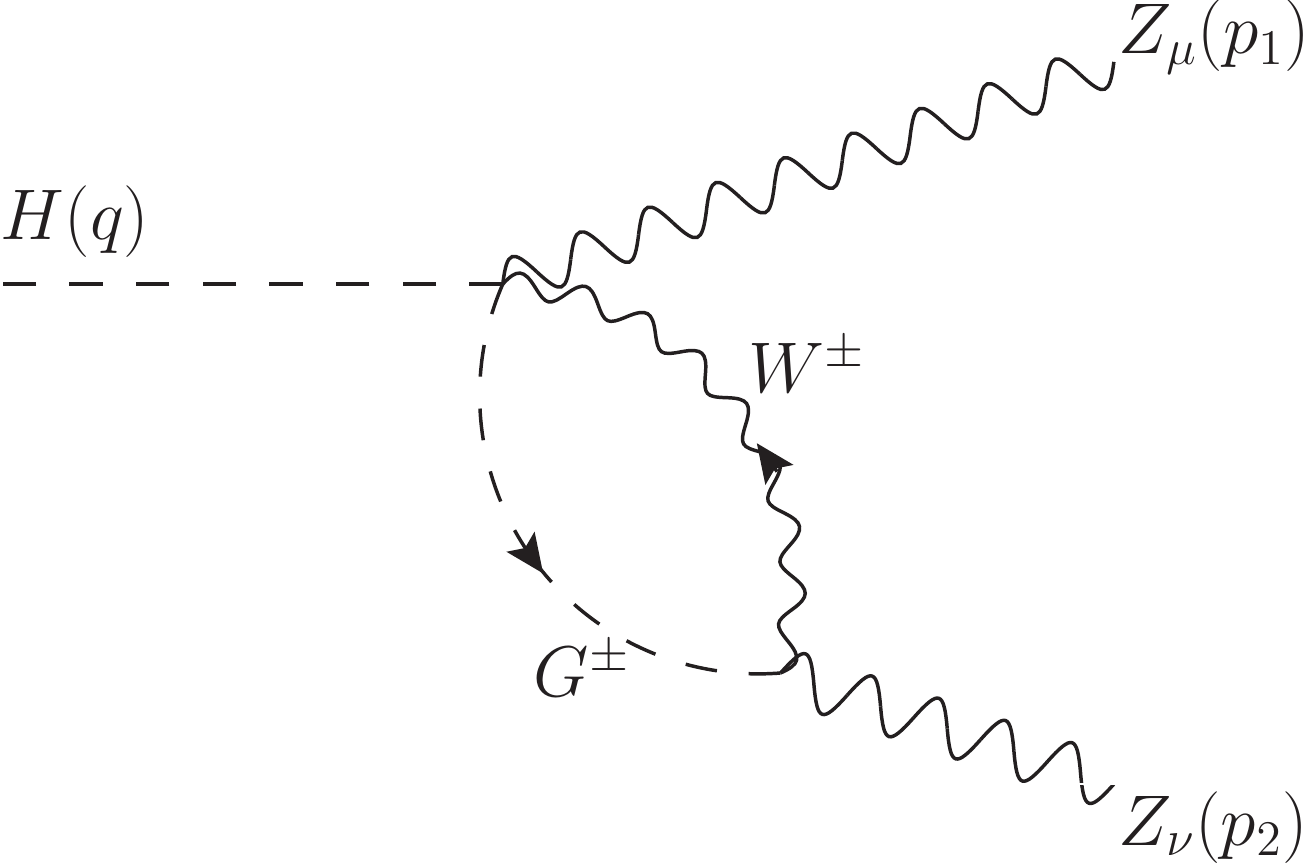}}
%\subfigure[]{\includegraphics[width=14cm]{diag3.eps}\label{Diagram3}}
\caption{Feynman diagrams for  the contribution of the $W$ gauge boson and its associated unphysical particles at the one-loop level in the Feynman-t'Hooft gauge of the BFM.} \label{Wc}
\end{center}
\end{figure}

\begin{figure}[!hbt]
\begin{center}
\subfigure[]{\includegraphics[width=5cm]{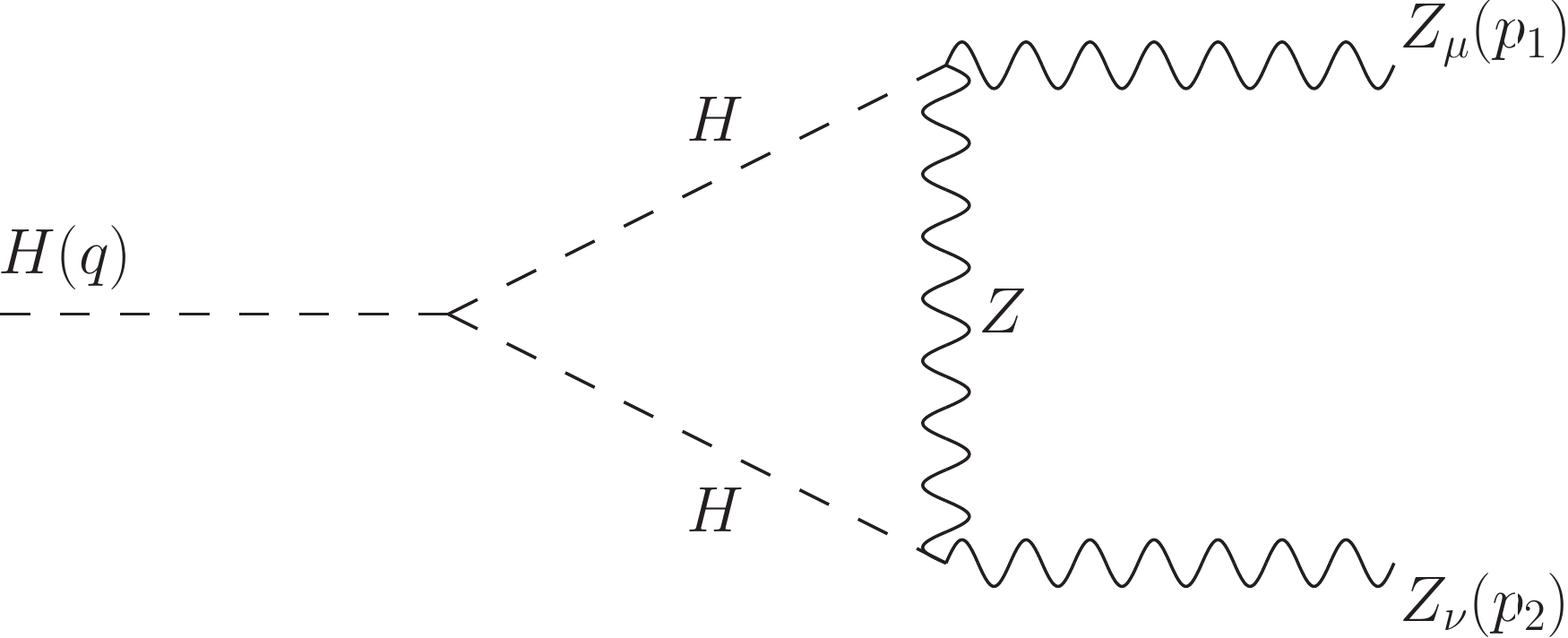}}
\subfigure[]{\includegraphics[width=5cm]{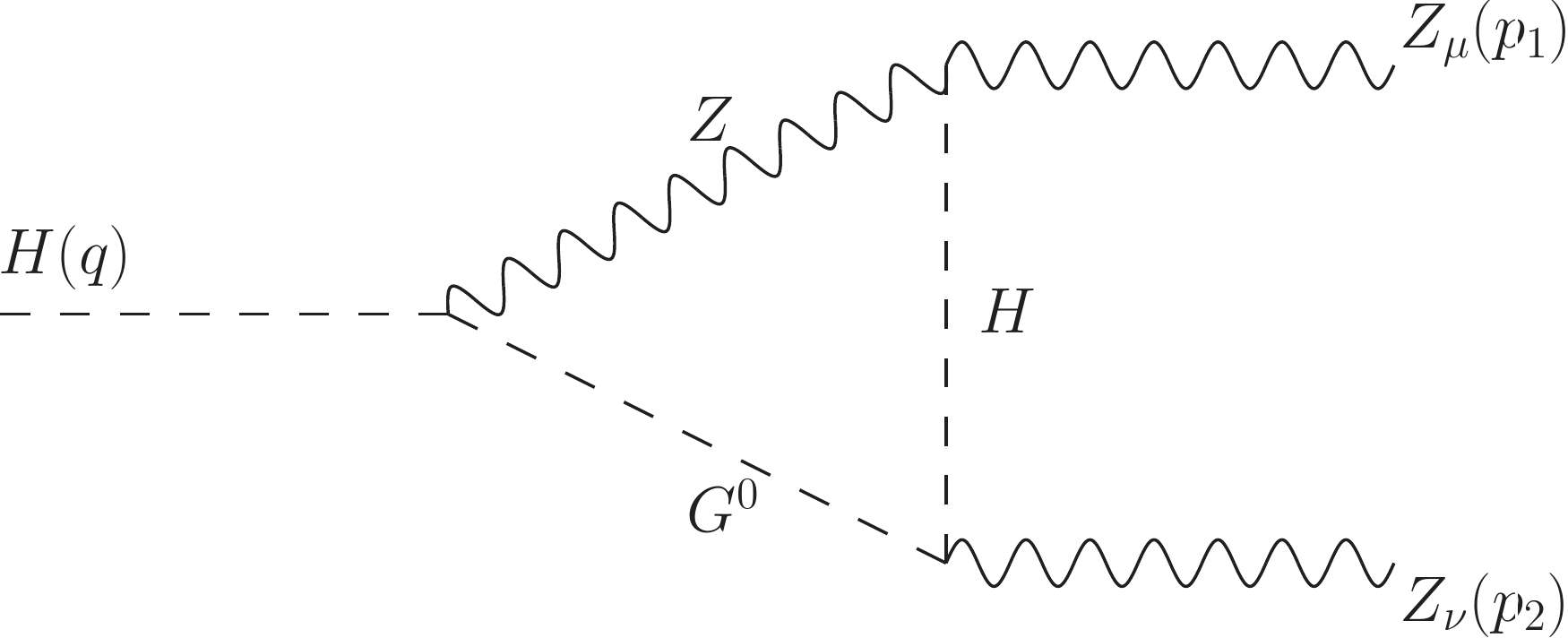}}
\subfigure[]{\includegraphics[width=5cm]{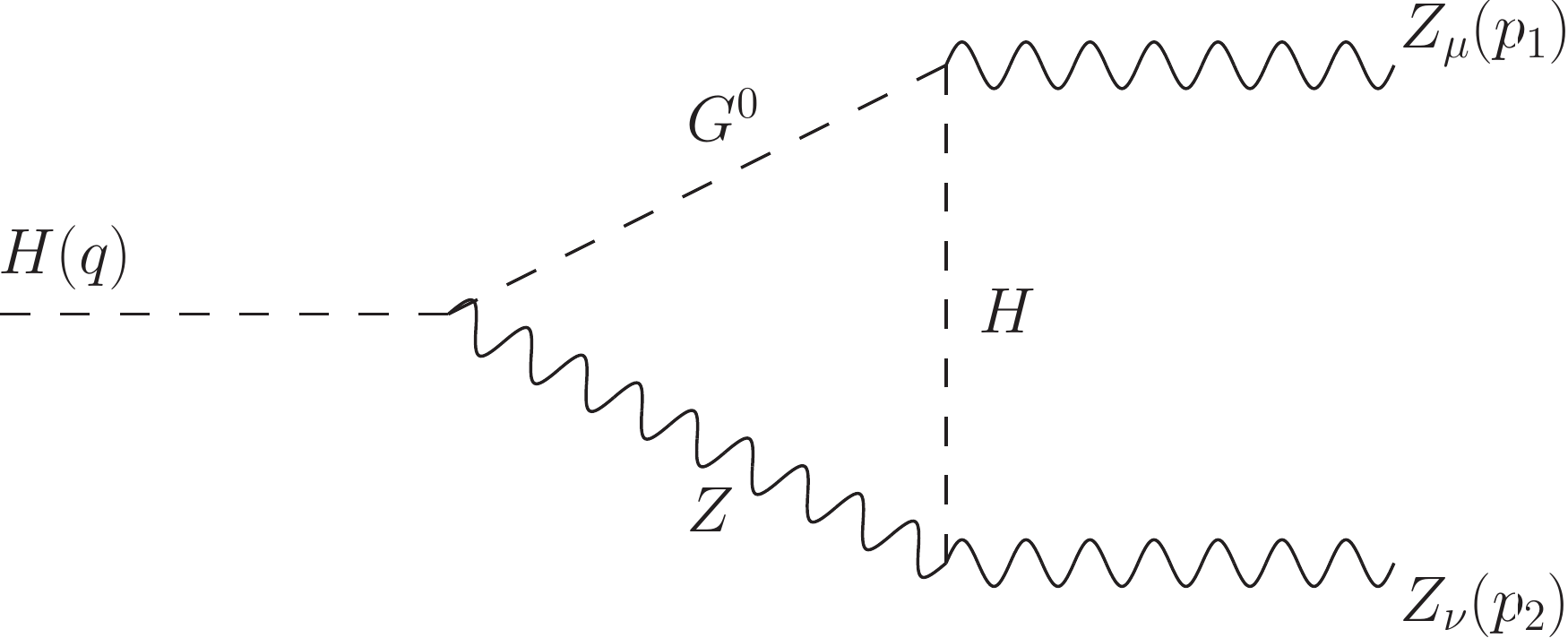}}
\subfigure[]{\includegraphics[width=5cm]{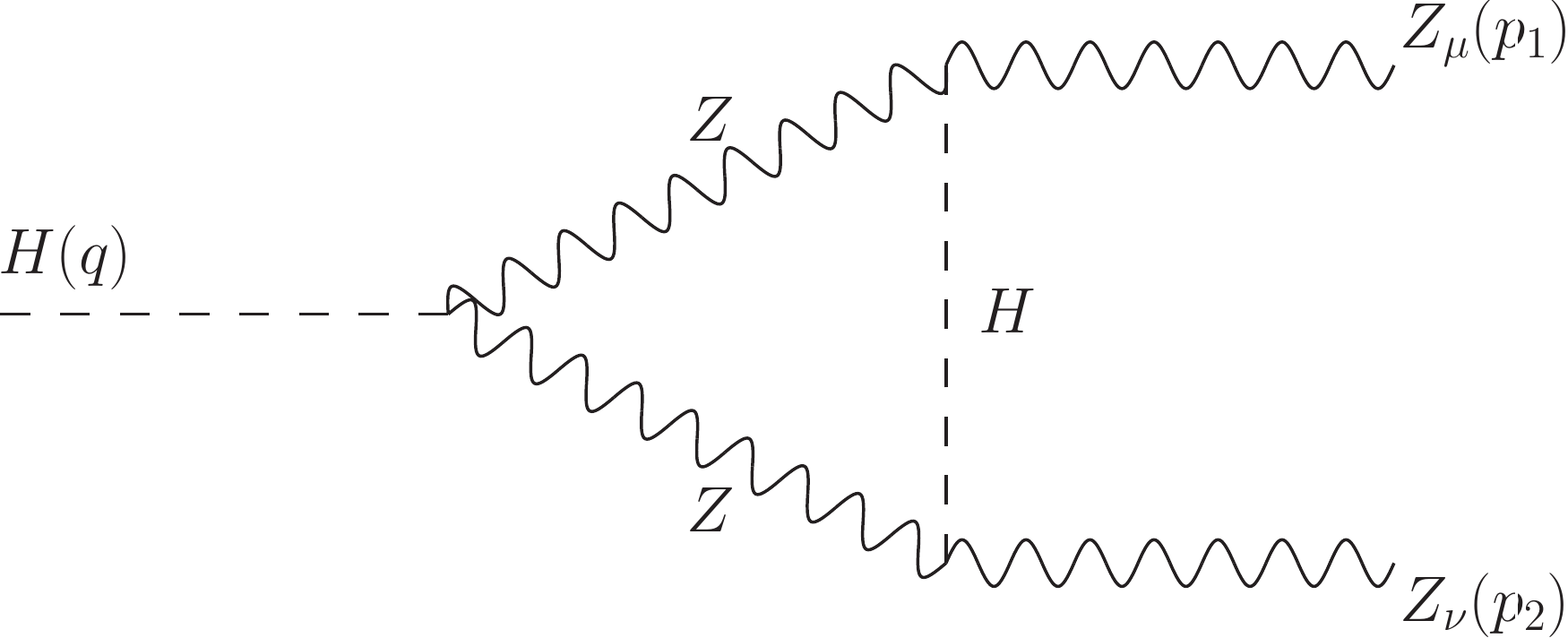}}
\subfigure[]{\includegraphics[width=5cm]{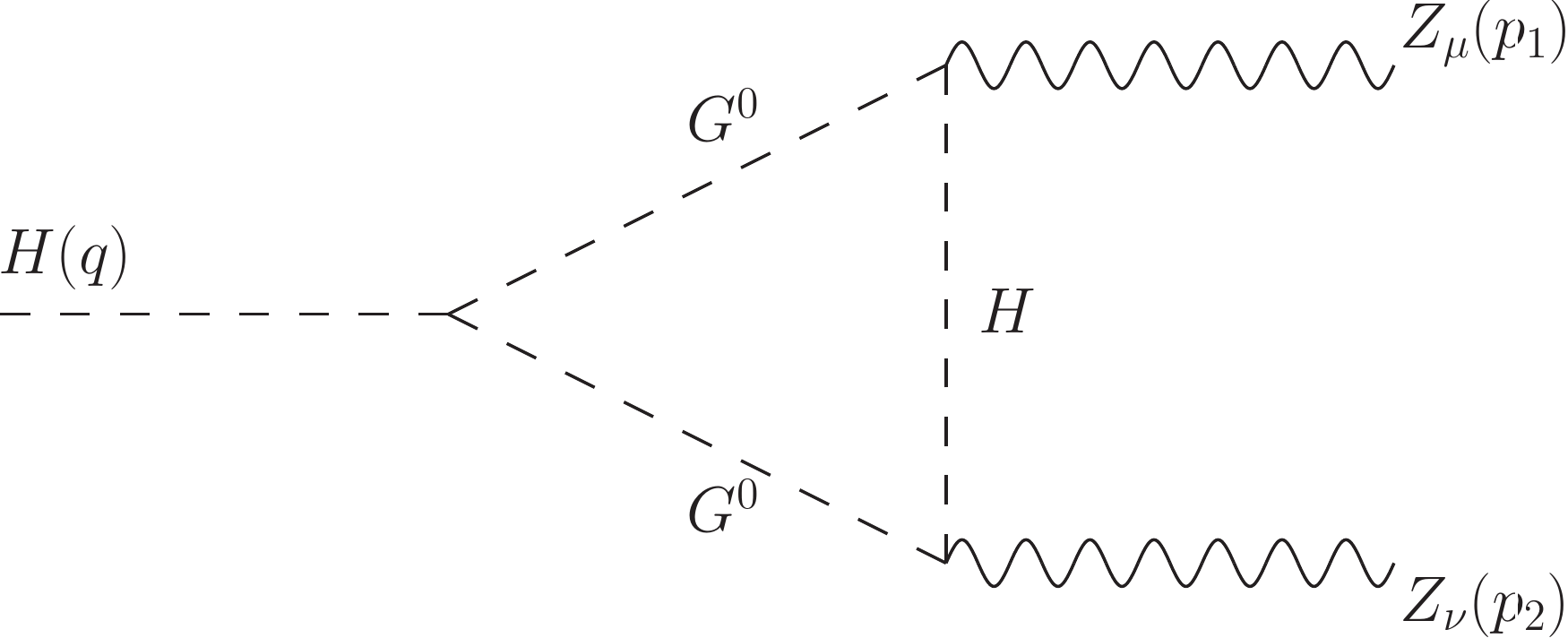}}
\subfigure[]{\includegraphics[width=5cm]{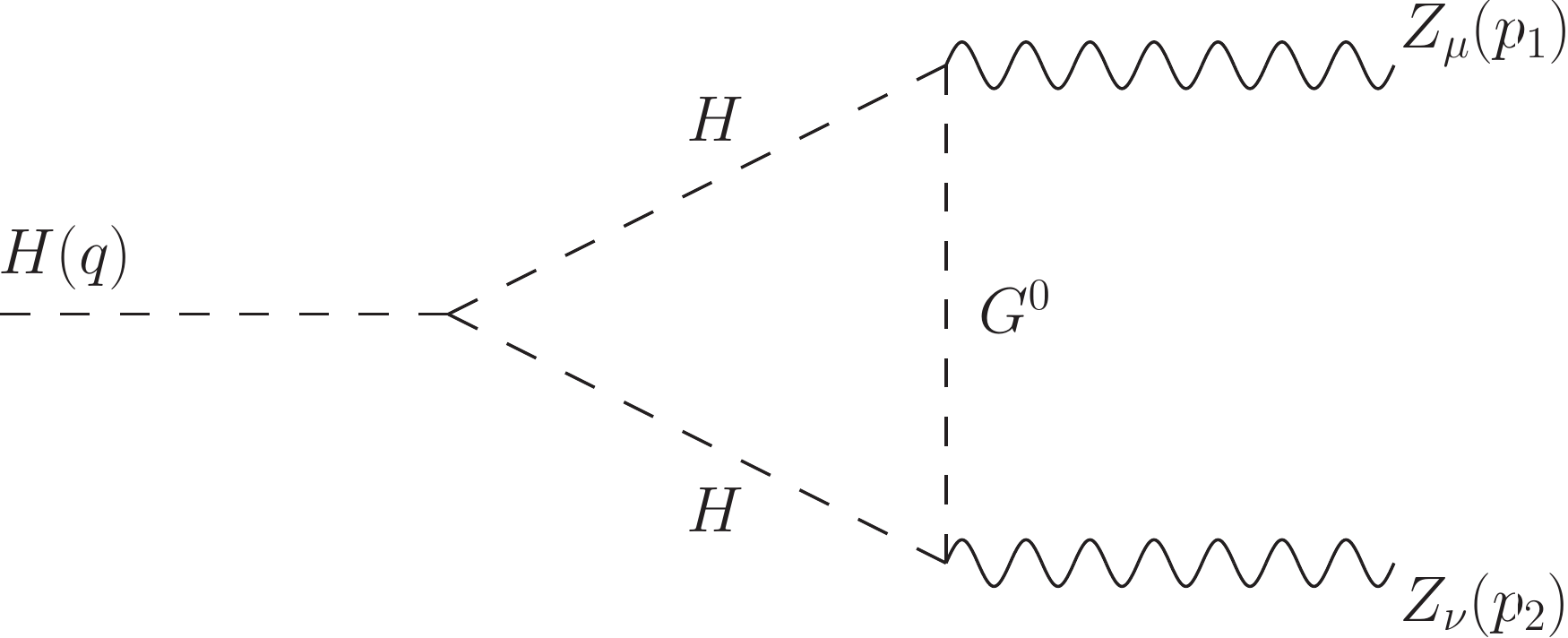}}
\subfigure[]{\includegraphics[width=5cm]{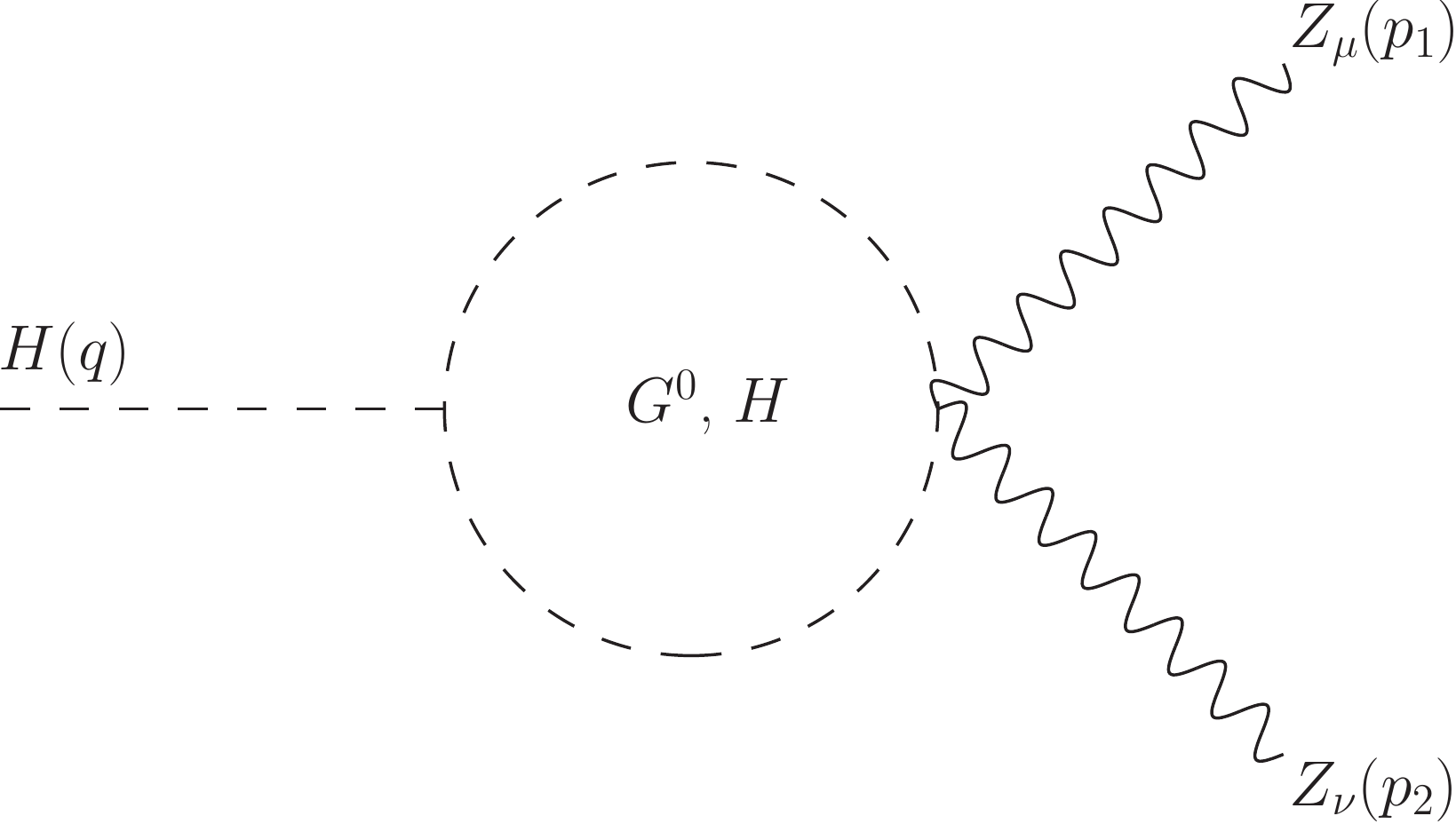}}
\subfigure[]{\includegraphics[width=5cm]{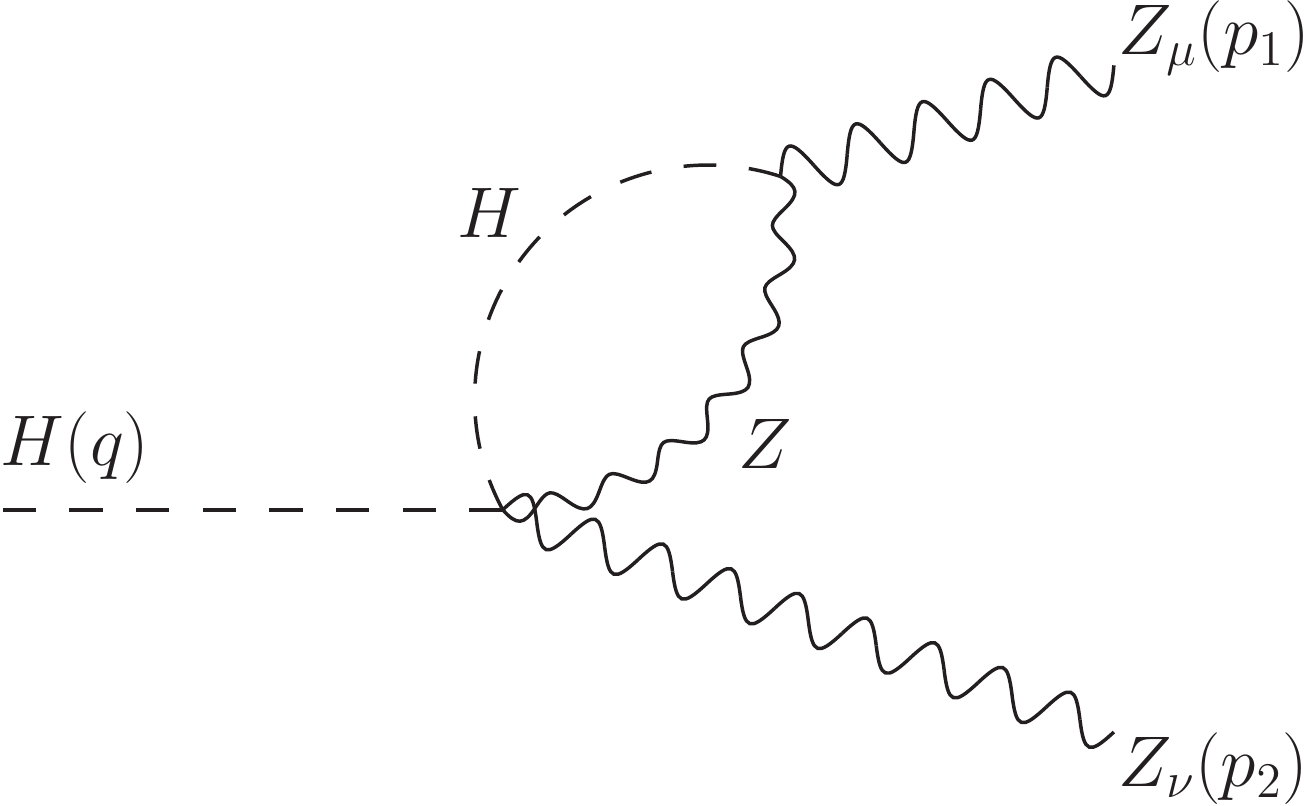}}
\subfigure[]{\includegraphics[width=5cm]{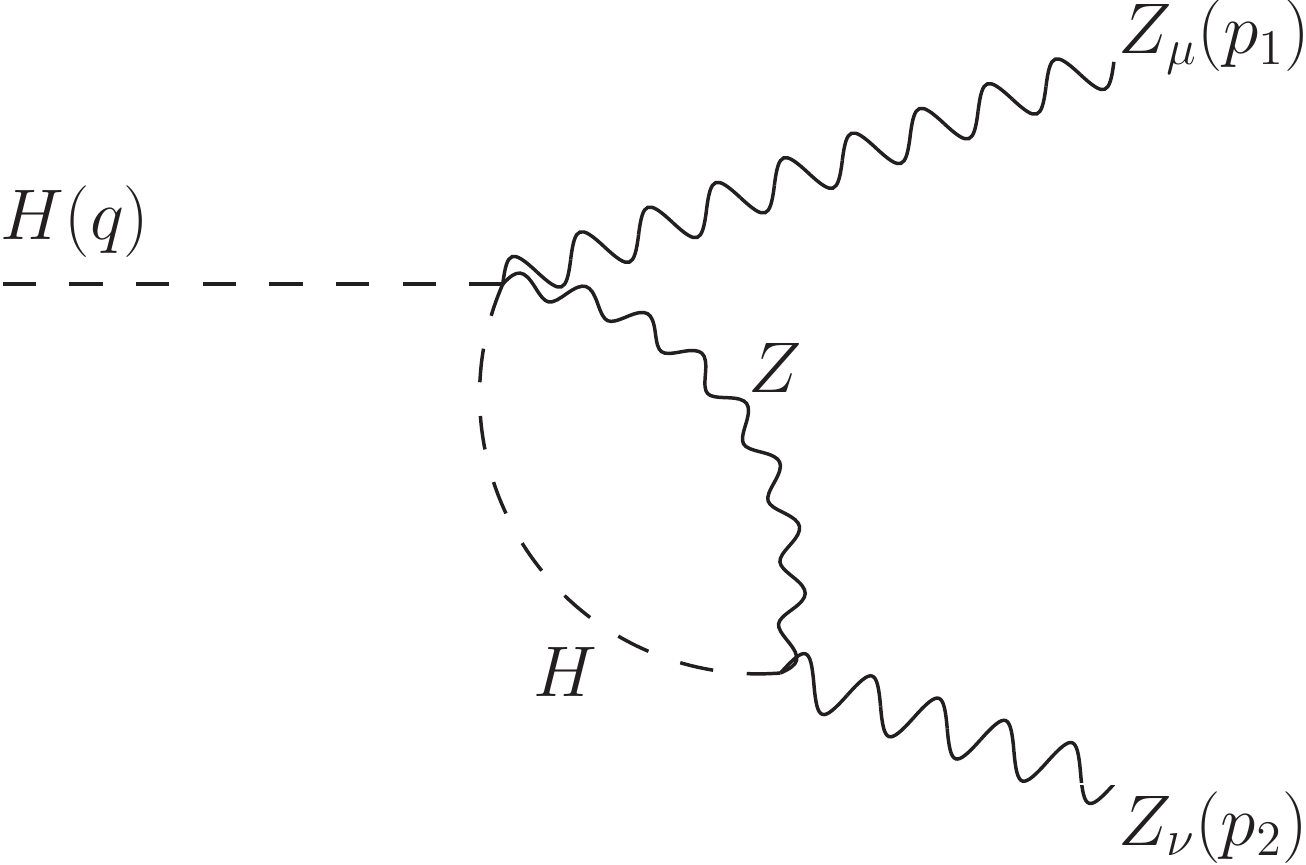}}
%\subfigure[]{\includegraphics[width=14cm]{diag3.eps}\label{Diagram3}}
\caption{The same as in Fig. \ref{Wc} but for the contribution of the Higgs boson and the $Z$ gauge boson and its associated unphysical particle.} \label{HZc}
\end{center}
\end{figure}

For the analytical calculation, we used the Mathematica  package FeynArts \cite{Hahn:2000kx}, which allows one to obtain the complete set of Feynman diagrams and their corresponding invariant amplitudes via the BFM. We then used the FeynHelpers and FeynCalc packages \cite{Mertig:1990an,Shtabovenko:2016sxi,Shtabovenko:2020gxv} to perform the tensor decomposition and obtain results  in terms of Passarino-Veltman scalar functions, which can be numerically evaluated via either the LoopTools \cite{Hahn:1998yk} or the  Collier \cite{Denner:2016kdg} packages. The  Feynman diagrams that contribute at the one-loop level to the $HZZ$ vertex can be classified into three types according to the particles circulating into the loop:
in Fig \ref{Fc} we show the fermion loop Feynman diagram, whereas the Feynman diagrams with $W$ gauge boson loops ($H-Z$ loops) are shown in   Fig. \ref{Wc}  (\ref{HZc}).  For simplicity,  these contributions will be denoted by the subscripts $\mathcal{F}$,  $\mathcal{W}$, and $\mathcal{HZ}$, respectively. 

Since the main aim of this work is the calculation of the SM contributions to the $HZZ$ anomalous couplings, we will focus on the $h_2^V$ form factor and use $a_Z=0$ throughout the rest of this work. Therefore, following the notation of Ref. \cite{Kniehl:1990mq} our results can be written as
\begin{align}
\label{h2}
h^V_2(q^2,p_1^2,m^2_Z)=m_Z\frac{g^2}{4 \pi^2c_W m_W}&\Big[\sum_f N_f m_f^2\big\{g_{Vf}^2 A^V_{Vf}(q^2,p_1^2,m^2_Z)+g^2_{Af}A^V_{Af}(q^2,p_1^2,m^2_Z)\big\} \\\nonumber &+ A^V_W(q^2,p_1^2,m^2_Z)+A^V_{ZH}(q^2,p_1^2,m^2_Z)\Big],
\end{align}
where 
$g_{Vf,Af}$ are the $Z$ gauge boson couplings to fermion pairs:
\begin{equation}
g_{Vf}=\frac{I_f}{2}-Q_f s_W^2\text{,} \quad\quad g_{Af}= \frac{I_f}{2},
\end{equation}
with $I_f$ and $Q_f$  the fermion weak isospin and electric charge. 

The analytic expressions for the $A^V_{Vf, Af}$,  $A^V_W$ and $A^V_{ZH}$ functions in terms of  Passarino-Veltman scalar functions are too lengthy and are presented in Appendix \ref{PassVel}. We  present explicit results for  the contributions to the $H^\ast ZZ$ ($p_1^2=m_Z^2$) and $HZZ^\ast$ ($q^2=m_H^2$) couplings. We note that for the $H^\ast ZZ$ coupling, our results agree with those reported in Ref. \cite{Kniehl:1990mq}, though there is an apparent change of sign in the coefficients of all the three-point scalar function $C_0$, which,  however, stems from the fact that there is an extra minus sign in the definition of such functions in \cite{Kniehl:1990mq} as compared to the usual definition presented in Appendix \ref{PassVel}. We would like to emphasize that, to our knowledge, the results for the  $HZZ^\ast$ ($q^2=m_H^2$) coupling have never been  reported in the literature. Thus, we present a more  comprehensive calculation, which could be helpful to assess the anomalous contributions to the $HZZ$ coupling in distinct scenarios. 
The Mathematica code for our analytical results  is available for the interested reader in \cite{urlcode}, where we also include  master formulas for the general case with three off-shell particles, which are too cumbersome to be reported in this work. The results presented in  Appendix \ref{PassVel} can be straightforwardly obtained from such master formulas.  It is interesting to note that  the  three contributions are free of ultraviolet divergences by their own.

\section{Behavior of the $HZZ$ coupling}\label{NA}

We now turn to present the numerical analysis. For the evaluation of the Passarino-Veltman scalar functions,  we used the LoopTools package \cite{Hahn:1998yk} and a cross-check was done via the Collier routines \cite{Denner:2016kdg}. A good agreement was found between both numerical evaluations. We will first study the behavior of the $H^\ast ZZ$ and $HZZ^\ast$ form factors in an energy region that allows for on-shell final bosons.

\subsection{$H^*ZZ$ form factor}
\begin{figure}[H]
\begin{center}
\subfigure[]{\includegraphics[width=9cm]{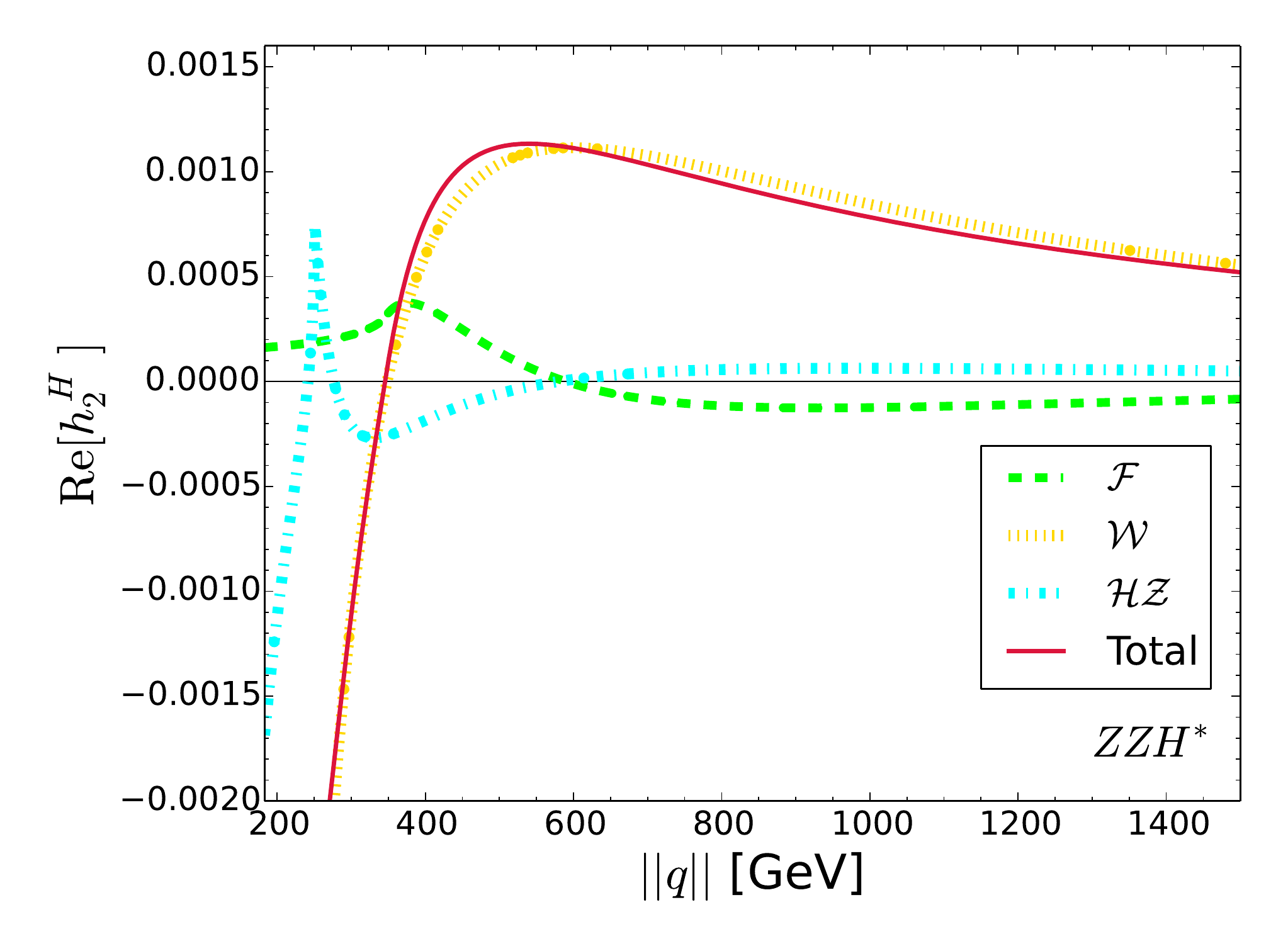}\label{plot1}}\hspace{-.2cm}
\subfigure[]{\includegraphics[width=9cm]{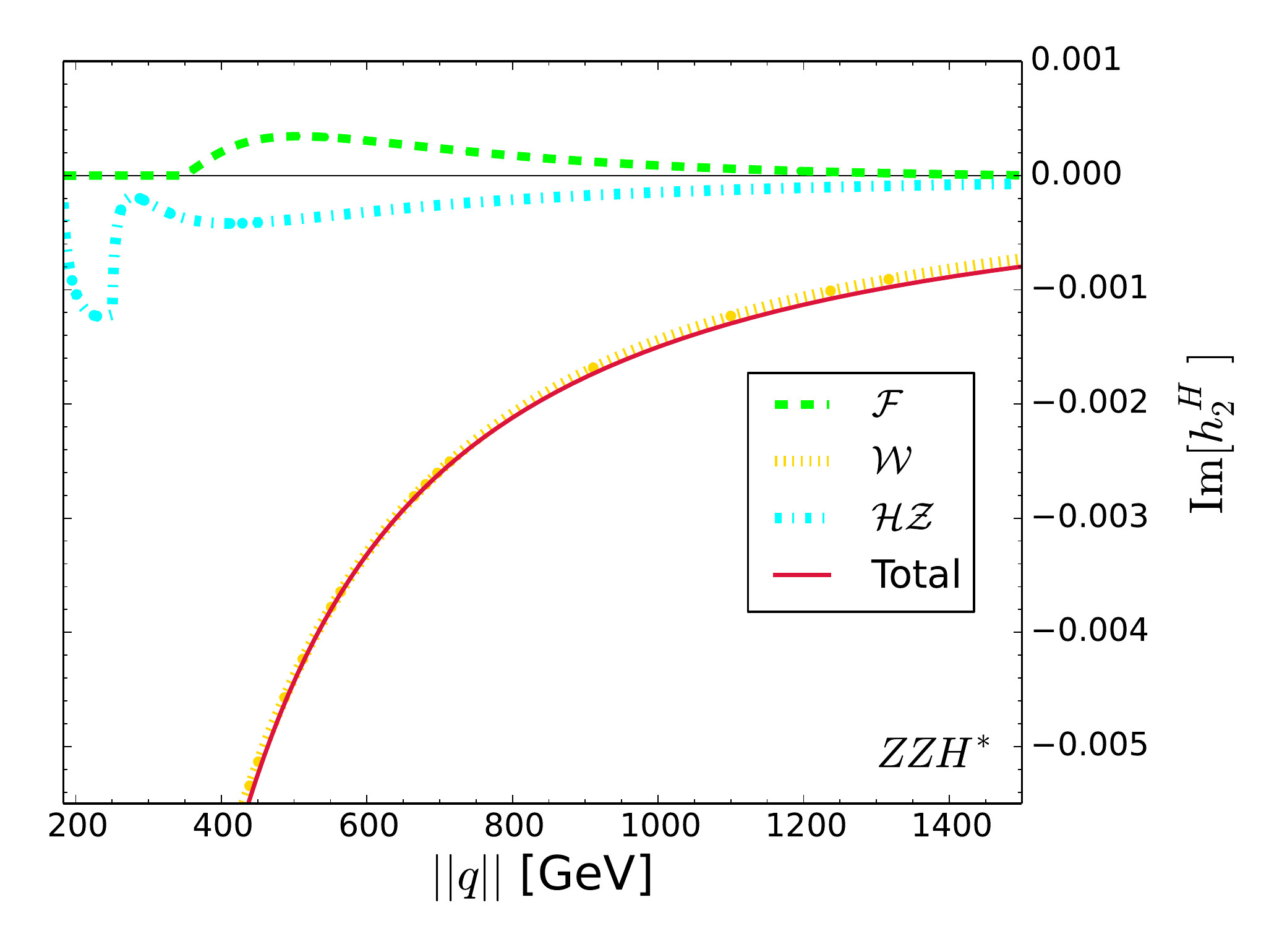}\label{plot2}}
%\subfigure[]{\includegraphics[width=14cm]{diag3.eps}\label{Diagram3}}
\caption{One-loop contributions to  the real (left plot) and absorptive (right plot) parts of the form factor $h_2^H$ as functions of the Higgs boson transfer momentum $\|q\|$: fermion ($\mathcal{F}$), $W$ gauge boson ($\mathcal{W}$),  $H$-$Z$ bosons ($\mathcal{HZ}$) and total contributions. \label{Figh2H}} 
\end{center}
\end{figure}

  We show in Figs. \ref{plot1} and \ref{plot2} the real and imaginary parts  of the fermion ($\mathcal{F}$), $W$ gauge boson ($\mathcal{W}$), and  $H$-$Z$ boson ($\mathcal{HZ}$) contributions, along with the total contributions to the $H^\ast ZZ$ form factor as functions of the Higgs boson transfer momentum $\|q\|$. As for the real part of $h_2^H$,  it is dominated by  the $\mathcal W$ contribution, whereas the $\mathcal{F}$ and $\mathcal{HZ}$ ones are sizeable at low energy only, where they are of similar order of magnitude than the one of the $\mathcal{W}$ contribution. Furthermore, from $\|q\|=300$ GeV onwards,   the $\mathcal{F}$ and $\mathcal{HZ}$  contributions are of similar size but  of opposite sign and tend to cancel each other out. It is also interesting to note that the fermion contribution is mainly dominated by the  top quark, whereas all other fermions yield negligible contributions. As far as the imaginary part of the  $h_2^H$ form factor is concerned,  we observe  that  the  $\mathcal{W}$ contribution is also the dominant one, whereas the $\mathcal{F}$  and $\mathcal{HZ}$ contributions  are one order of magnitude smaller. As expected,  the absorptive part of the $\mathcal{F}$ contribution  is non-vanishing only above the threshold $\|q\|=2m_t$,  where the two top quarks attached to the off-shell Higgs boson can be on-shell.  In general, the real and imaginary parts of $h_2^H$ are of similar order of magnitude, about $10^{-2}-10^{-3}$, but at high energies the absorptive part can be larger than the real part. This behavior was also  observed in other off-shell  form factors: see for instance \cite{Gounaris:2000tb,Hernandez-Juarez:2020drn,Hernandez-Juarez:2020gxp,Hernandez-Juarez:2021mhi,Hernandez-Juarez:2022kjx}.

For illustration purposes, the values of the  real and imaginary parts of the $h_2^H$  form factors at a few $\|q\|$ values are presented in Table \ref{TabNum}, along with the values for $\hat{b}_Z$, $a_2^{ZZ}$ and $c_{zz}$, which can be obtained from Eqs. \eqref{H22}, \eqref{h222} and \eqref{czzEq}, respectively.
In effective field theories, the couplings are taken as constant and do not depend on $\|q\|$, but our results can be useful to constrain the energy scale $\Lambda$ of the model.

\begin{table}[H]
\begin{center}
  \caption{Total contributions to the $h_2^H$ form factor for a few values of $\|q\|$. The respective values of  $\hat{b}_Z$, $a_{2}^{ZZ}$ and $c_{zz}$ are also shown. All these results are  in units of $10^{-3}$. \label{TabNum}}
  \begin{tabular}{ccccc}
\hline \hline
  $\big\| q \big\|$ & $ h_2^{H}$ & $\hat{b}_Z$ & $a_{2}^{ZZ}$ & $c_{zz}$  \\
  \hline
  190&$-12.99-14.02\ i$&$-6.49-7.01\ i$&$12.99+14.02\ i$&$-48.16-51.98\ i$\\
  220&$ -6.82 -13.86\ i$&$ -3.42-6.93\ i$ &$6.82+13.86\ i$&$ -25.3-51.4\ i$\\
350  &$0.09  -7.77\ i$&$0.04-3.88\ i$ & $-0.09+7.77\ i$&$0.35-28.8\ i$   \\
450&$1.02-5.24\ i$&$0.51-2.62\ i$&$-1.02+5.24\ i$&$3.81-19.43\ i$\\
600&$1.11-3.31\ i$&$0.55-1.65\ i$&$-1.11+3.31\ i$&$ 4.12-12.29\ i$\\
1000&$0.78-1.5\ i$&$0.39-0.75\ i$&$-0.78+1.5\ i$&$2.9-5.56\ i$\\
1500&$0.52-0.79\ i$&$0.26-0.39\ i$&$-0.52+0.79\ i$&$1.92-2.95\ i$\\
  \hline\hline
\end{tabular}
\end{center}
\end{table}

\subsection{$HZZ^*$ form factor}

\begin{figure}[H]
\begin{center}
\subfigure[]{\includegraphics[width=9cm]{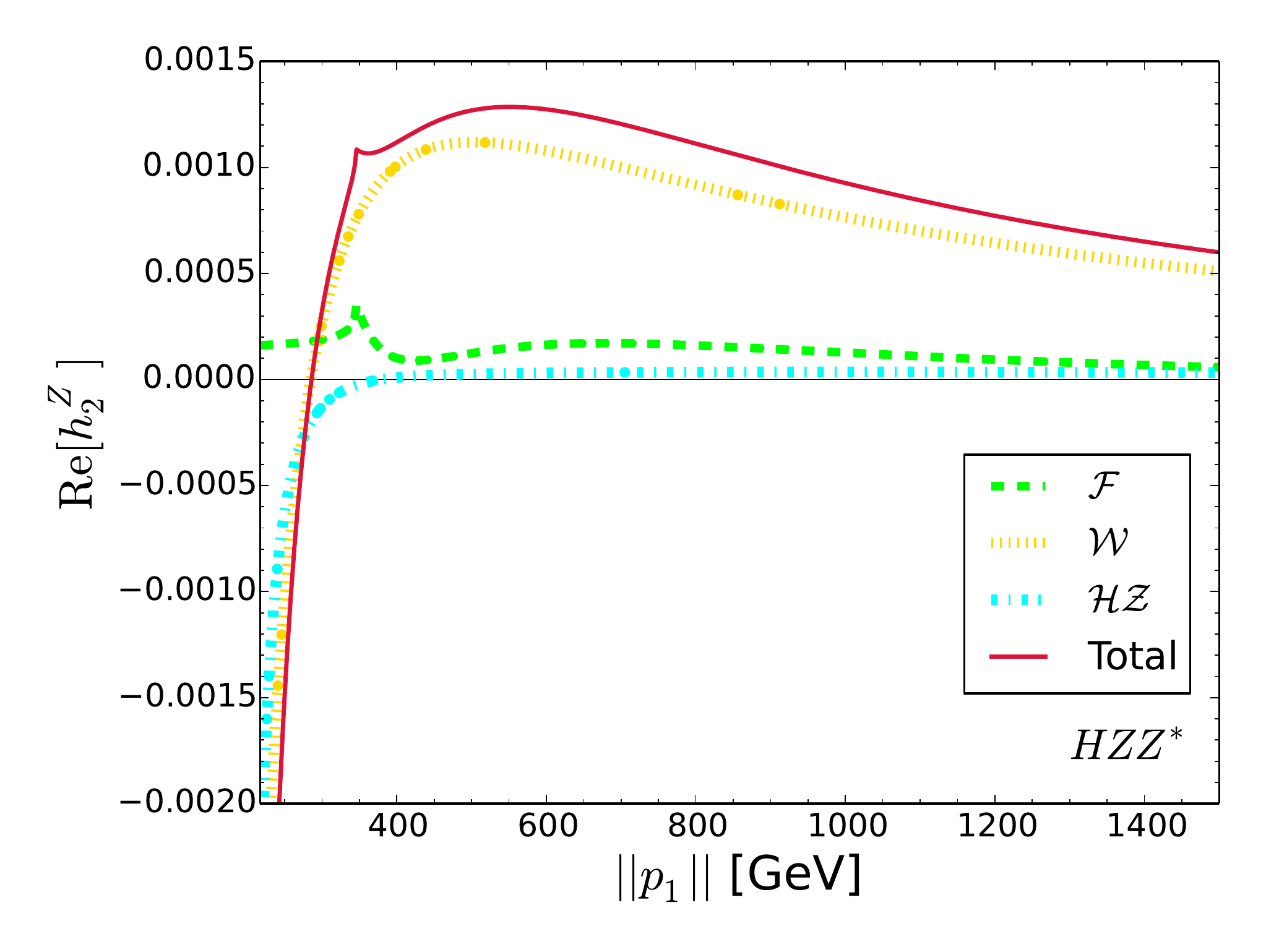}}\hspace{-.2cm}
\subfigure[]{\includegraphics[width=9cm]{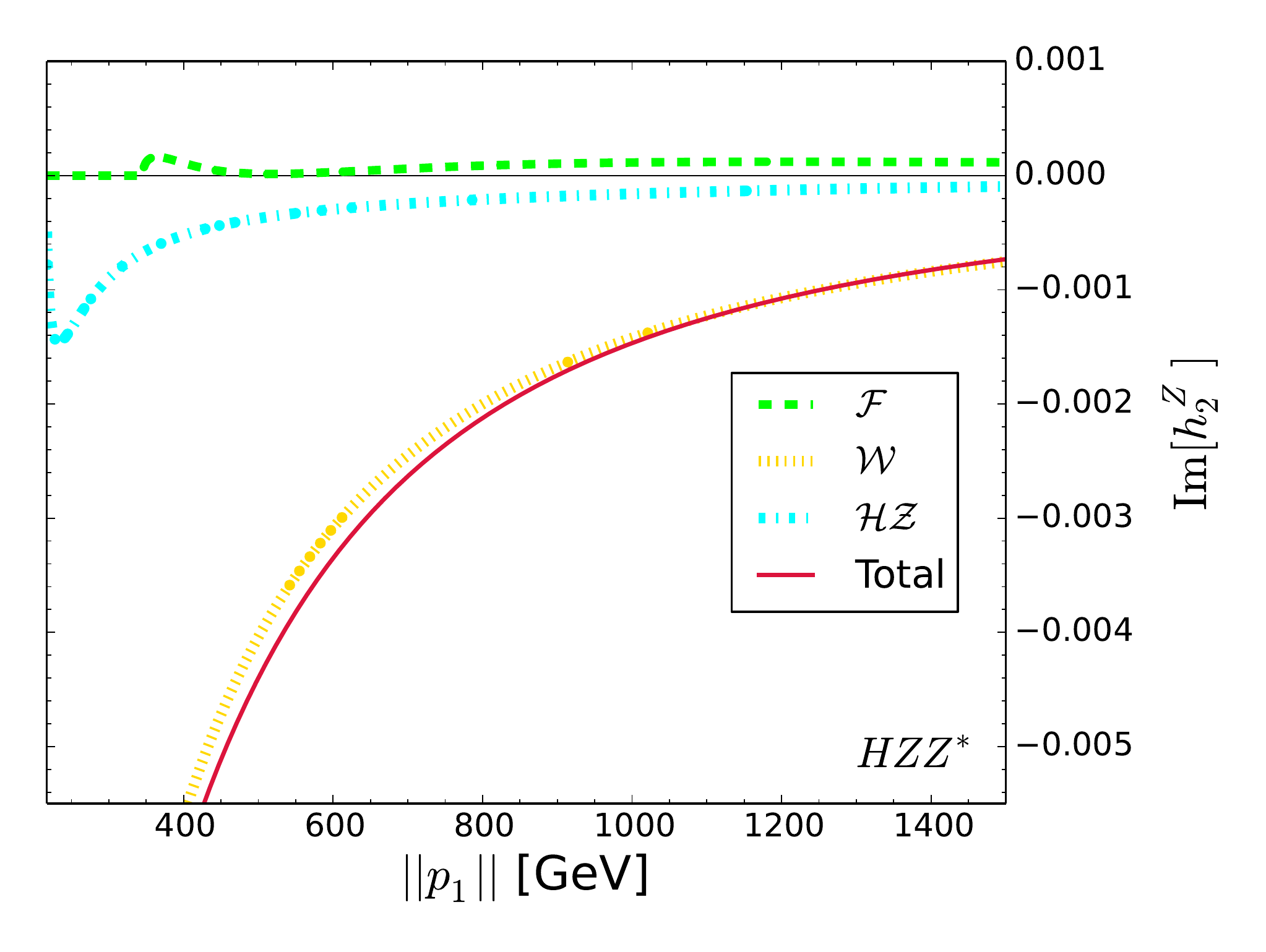}}
%\subfigure[]{\includegraphics[width=14cm]{diag3.eps}\label{Diagram3}}
\caption{The same as in Fig. \ref{Figh2H}, but for the form factor $h_2^Z$ as a function of off-shell $Z$ gauge boson transfer momentum $\|p_1\|$. \label{Plot22}}
\end{center}
\end{figure}

We now show in Fig. \ref{Plot22} the behavior of the real and imaginary parts of the partial and total contributions to the $HZZ^\ast$ form factor  as functions of the off-shell $Z$ gauge boson transfer momentum $\|p_1\|$. We observe that the $h_2^Z$  form factor has a similar  behavior to that of  $h_2^H$. In both cases the $\mathcal{W}$ contribution dominates, but at  low energies, the $\mathcal{F}$ and $\mathcal{HZ}$ contributions may be  of similar order of magnitude than the $\mathcal{W}$ contribution, whereas  at high energies they are negligible. 
%However, there is a slight difference with the behavior of $h_2^H$, since  the values of both the real and imaginary parts are different for $h_2^Z$ than those of $h_2^H$. 

In Table \ref{TabNum1}, we present numerical values for the real and absorptive parts of $h_2^Z$ at a few values of  $\|p_1\|$. We also present the respective values for $\hat{b}_Z$, $a_2^{ZZ}$ and $c_{zz}$. We can observe that at some energy values,  the $h_2^Z$ form factor is larger than the $h_2^H$ one, reaching values of the order of $10^{-2}-10^{-3}$. It is also noted that for both $H^* ZZ$ and $HZZ^*$ couplings, the real and absorptive parts of the anomalous coupling $\hat{b}_Z$ are of the order of $10^{-3}-10^{-4}$, which agrees with \cite{Rao:2020hel}. On the experimental side, the current bounds on $\hat{b}_Z$ are of the same order of magnitude, and thus this anomalous coupling could be  at reach of the LHC in the near future.

\begin{table}[H]
\begin{center}
  \caption{The same as in Table \ref{TabNum}, but for  the form factor $h_2^Z$ for as a function of $\|p_1\|$.  }\label{TabNum1}
  \begin{tabular}{ c   c c c c}
\hline \hline
  $\big\| p_1 \big\|$ &$h_2^{Z}$ & $\hat{b}_Z$ & $a_{2}^{ZZ}$ & $c_{zz}$  \\
  \hline
  220&$ -4.84 -15.16\ i$& $2.42+7.58\ i$ & $4.84+15.16\ i$&$ 17.93+56.19\ i$\\
350  &$1.07 -7.37\ i$&$-0.53+3.68\ i$ & $-1.07+7.37\ i$&$-3.98+27.33\ i$   \\
450&$1.21-5.11\ i$&$-0.6+2.55\ i$&$-1.21+5.11\ i$&$-4.49+18.95\ i$\\
600&$1.27-3.35\ i$&$-0.63+1.67\ i$&$-1.27+3.35\ i$&$ -4.72+12.41\ i$\\
1000&$0.92-1.46\ i$&$-0.46+0.73\ i$&$-0.92+1.46\ i$&$-3.43+5.43\ i$\\
1500&$0.59-0.73\ i$&$-0.29+0.36\ i$&$-0.59+0.73\ i$&$-2.22+2.71\ i$\\
  \hline\hline
\end{tabular}
\end{center}
\end{table}

\section{Bounds on the  $H^*ZZ$ anomalous couplings}\label{BOTAC}

Until now we have not focused on the $\hat{c}_Z$ and $\widetilde{b}_Z$ anomalous couplings yet. Limits on their real and absorptive parts have been set in the past \cite{Rao:2020hel}, though their energy dependence  has not been considered. Also, the current CMS bounds \cite{CMS:2022ley} only consider constraints on the ratios of such couplings, which cannot be used to assess the corresponding contributions to physical observables \cite{Rao:2020hel}. These couplings are expected to play a relevant role in LHC processes given the recent measurement of the $H^\ast\rightarrow ZZ$ process, therefore, independent constraints on each one of these anomalous  couplings are necessary and we will address this issue below.

\subsection{Bounds from LHC data}

By combining  the LHC bounds on the effective ratios $f_{ai}$ of Table \ref{TabBound1} through their definitions of Eq. \eqref{aiaj} \cite{CMS:2022ley}, it is possible to  obtain limits on the ratios $a_{i}^{ZZ}/a_{2}^{ZZ}$, which   are presented in Table \ref{Tabaia2}, where we only consider the scenario with $\Gamma_H=\Gamma_H^{SM}$ (the bounds  obtained in the unconstrained case are of a similar order of magnitude). In Table \ref{Tabaia2} we also present the bounds  in terms of the $\hat{b}_Z$, $\hat{c}_Z$ and $\widetilde{b}_Z$ form factors, which can be easily obtained via the mappings of Table \ref{mappings}. In the following, we will use the parametrization of Eq. \eqref{Lag2}, but our bounds can be easily translated into the remaining parametrizations discussed in Sec. \ref{theofram}.

\begin{table}[!hbt]
  \centering 
  \caption{Allowed intervals at 95 \% CL for the ratios defined in Eq. \eqref{aiaj}, where we consider the case $\Gamma_H=\Gamma_H^{SM}$ and the limits of Table \ref{TabBound1}. }\label{Tabaia2}
  \begin{tabular}{ c  c    }
\hline \hline
% after \\ : \hline or \cline{col1-col2} \cline{col3-col4} ...
  \text{Ratio} & \text{Allowed values}  \\
  \hline
  $a_3^{ZZ}/a_2^{ZZ}$&$\big[-1.84,0.70\big]$\\
  $\kappa_1^{ZZ}/a_2^{ZZ}$& $\big[-0.35,0.18\big]$ \\
$\widetilde{b}_Z/\hat{b}_Z$&$\big[-1.84,0.70\big]$\\
  $\hat{c}_Z/\hat{b}_Z$& $\big[-0.36,0.70\big]$ \\
  \hline\hline
\end{tabular}
\end{table}

We now use the constraints of Table \ref{Tabaia2} and the expressions for $\hat{b}_Z$ from Sec. \ref{AnRe} to find the allowed areas of the real parts of $\widetilde{b}_Z$ and $\hat{c}_Z$ as functions of the Higgs boson transfer momentum. Furthermore, since the results of Table \ref{TabNum} hint that the real and absorptive parts of the $H^*ZZ$  form factors are of similar size, we will assume that the limits of Table  \ref{Tabaia2} are also valid for the imaginary parts of $\widetilde{b}_Z$ and $\hat{c}_Z$, which  allows us to set constraints on ${\rm Im}\big[\widetilde{b}_Z\big]$ and ${\rm Im}\big[\hat{c}_Z\big]$.  For our calculations, we use the expressions of the form factor $h_2^H=2\hat{b}_Z$, but the same results are expected for  $h_2^Z$. Moreover, only energy regions where both $Z$ gauge bosons can be on-shell are considered in our analysis.

\begin{figure}[!hbt]
\begin{center}
\subfigure[]{\includegraphics[width=8cm]{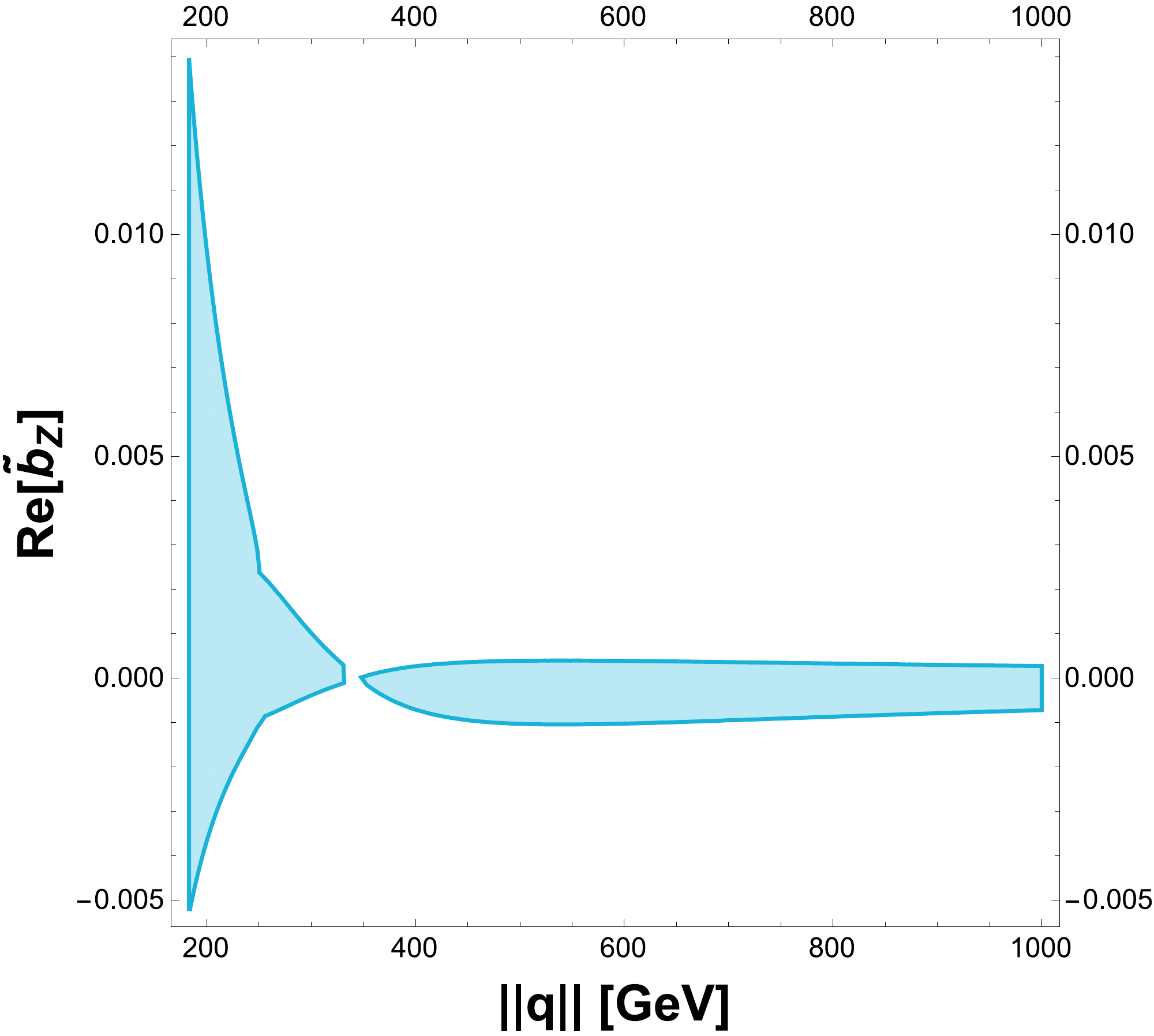}\label{limit1}}
\subfigure[]{\includegraphics[width=8cm]{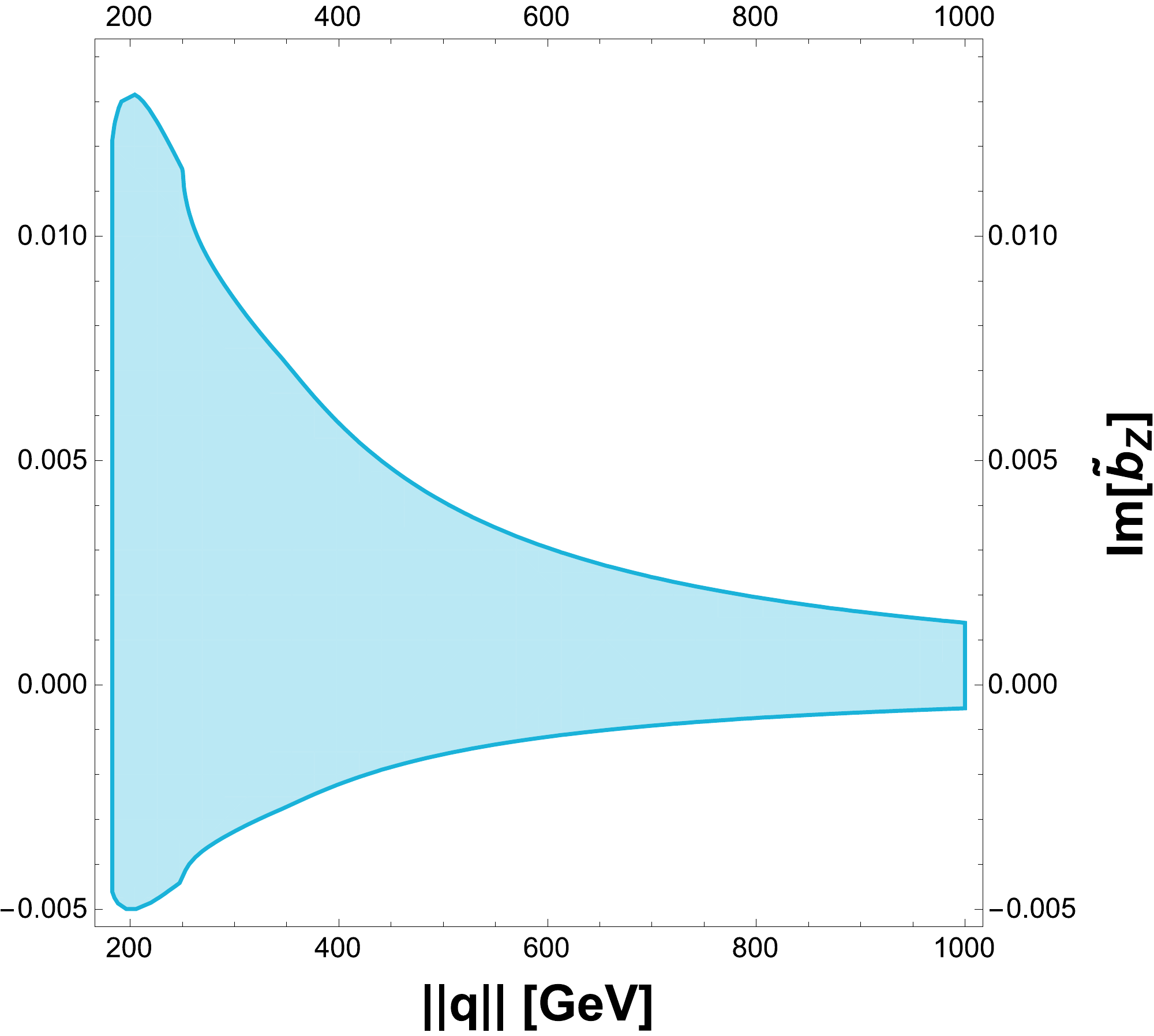}\label{limit2}}
%\subfigure[]{\includegraphics[width=14cm]{diag3.eps}\label{Diagram3}}
\caption{Allowed area  at 95\% CL for the the real (left plot) and absorptive  (right plot) parts of $\widetilde{b}_Z$ as  functions of $\|q\|$. These results are compatible with the bounds of Table  \ref{Tabaia2}.\label{a3ZZlimits}} 
\end{center}
\end{figure}

Our results for the allowed area on the $\|q\|$ vs ${\rm Re}\big[\widetilde{b}_Z\big]$ plane are shown in Fig. \ref{limit1}.  It is observed that for a fixed small value of $\|q\|$, the allowed   ${\rm Re}\big[\widetilde{b}_Z\big]$ values lie in the  $[-10^{-3},10^{-2}]$ interval, which becomes narrower as  $\|q\|$ increases: at high  $\|q\|$,  ${\rm Re}\big[\widetilde{b}_Z\big]$ is allowed to have values of the order of $10^{-4}$. It is also worth noting that around $\|q\|=320$ GeV,  the allowed  ${\rm Re}\big[\widetilde{b}_Z\big]$ area collapses, which stems from the fact that in such a region the real part of $\hat{b}_Z$ changes  sign and  ${\rm Re}\big[\hat{b}_Z\big]\approx 0$ [see Fig. \ref{plot1}]. Therefore, very small values of ${\rm Re}\big[\widetilde{b}_Z\big]$ are required to get $\widetilde{b}_Z/\hat{b}_Z$ inside the allowed region. 
 As for ${\rm Im}\big[\widetilde{b}_Z\big]$, the corresponding limits are shown in Fig. \ref{limit2}, where we observe that for small  $\|q\|$ there is a relatively large allowed area, which becomes smaller as  $\|q\|$ increases, with the allowed ${\rm Im}\big[\widetilde{b}_Z\big]$ values being of the order of $10^{-2}-10^{-3}$. In this case, there is no abrupt collapse of the allowed area  as  the absorptive part of $\widetilde{b}_Z$ does not change the sign in this energy range. The current bounds on the real and absorptive parts of  $\widetilde{b}_Z$ are of the order of $10^{-3}$ \cite{Rao:2020hel}, therefore  our limits obtained for high energies are more stringent. 

\begin{figure}[!hbt]
\begin{center}
\subfigure[]{\includegraphics[width=8cm]{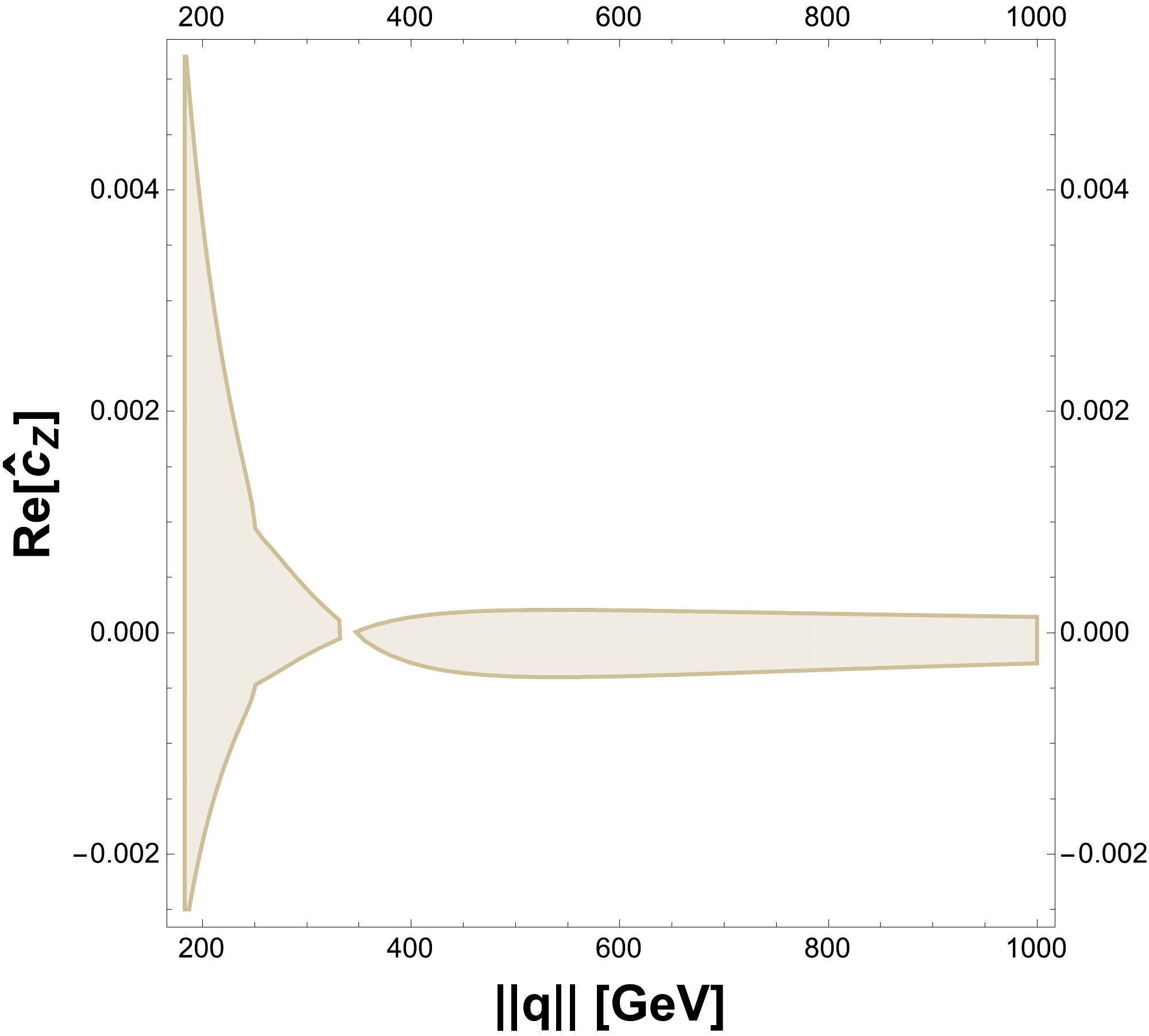}\label{limit3}}
\subfigure[]{\includegraphics[width=8cm]{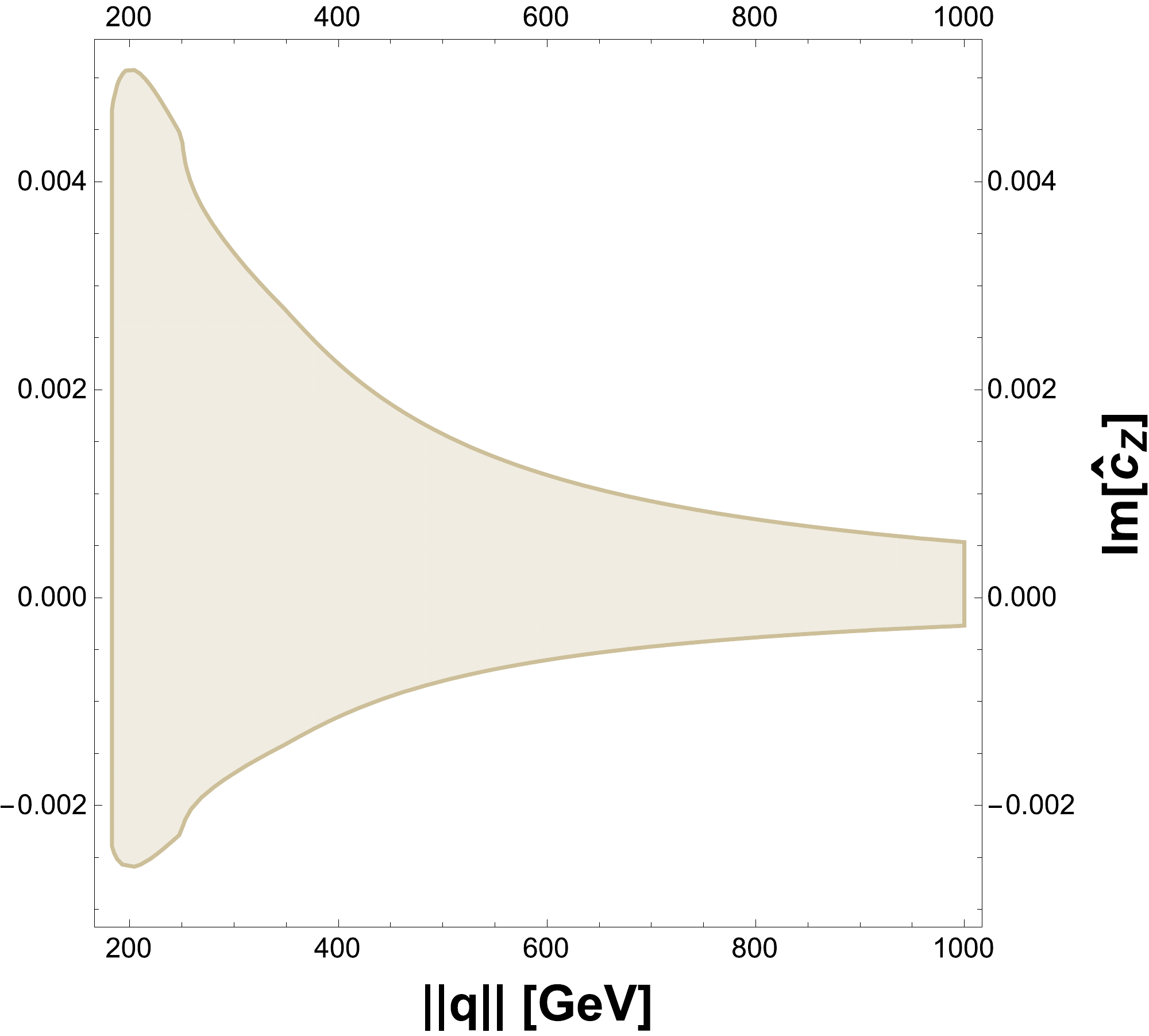}\label{limit4}}
%\subfigure[]{\includegraphics[width=14cm]{diag3.eps}\label{Diagram3}}
\caption{The same as in Fig. \ref{a3ZZlimits}, but for the $\hat{c}_Z$ form factor. \label{k1ZZlimits}}
\end{center}
\end{figure}

We now show the allowed areas of the real and absorptive parts of $\hat{c}_Z$ as functions of $\|q\|$ in Figs. \ref{limit3} and \ref{limit4}, respectively. The results are  similar to those obtained for $\widetilde{b}_Z$. Nevertheless, in this case, the  allowed upper values of both the real and imaginary parts of $\hat{c}_Z$  are of the order of $10^{-3}$ at low energies, which is one order of magnitude smaller than previous results \cite{Hagiwara:2000tk}, and become smaller as the energy increases. It is noted that in general, the constraints on the $CP$ violating form factor $\widetilde{b}_Z$ are less tight than those on the $CP$ conserving one  $\hat{c}_Z$, although  both can be of the same order of magnitude in some energy regions.

As already mentioned, our constraints on  $\widetilde{b}_Z$ and $\hat{c}_Z$ can be translated into other parametrizations via the mappings of Table \ref{mappings}. Therefore, we can easily obtain the allowed regions for the couplings  $a_3^{ZZ}$, $\kappa_1^{ZZ}$, $\widetilde{c}_{zz}$ and $c_{z\Box}$, which are used in other analyses of  the $HZZ$ vertex. For instance, in the SMEFT the anomalous couplings do not depend on the  off-shell boson transfer momenta \cite{Degrande:2012wf}, and  the scale $\Lambda$, where the effective model remains valid, has been absorbed in the definition of the  Wilson coefficients $\widetilde{c}_{zz}$ and $c_{z\Box}$. Hence, our results may be used to set a bound on  $\Lambda$. The corresponding limits are presented in Tables \ref{b1} and \ref{b3}. We find that the allowed values of the real and imaginary parts of the $CP$ violating form factors $a_3^{ZZ}$ and $\widetilde{c}_{zz}$ are in general of the order of $10^{3}-10^{-4}$, but they can be as large as $10^{-2}$ for some  $\|q\|$ values.  On the other hand, the  bounds on the $CP$ conserving form factor $k_1^{ZZ}$ have  similar behavior  to the ones obtained for $\hat{c}_Z$, and the allowed values of this form factor can be as large as $10^{-3}$ at most, whereas the SMEFT coefficient $c_{z\Box}$ can be in general of the order of $10^{-3}-10^{-4}$. All these values decrease at high energies.
 
\begin{table}[!hbt]
\centering  \caption{Allowed intervals of the real and absorptive  parts of the $CP$ violating form factor of the $H^*ZZ$ coupling for a few values of the transfer momentum. We consider the parametrizations of Eq. \eqref{Lag2} ($\widetilde{b}_Z$), the LHC framework ($a_3^{ZZ}$) and the SMEFT ($\widetilde{c}_{zz}$).\label{b1}}
\resizebox{17cm}{!} {
  \begin{tabular}{ c  c  c  c   c c c}
\hline \hline
  $\big\| q \big\|$ & ${\rm Re}\big[\widetilde{b}_Z\big]$&${\rm Re}\big[ a_3^{ZZ}\big]$ &  ${\rm Re}\big[\widetilde{c}_{zz}\big]$& ${\rm Im}\big[ \widetilde{b}_Z\big]$ & ${\rm Im}\big[a_3^{ZZ}\big]$& ${\rm Im}\big[\widetilde{c}_{zz}\big]$ \\
  \hline
  190&$\big[-0.0045,0.012\big]$&$\big[-0.024,0.009\big]$&$\big[-0.033,0.088\big]$&$\big[-0.005,0.013\big]$&$\big[-0.026,0.01\big]$&$\big[-0.037,0.096\big]$\\
  285&$\big[-0.00055,0.0014\big]$& $\big[-0.0029,0.0011\big]$ &  $\big[-0.004,0.01\big]$&$\big[-0.0034,0.009\big]$&$\big[-0.018,0.0069\big]$&$\big[-0.025,0.066\big]$\\
400  &$\big[-0.0007,0.00026\big]$&$\big[-0.00053,0.0014\big]$& $\big[-0.0051,0.0019\big]$&$\big[-0.0022,0.006\big]$&$\big[-0.012,0.0044\big]$&$\big[-0.016,0.044\big]$  \\
800&$\big[-0.0009,0.00034\big]$&$\big[-0.00069,0.0018 \big]$&$\big[-0.0066,0.0025\big]$&$\big[-0.00075,0.0019\big]$& $\big[-0.0039,0.0015\big]$ &  $\big[-0.0055,0.014\big]$\\
1500&$\big[-0.00047,0.00018\big]$&$\big[-0.00036,0.00095\big]$&$\big[-0.0034,0.0013\big]$&$\big[-0.00028,0.00075\big]$&$\big[-0.0015,0.00057\big]$& $\big[-0.002,0.0055\big]$\\
  \hline\hline
\end{tabular}
}
\end{table}

\begin{table}[H]
  \centering 
  \caption{ Allowed intervals for the real and absorptive parts of one of the $CP$ conserving form factors of the $H^*ZZ$ coupling for a few values of the transfer momentum. We consider the parametrizations of Eq. \eqref{Lag2} ($\hat{c}_Z$), the LHC framework ($\kappa_1^{ZZ}$) and the SMEFT ($\widetilde{c}_{z\Box}$). From Eq. \eqref{chat} we note that $\kappa_1^{ZZ}=\hat{c}_Z$.\label{b3}}
    \resizebox{17cm}{!} {
  \begin{tabular}{ c  c  c c cc}
\hline \hline
% after \\ : \hline or \cline{col1-col2} \cline{col3-col4} ...
  $\big\| q \big\|$ & ${\rm Re}\big[ \hat{c}_Z\big]$ $\big({\rm Re}\big[k_1^{ZZ} \big]\big)$ & ${\rm Re}\big[c_{z\Box}\big]$ & ${\rm Im}\big[\hat{c}_Z \big]$ $\big({\rm Im}\big[k_1^{ZZ}\big]\big)$ & ${\rm Im}\big[c_{z\Box}\big]$\\
  \hline
  190&$\big[-0.0024, 0.0046\big]$&$\big[-0.0058,0.011\big]$&$\big[-0.0026, 0.005\big]$&$\big[-0.0063,0.012\big]$\\
  285& $\big[-0.00028, 0.00055\big]$ & $\big[-0.00068,0.0013\big]$& $\big[-0.0018, 0.0035\big]$ & $\big[-0.0043,0.0085\big]$\\
400  &$\big[-0.00027, 0.00014\big]$& $\big[-0.00065,0.00034\big]$ &$\big[-0.0012, 0.0023\big]$& $\big[-0.0029,0.0055\big]$ \\
800&$\big[-0.00034, 0.00017 \big]$&$\big[-0.00082,0.00041\big]$&$\big[-0.00038, 0.00075 \big]$&$\big[-0.00092,0.0018\big]$\\
1500&$\big[-0.00019, 0.0001\big]$&$\big[-0.00046,0.00024\big]$&$\big[-0.00015, 0.00029\big]$&$\big[-0.00036,0.0007\big]$\\
  \hline\hline
\end{tabular}
}
\end{table}

\subsection{Bounds from weak dipole moments}
$CP$ violating effects can be severely constrained by limits on the electric dipole moments of the electron and neutron. Nevertheless, the $HZZ$ coupling does not contribute to these observable quantities, which can be induced, for instance, by anomalous contributions to the $H\gamma\gamma$ and $HZ\gamma$ couplings \cite{Chang:1998uc,Chang:2002ex,Arkani-Hamed:2004zhs}. On the other hand, the  $HZZ$ vertex does contribute to the weak magnetic and weak electric dipole moments (WMDM and WEDM) of  fermions, which are determined by the anomalous contributions to the $Zf\bar{f}$ vertex function \cite{Arroyo-Urena:2016ygo}
\begin{align}
\Gamma^\mu\subset\,ie\left(i\sigma^{\mu\nu}p_\nu F_2(p^2)+F_3(p^2)\sigma^{\mu\nu}p_\nu\gamma^5\right),
\end{align}
with  the  WMDM and  WEDM being given by $a^W_f=-2m_f F_2(m_Z^2)$ and $d^W_f=-e F_3(m_Z^2)$. The latter is $CP$ violating and can be induced up to the three-loop level in the SM.

\begin{figure}[!hbt]
\begin{center}
\subfigure[]{\includegraphics[width=6cm]{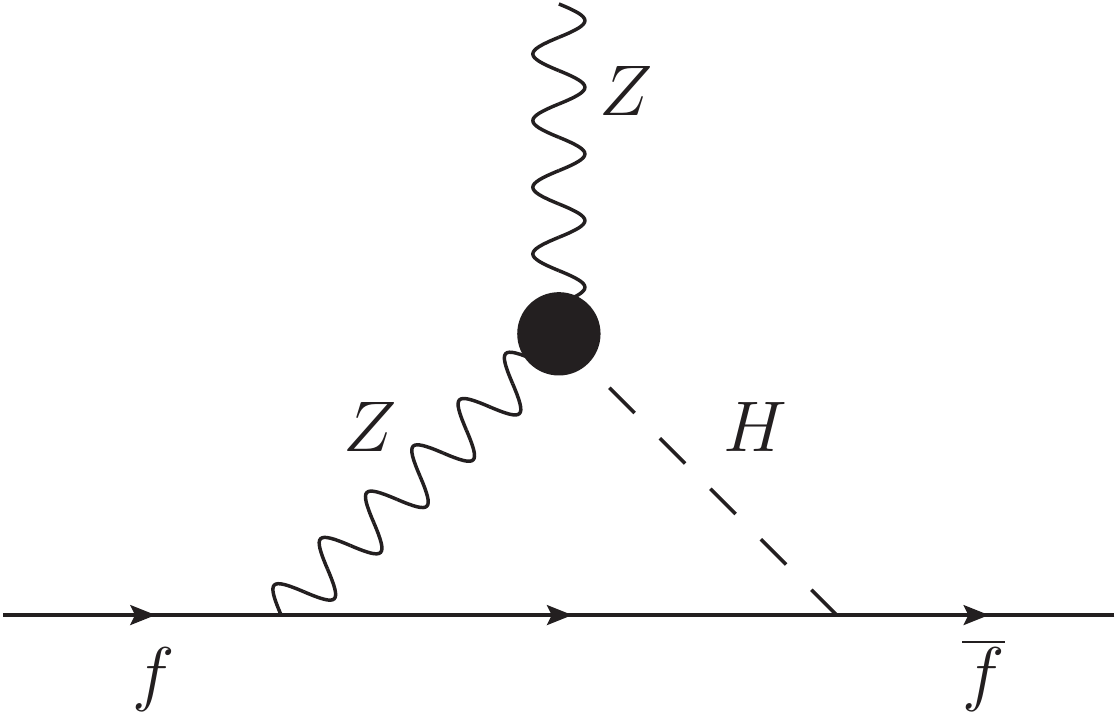}\label{wdm1}}
\subfigure[]{\includegraphics[width=6cm]{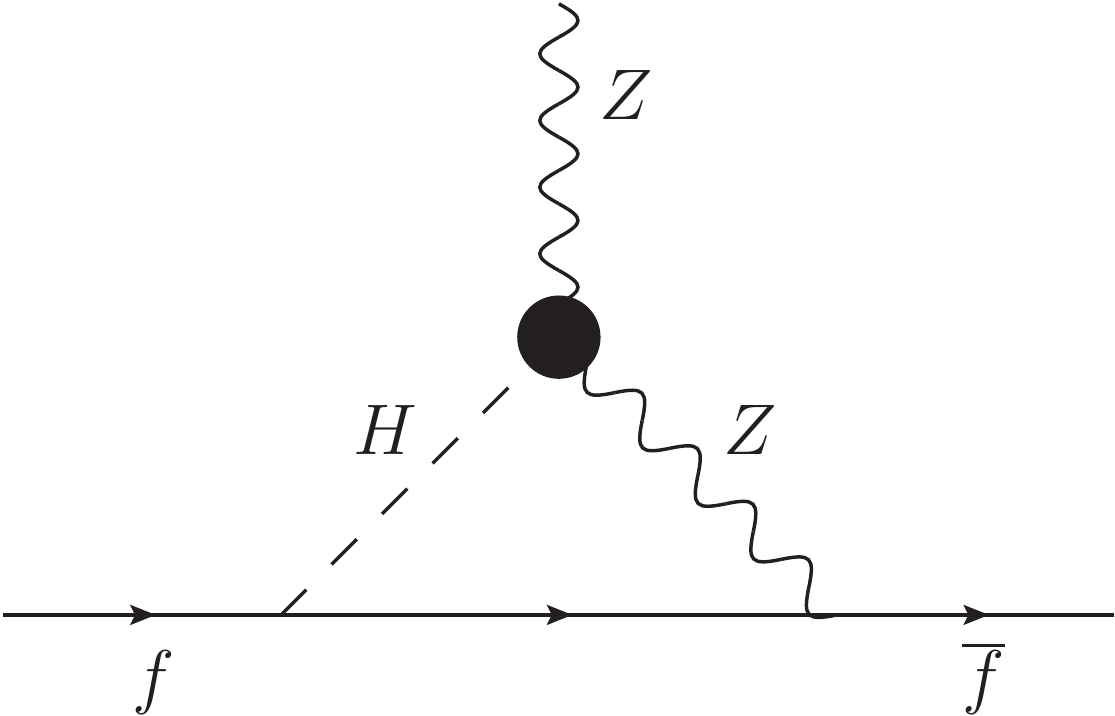}\label{wdm2}}
\caption{Feynman diagrams for the contribution of the anomalous $HZZ$ coupling to the fermion weak dipole moments.} \label{WDMdiag}
\end{center}
\end{figure}

In the SM, the  $HZZ$ coupling contributes at the one-loop level to the WMDM of fermions  via the Feynman diagrams of Fig. \ref{WDMdiag}, which have already been calculated  \cite{DeRujula:1990db,Arroyo-Urena:2017sfb}. The corresponding contribution from those diagrams to the  WEDM via $CP$-violating contributions to the $Hf\bar{f}$ vertex have also been calculated  in the minimal supersymmetric standard model (MSSM) \cite{Hollik:1998vz} and models with an extended scalar sector \cite{Arroyo-Urena:2017sfb}.  However, the contribution of the  anomalous $HZZ$ $CP$-violating form factor  has not been studied yet to our knowledge. 

We can obtain a rough estimate for the contribution to the WEDM from the interaction Lagrangian of Eq. \eqref{Lag2}:
\begin{align}
d^W_f\simeq \frac{e\alpha}{4\pi s_W^3 c_W}\frac{m_f}{m_Z^2}{\rm Im}\left(\widetilde{b}_Z\right)I\left(\frac{m_\tau^2}{m_Z^2},\frac{m_H^2}{m_Z^2}\right),
\end{align}
where $I(x,y)$ is the loop integral function, which we will assume to be of the order of $O(1)$. For the tau lepton we obtain
\begin{equation}
d^W_\tau\approx  {\rm Im}\left(\widetilde{b}_Z\right)\times 10^{-20}\quad \text{e-cm},
\end{equation}
which is well beyond the current best limit of $d^W_\tau <10^{-18}$ e-cm for both the real and imaginary parts \cite{ALEPH:2002kbp}. Therefore, our bounds on  $\widetilde{b}_Z$ presented in Table \ref{b1} are not in conflict with the WEDM of the tau lepton.  

The $CP$-violating effects induced by the $HZZ$ couplings have not been studied profusely, thus there are no tight bounds on the $h_3^V$ form factor currently.

\section{$ZZ$ production via the $H^\ast ZZ$ coupling}\label{WidthSec}
We now study the effects of the anomalous $HZZ$ couplings on the partial decay width  of the $H^\ast\rightarrow ZZ$ process.  The virtuality of the Higgs boson in  physical observables such as cross-sections and partial decay widths have been established in Ref. \cite{Goria:2011wa}, though the implications of an absorptive part of the $HZZ$ coupling  on such observables  have not been discussed in the literature yet. To study their role in $ZZ$ production, we   consider complex $HZZ$ anomalous couplings  and express  the $h_i^H$ form factors  as
\begin{equation}
h_i^H={\rm Re}\big[h_i^H\big]+i {\rm Im}\big[h_i^H\big],
\end{equation}
which lead to the following invariant amplitude for the production of two on-shell $Z$ gauge bosons  from an off-shell Higgs boson:  
\begin{align}
\label{amp1}
\mathcal{M}(\lambda_1,\lambda_2)=&\frac{g }{c_W}m_Z\Big\{g^{\mu \nu}\Big({\rm Re}\big[h_1^H\big]+i {\rm Im}\big[h_1^H\big] \Big) +\frac{p_2^{\mu }{p_1}^{\nu }}{m_Z^2} \Big(  {\rm Re}\big[h_2^H\big]+i 
   {\rm Im}\big[h_2^H\big]\Big) \nonumber\\
 &+\frac{\epsilon^{\mu \nu \alpha \beta }p_{1\alpha} p_{2\beta} }{m_Z^2}\Big({\rm Re}\big[h_3^H\big]
  +i {\rm Im}\big[h_3^H\big] \Big)\Big\}\epsilon^{\ast}_\mu(p_1,\lambda_1)\epsilon^{\ast}_\nu(p_2,\lambda_2),
\end{align}
where $\lambda_i$ and $\epsilon_\mu (p_i,\lambda_i)$ ($i=1,2$) stand for the polarization and polarization vector of the $Z$ gauge bosons. We would like to stress that that the absorptive parts  ${\rm Im}\big[h_k^H\big]$ ($k=1,2,3$) have not been  considered in previous studies of the   $H^\ast\rightarrow ZZ$ process \cite{Kniehl:1990mq,Chen:2022mre}.

\subsection{The unpolarized case}
From the amplitude \eqref{amp1}, the following partial decay width  can be obtained:
\begin{align}
\label{WidthHZZ}
    \Gamma_{H^\ast\rightarrow ZZ}=& \frac{g^2 \sqrt{q^2-4  m_Z^2}}{512 \pi 
   q^2 c_W^2 m_Z^6}   \Big\{4 q^6 m_Z^2 \Big(\left({\rm Im}\big[h_1^H\big]-2
   {\rm Im}\big[h_2^H\big]\right) {\rm Im}\big[h_2^H\big]+\left({\rm Re}\big[h_1^H\big]-2
   {\rm Re}\big[h_2^H\big]\right) {\rm Re}\big[h_2^H\big]\Big)\nonumber\\
   &+4 q^4 m_Z^4 \Big(2
   \left(2 {\rm Im}\big[h_2^H\big]{}^2+{\rm Im}\big[h_3^H\big]{}^2+2
   {\rm Re}\big[h_2^H\big]{}^2+{\rm Re}\big[h_3^H\big]{}^2\right) +{\rm Im}\big[h_1^H\big]{}^2
   -6 {\rm Im}\big[h_2^H\big] {\rm Im}\big[h_1^H\big]  \nonumber\\
   &+{\rm Re}\big[h_1^H\big]{}^2-6
   {\rm Re}\big[h_1^H\big] {\rm Re}\big[h_2^H\big]\Big)-16 q^2 m_Z^6 \Big(2
   \left({\rm Im}\big[h_3^H\big]{}^2+{\rm Re}\big[h_3^H\big]{}^2\right)+{\rm Im}\big[h_1^H\big]{}^2-2 {\rm Im}\big[h_2^H\big] {\rm Im}\big[h_1^H\big]\nonumber\\
   &+{\rm Re}\big[h_1^H\big]{}^2-2
   {\rm Re}\big[h_1^H\big] {\rm Re}\big[h_2^H\big]\Big)+48 m_Z^8
   \left({\rm Im}\big[h_1^H\big]{}^2+{\rm Re}\big[h_1^H\big]{}^2\right)+q^8
   \left({\rm Im}\big[h_2^H\big]{}^2+{\rm Re}\big[h_2^H\big]{}^2\right)\Big\}, 
\end{align}
which reduces to the SM tree-level result \cite{Kniehl:1990mq} when ${\rm Re}\big[h_1^H\big]=1$, ${\rm Re}\big[h_{2,3}^H\big]={\rm Im}\big[h_{1,2,3}^H\big]=0$:
\begin{equation}
 \Gamma^{\text{Tree}}_{H^\ast\rightarrow ZZ}=\frac{g^2 \sqrt{q^2-4  m_Z^2}}{512
      \pi  c_W^2   m_Z^6} \left(4 q^2 m_Z^4-16  m_Z^6+48 \frac{m_Z^8}{q^2}\right).
\end{equation}

The  SM contributions to the $H^\ast\rightarrow ZZ$ decay was studied long ago up to the one-loop level \cite{Kniehl:1990mq,Chen:2022mre}, but the entire set of  anomalous couplings has not been considered yet and their implications remain unexplored. Thus, a complete analysis is required since any non-standard contribution to the anomalous $HZZ$ couplings may be at the reach of measurement in a near future. To examine the corresponding  effects  on $Z$ pair production, we define the ratio $\mathcal{R}$ between the full (tree-level SM plus one-loop SM plus non-standard anomalous contributions) and the SM tree-level partial decay widths
\begin{equation}
\label{ratioR}
\mathcal{R}=\frac{ \Gamma_{H^\ast\rightarrow ZZ}}{ \Gamma^{\text{Tree}}_{H^\ast\rightarrow ZZ}}.
\end{equation}
For the numerical evaluation of  Eq. \eqref{ratioR} we write the $h_i^H$ form factors in terms of  the $\hat{b}_Z$, $\hat{c}_Z$  and $\widetilde{b}_Z$ anomalous couplings via   Eqs. \eqref{H11}, \eqref{H22}, \eqref{H33}. Furthermore, for $\hat{b}_Z$ we use the SM one-loop results obtained in Sec. \ref{AnRe}, whereas for the remaining anomalous couplings we use values consistent with the bounds of Tables \ref{b1} and \ref{b3}. To asses the effects of such anomalous couplings, we consider in our analysis the following illustrative scenarios:

\begin{itemize}
  \item Scenario I: ${\rm Re}\big[ \hat{c}_Z\big]\sim {\rm Re}\big[ \widetilde{b}_Z\big]= 10^{-4}$ and ${\rm Im}\big[ \hat{c}_Z\big]= {\rm Im}\big[ \widetilde{b}_Z\big]=10^{-3}$.
  \item Scenario II: ${\rm Re}\big[ \hat{c}_Z\big]={\rm Re}\big[ \widetilde{b}_Z\big]\sim-10^{-3}$ and ${\rm Im}\big[ \hat{c}_Z\big]={\rm Im}\big[ \widetilde{b}_Z\big]\sim 10^{-3}$.
  \item Scenario III: $-{\rm Im}\big[ \widetilde{b}_Z\big]={\rm Re}\big[ \hat{c}_Z\big]=10^{-3}$ and $-{\rm Im}\big[ \hat{c}_Z\big]={\rm Re}\big[ \widetilde{b}_Z\big]\sim 10^{-4}$.
\end{itemize}

\begin{figure}[H]
\begin{center}
{\includegraphics[width=10cm]{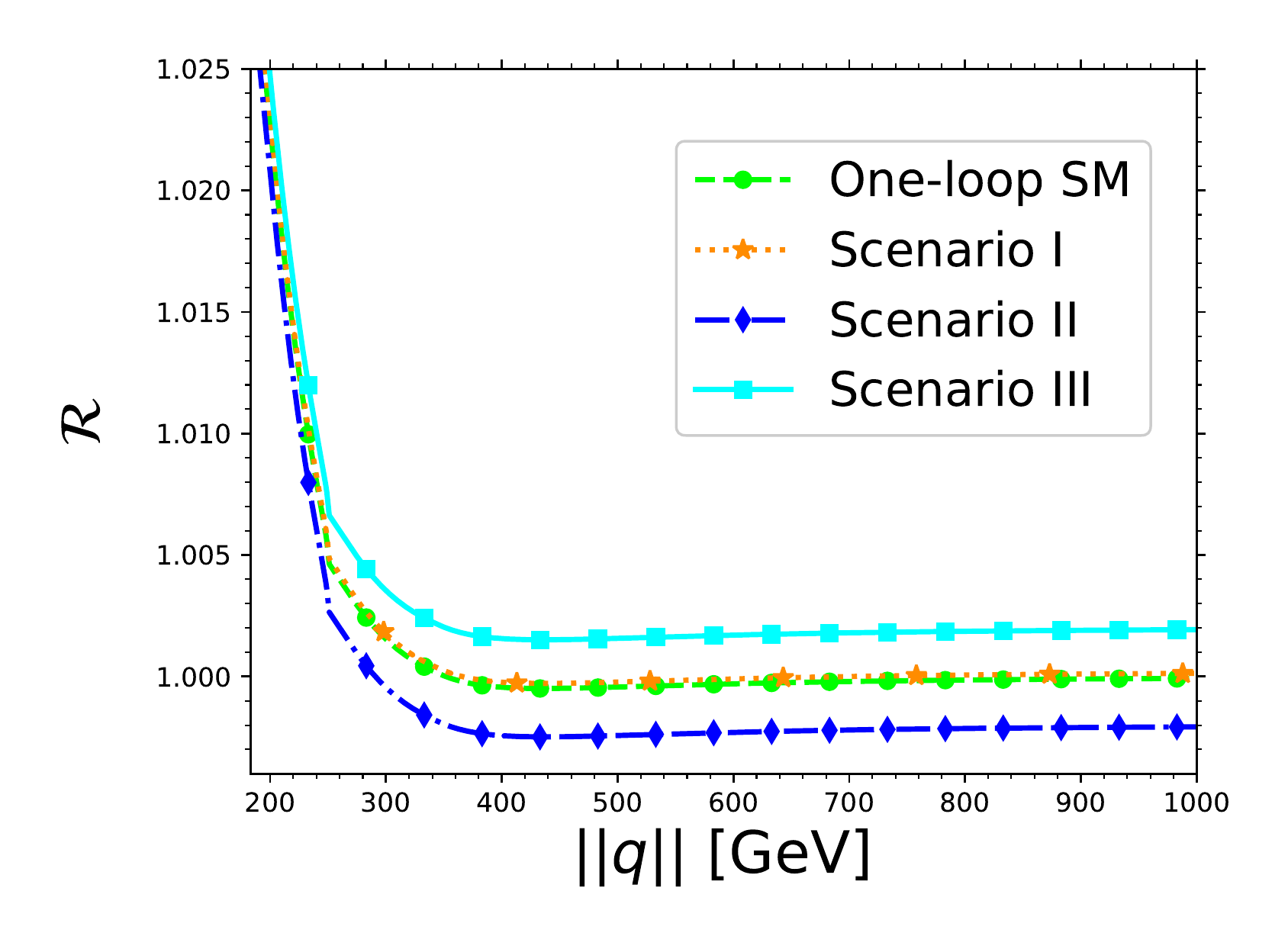}}
%\subfigure[]{\includegraphics[width=14cm]{diag3.eps}\label{Diagram3}}
\caption{Behavior of the ratio $\mathcal{R}$ as function of the transfer momentum of the Higgs boson $\|q\|$ for the  SM contribution up to one-loop leval, namely, $\hat{c}_Z=\widetilde{b}_Z=0$, and also in the three scenarios discussed in the text for  $\hat{c}_Z$ and $\widetilde{b}_Z$. For $\hat{b}_Z$ we use the expressions obtained in Sec. \ref{AnRe}.\label{R}  }
\end{center}
\end{figure}

We now show in Fig. \ref{R} the behavior of $\mathcal{R}$ as a function of the Higgs boson transfer momentum $\|q\|$ for the  tree-level plus one-loop level SM contribution, namely, $\hat{c}_Z=\widetilde{b}_Z=0$, and also in the three above scenarios. We observe that at high energies $\mathcal{R}$ becomes constant: the  curves for the one-loop SM contribution and that for scenario II approaches the unity, whereas the curves for  scenarios I and III show a slight deviation up to about 0.2\%. On the other hand,  at low energies, the deviation  of $\mathcal R$ from the unity can be as large as 3\% in the four scenarios, which could put  the anomalous couplings at the reach of measurement. Nevertheless, discriminating  between the SM one-loop contribution and those arising from  the $\hat{c}_Z$ and $\widetilde{b}_Z$ anomalous  couplings could represent a hard task.

\subsection{The polarized case}
From the amplitude \eqref{amp1} it is also possible to obtain the partial  decay width  $ \Gamma_{H^\ast\rightarrow ZZ}$ for polarized $Z$ gauge bosons.  In the frame in which the vector bosons propagate along the $x-z$ plane, the polarization vectors can be written as
\begin{align}
  & \epsilon^\mu_{1,2}(0)=\frac{1}{m_Z}\Big( \| \vec{p} \|\text{,} \pm E\sin{\theta} \text{,}0  \text{,} \pm E\cos{\theta}  \Big), \nonumber\\
    &  \epsilon^\mu_{1}(L/R)=\frac{1}{\sqrt{2}}\Big(0 \text{,} \pm \cos{\theta} \text{,} - i  \text{,} \mp \sin{\theta}  \Big) ,
\end{align}
where the energy and magnitude of the three-momentum can be written as $E=\sqrt{q^2}/2$ and $\|p\|=\sqrt{q^2-4m_Z^2}/2$. In terms of the three contributing polarized amplitudes $\mathcal{M}(0,0)$, $\mathcal{M}(L,L)$, $\mathcal{M}(R,R)$ the squared amplitude is
\begin{align}
   \mathcal{M} ^2=& \Big(\frac{g }{c_W}\Big)^2 m^2_Z\sum_{\lambda_i} \Big(    \mathcal{M}^2(\lambda_i,\lambda_i) \Big)\nonumber \\
      =&\Big(\frac{g }{c_W}\Big)^2 m^2_Z \Big(    \mathcal{M}^2_{LL}+ \mathcal{M}^2_{RR}+ \mathcal{M}^2_{00} \Big),
\end{align}
with $\mathcal{M}^2(\lambda_i,\lambda_i)\equiv\mathcal{M}^2_{\lambda_i,\lambda_i}$ given by
\begin{align}
\label{MLL}
 \mathcal{M}^2_{LL}=& \frac{1}{4 m_Z^4}\Big\{4  m_Z^2  \sqrt{q^4-4 q^2 m_Z^2} \Big({\rm Re}\big[h_1^H\big]{\rm Im}\big[h_3^H\big] -
   {\rm Im}\big[h_1^H\big] {\rm Re}\big[h_3^H\big] \Big)\nonumber\\
   &+q^2 \left(q^2-4
   m_Z^2\right) \left({\rm Re}\big[h_3^H\big]{}^2+{\rm Im}\big[h_3^H\big]{}^2\right)+4 m_Z^4
   \Big({\rm Re}\big[h_1^H\big]{}^2+{\rm Im}\big[h_1^H\big]{}^2\Big)\Big\},
\end{align}
\begin{align}
\label{MRR}
 \mathcal{M}^2_{RR}=&\frac{1}{4 m_Z^4}\Big\{4  m_Z^2  \sqrt{q^4-4 q^2 m_Z^2} \Big(-{\rm Re}\big[h_1^H\big]{\rm Im}\big[h_3^H\big] +
   {\rm Im}\big[h_1^H\big] {\rm Re}\big[h_3^H\big] \Big)\nonumber\\
   &+q^2 \left(q^2-4
   m_Z^2\right) \left({\rm Re}\big[h_3^H\big]{}^2+{\rm Im}\big[h_3^H\big]{}^2\right)+4 m_Z^4
   \Big({\rm Re}\big[h_1^H\big]{}^2+{\rm Im}\big[h_1^H\big]{}^2\Big)\Big\},
   \end{align}
and
\begin{align}
 \mathcal{M}^2_{00}=&\frac{1}{16 m_Z^8}\Big\{4 q^6 m_Z^2 \Big[\left({\rm Im}\big[h_1^H\big]-2 {\rm Im}\big[h_2^H\big]\right)
   {\rm Im}\big[h_2^H\big]+\left({\rm Re}\big[h_1^H\big]-2 {\rm Re}\big[h_2^H\big]\right)
   {\rm Re}\big[h_2^H\big]\Big]\nonumber\\
   &+4 q^4 m_Z^4 \Big[4
   \left({\rm Im}\big[h_2^H\big]{}^2+{\rm Re}\big[h_2^H\big]{}^2\right)+{\rm Im}\big[h_1^H\big]{}^2-6 {\rm Im}\big[h_2^H\big] {\rm Im}\big[h_1^H\big]+{\rm Re}\big[h_1^H\big]{}^2-6
   {\rm Re}\big[h_1^H\big] {\rm Re}\big[h_2^H\big]\Big]\nonumber\\
   &-16 q^2 m_Z^6
   \Big[{\rm Im}\big[h_1^H\big]{}^2-2 {\rm Im}\big[h_2^H\big]
   {\rm Im}\big[h_1^H\big]+{\rm Re}\big[h_1^H\big] \left({\rm Re}\big[h_1^H\big]-2
   {\rm Re}\big[h_2^H\big]\right)\Big]\nonumber\\
   &+16 m_Z^8
   \left({\rm Im}\big[h_1^H\big]{}^2+{\rm Re}\big[h_1^H\big]{}^2\right)+q^8
   \left({\rm Im}\big[h_2^H\big]{}^2+{\rm Re}\big[h_2^H\big]{}^2\right)\Big\},
   \end{align}
which  agree with the results of Ref. \cite{Bolognesi:2012mm} in the case of real form factors, though  there is a difference of sign in $\mathcal{M}(0,0)$, which in our case is absorbed in the corresponding polarization vector. 

The polarized partial width  can be defined as 
\begin{equation}\label{gammaHZZp}
\Gamma_{H^\ast\rightarrow Z_{\lambda_i} Z_{\lambda_i}}=\frac{g^2 m_Z^2 \sqrt{q^2-4  m_Z^2}}{32 \pi  q^2 c_W^2}\mathcal{M}^2_{\lambda_i \lambda_i}.
\end{equation}

We note that the $CP$-violating form factor $h_3^H$ only enters into  $\mathcal{M}^2_{LL}$ and $ \mathcal{M}^2_{RR}$, which only differ by a change  of sign of $ h_3^H$.  This peculiarity allows us to introduce an asymmetry, which, to our knowledge, has  never been reported in the literature. This left-right asymmetry can be written as  
\begin{align}
\mathcal{A}_{LR}=\frac{\Gamma_{H^\ast\rightarrow Z_{L} Z_{L}}-\Gamma_{H^\ast\rightarrow Z_{R} Z_{R}}}{\Gamma_{H^\ast\rightarrow Z_{L} Z_{L}}+\Gamma_{H^\ast\rightarrow Z_{R} Z_{R}}}, 
   \end{align}
and can be expressed in terms of the real and imaginary parts of the form factors via  Eqs. \eqref{MLL} and \eqref{MRR}:
\begin{align}
\label{ALRe}
\mathcal{A}_{LR}= \frac{4  m_Z^2 \|q\|\sqrt{q^2-4 m_Z^2} \left({\rm Re}\big[h_1^H\big]{\rm Im}\big[h_3^H\big]
   -{\rm Re}\big[h_3^H\big]{\rm Im}\big[h_1^H\big] \right)}{q^2
   \left(q^2-4 m_Z^2\right)
   \left({\rm Re}\big[h_3^H\big]{}^2+{\rm Im}\big[h_3^H\big]{}^2\right)+4 m_Z^4\big(
   {\rm Im}\big[h_1^H\big]{}^2+ {\rm Re}\big[h_1^H\big]{}^2\big)}.
   \end{align}
It is worth noting  that the size of this asymmetry is dominated by the $CP$-violating form factor $h_3^H$ as $\mathcal{A}_{LR}$ is proportional to  its real and absorptive parts. Therefore, $CP$-violating effects can be explored via polarized $Z$ gauge bosons through the $H^\ast\rightarrow ZZ$ process. Also, from Eq. \eqref{ALRe} we observe  that   a non-vanishing $\mathcal{A}_{LR}$ requires the presence of   absorptive parts of both $h_1^H$ and $h_3^H$, which are only induced for an off-shell Higgs boson. In summary, the left-right asymmetry is a consequence of  the presence of  $CP$ violation and complex form factors. Other asymmetries have been studied in the past through the $HZZ$ couplings \cite{Soni:1993jc,Godbole:2007cn},  though they are of  different nature as are given in terms of the energy of the four-lepton final state. Our result stress the relevance of the absorptive parts of the anomalous couplings in processes where off-shell particles are involved, though such terms have been overlooked in the past despite they could yield sizeable deviations in several observables. 

In the SM, the $\mathcal{A}_{LR}$ asymmetry would vanish  up to the two-loop level as the $CP$-violating form factor $h_3^H$ has been estimated to arise up to the three-loop level, with a value of the order of $10^{-11}$ \cite{Soni:1993jc}. Therefore, if we take  ${\rm Re}\big[h_3^H\big]={\rm Im}\big[h_3^H\big]\approx 10^{-11}$ and use the SM contributions to the $CP$ conserving form factors including only one-loop corrections, i.e., $\hat{c}_Z=0$, we obtain the following  estimate   
\begin{equation}
\mathcal{A}_{LR}^{SM}\approx 10^{-8}-10^{-9},
\end{equation}
for energies up to $\|q\|=3000$ GeV. 

To assess the importance of the $CP$-violating form factor, together with the complete   anomalous couplings  contributions and their allowed values obtained in Sec. \ref{BOTAC}, we fix ${\rm Re}\big[ \hat{c}_Z\big]={\rm Im}\big[ \hat{c}_Z\big]=10^{-3}$ and consider the following four scenarios: 
\begin{itemize}
  \item Scenario i:  ${\rm Re}\big[ \widetilde{b}_Z\big]=10^{-3}$ and ${\rm Im}\big[ \widetilde{b}_Z\big]=10^{-2}$.
  \item Scenario ii:  ${\rm Re}\big[ \widetilde{b}_Z\big]=10^{-3}$ and ${\rm Im}\big[ \widetilde{b}_Z\big]=-10^{-2}$.
  \item Scenario iii:  ${\rm Re}\big[ \widetilde{b}_Z\big]=10^{-4}$ and ${\rm Im}\big[ \widetilde{b}_Z\big]=10^{-3}$.
    \item Scenario iv:  ${\rm Re}\big[ \widetilde{b}_Z\big]={\rm Im}\big[ \widetilde{b}_Z\big]=10^{-4}$.
\end{itemize}

\begin{figure}[H]
\begin{center}
{\includegraphics[width=12.5cm]{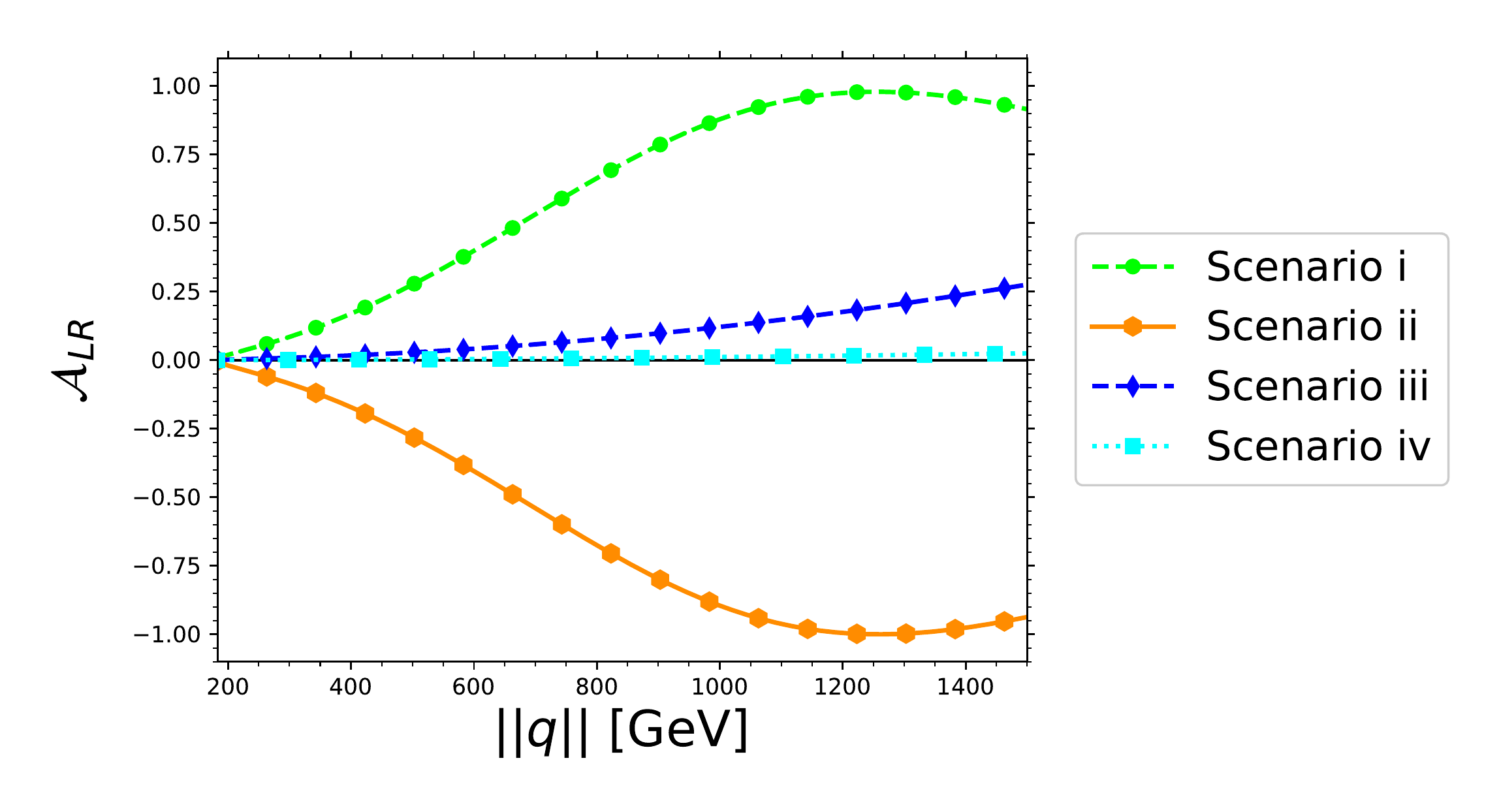}}
%\subfigure[]{\includegraphics[width=14cm]{diag3.eps}\label{Diagram3}}
\caption{ $\mathcal{A}_{LR}$ asymmetry as a function of the transfer momentum of the Higgs boson $\|q\|$ for the four scenarios described in the text  for the real and absorptive parts of the anomalous couplings $\hat{c}_Z$ and $\widetilde{b}_Z$. For $\hat{b}_Z$ we use the one-loop SM results presented in Sec. \ref{AnRe}. \label{ALR}}
\end{center}
\end{figure}

We show in Fig. \ref{ALR} the behavior of the   $\mathcal{A}_{LR}$ asymmetry as a function of the Higgs boson transfer momentum  $\|q\|$ in such scenarios. We first consider scenarios i and ii, where the real and imaginary parts of the $CP$-violating form factor $h_3^H$ are larger than ${\rm Re}\big[ \hat{c}_Z\big]$ and ${\rm Im}\big[ \hat{c}_Z\big]$. We observe that, except for a change of sign,  $\mathcal{A}_{LR}$ shows a similar behavior and can reach values of the order of $O(1)$, which means that the difference between the production of left-handed and right-handed $Z$ gauge bosons can be as large as the production of transverse $Z$ gauge bosons. Thus, if we only consider the total $Z$ gauge boson production, a lot of information about the properties of $Z_L$ and $Z_R$  is lost as the terms responsible for the asymmetry are cancelled out.
As far as  scenarios iii and iv are concerned,  $\mathcal{A}_{LR}$  can reach values of the order of $10^{-3}$ and $10^{-4}$, respectively, at low energies, which are still much larger, by about four orders of magnitude, than the SM contribution. In general, we note that in  the four scenarios  $\mathcal{A}_{LR}$  is  smaller at low energies and  reaches its highest values at high energy. This behavior is contrary to that  observed for the total $Z$ gauge bosons pair production of Fig. \ref{R}, where the anomalous coupling effects can only be observed in the region of low energy. Therefore, the study of polarized $Z$ gauge bosons offers a good opportunity to detect non-SM contributions to the $HZZ$ vertex, since the energy region where such effects can be discriminated is broader than for the production of unpolarized  $Z$ gauge boson pairs. Furthermore, our results could be at the reach of scrutiny at the LHC, where the role of polarized $Z$ gauge bosons in the $H\rightarrow ZZ^\ast\rightarrow4\ell$ process has already been explored  under the context of the SM and new physics scenarios \cite{Maina:2020rgd,Maina:2021xpe}. 

\begin{figure}[H]
\begin{center}
\subfigure[]{\includegraphics[width=9cm]{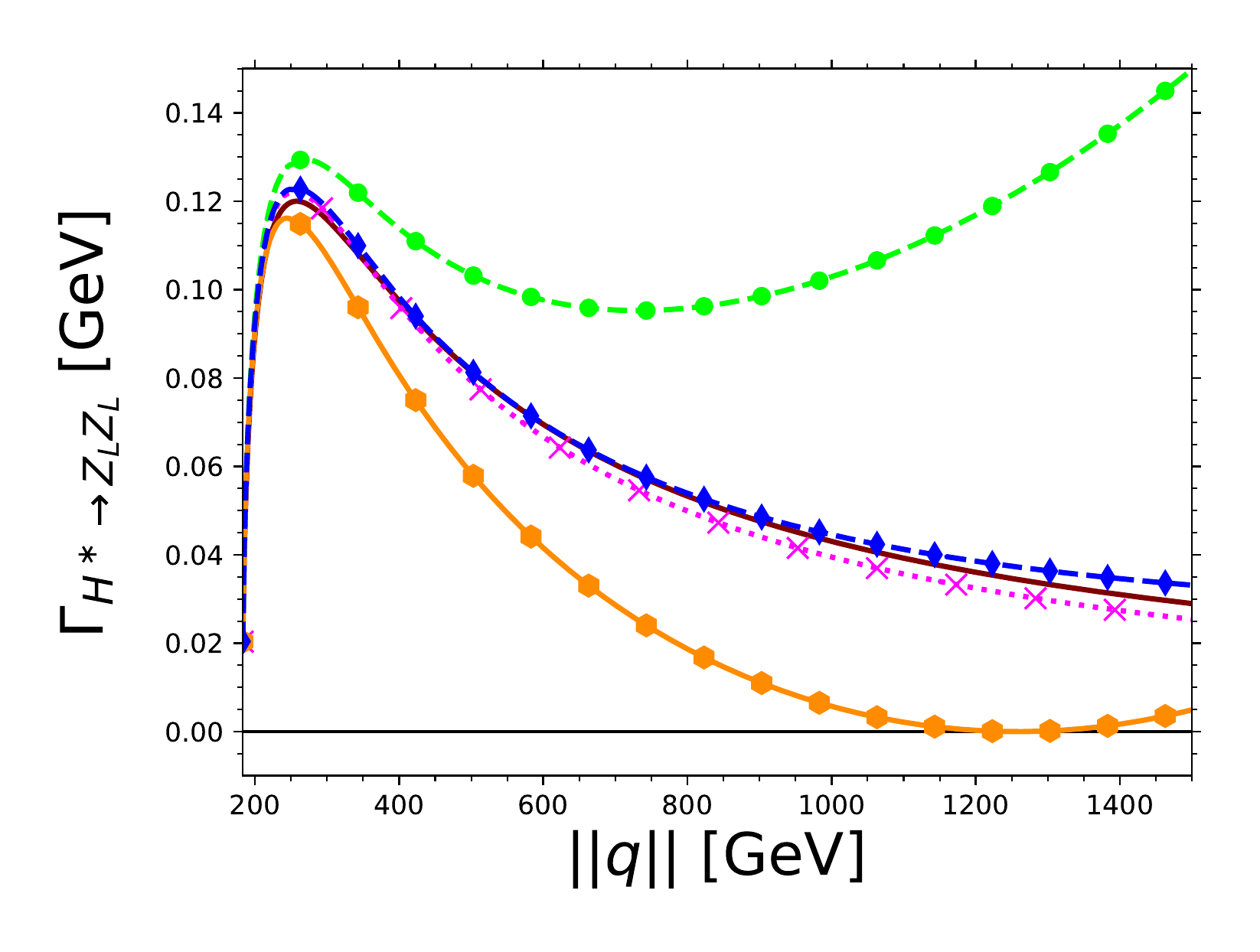}\label{plotLL}}\hspace{-.2cm}
\subfigure[]{\includegraphics[width=9cm]{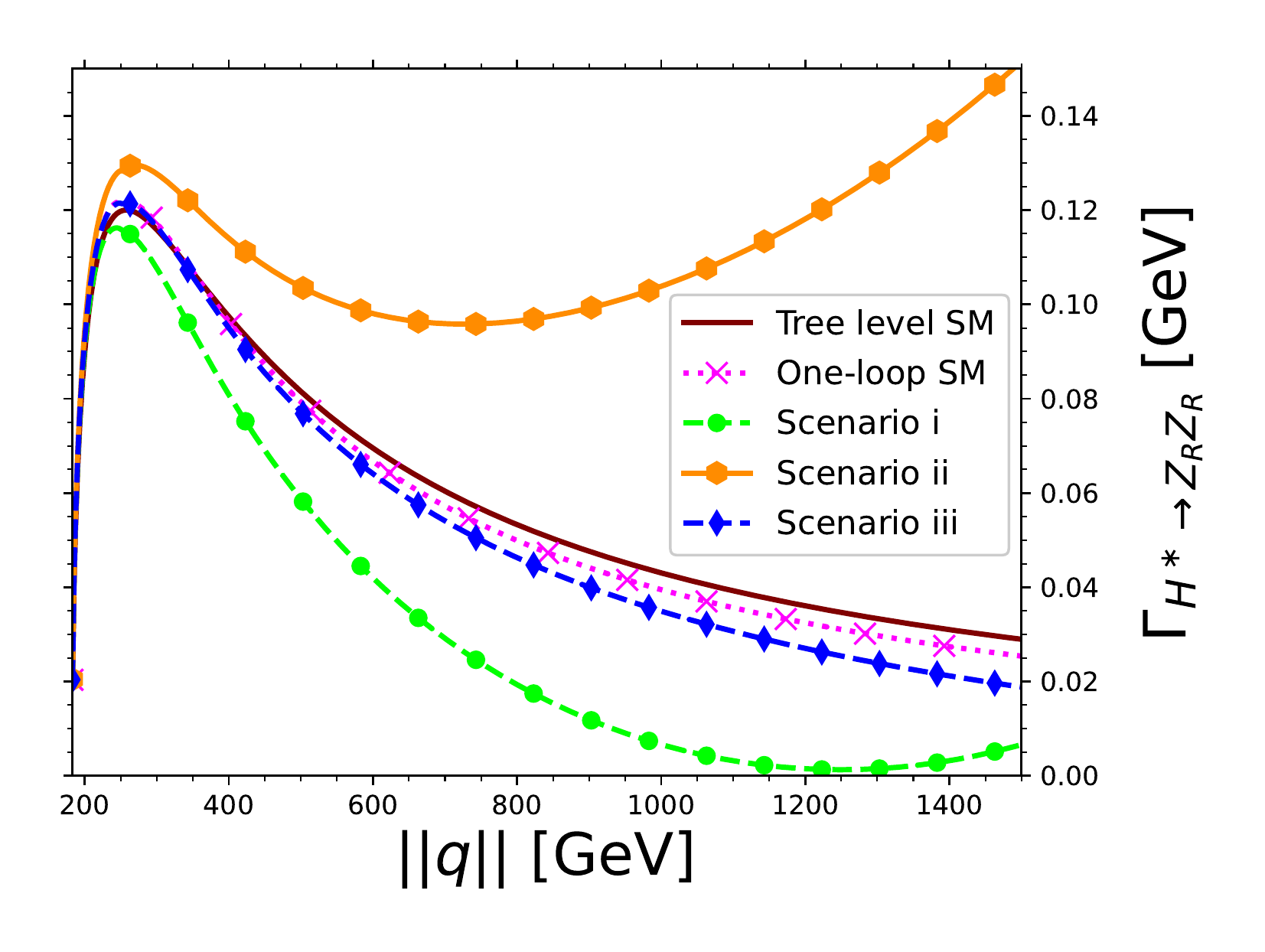}\label{plotRR}}
%\subfigure[]{\includegraphics[width=14cm]{diag3.eps}\label{Diagram3}}
\caption{Partial decay widths of the processes $H^\ast \rightarrow Z_L Z_L/Z_RZ_R$ as functions of the Higgs boson transfer momentum $\|q\|$ in the first three scenarios of  Fig. \ref{ALR} along with the tree and one-loop level SM predictions. \label{PlotPol}}
\end{center}
\end{figure}

For completeness, we show in Fig. \ref{PlotPol} the behavior of the partial decay widths $\Gamma_{H^\ast\rightarrow Z_L Z_L\text{,}Z_R Z_R}$ as functions of $\|q\|$ in the same scenarios of Fig. \ref{ALR} except for scenario iv, which does not yield interesting results. We also include the tree  and one-loop level SM results. Again,  the anomalous coupling contributions become more significant as the energy increases, which stems from the fact  that large deviations from the SM results arising from the $CP$-violating form factor (scenarios i and ii) are larger at very high   energies. This can be explained if we analyze the high energy behavior of  Eq.  \eqref{gammaHZZp}, which for  transverse polarizations reads as
\begin{equation}
\label{ }
\Gamma_{H^\ast\rightarrow Z_L Z_L/ Z_R Z_R}=\frac{g^2 \|q\|}{128\pi c_W^2 m_Z^2}\Big\{\pm4m_Z^2\Big({\rm Re}\big[h_1^H\big]{\rm Im}\big[h_3^H\big] -
   {\rm Im}\big[h_1^H\big] {\rm Re}\big[h_3^H\big] \Big)+q^2\left({\rm Re}\big[h_3^H\big]{}^2+{\rm Im}\big[h_3^H\big]{}^2\right)\Big\},
\end{equation}
where the term related to $\mathcal{A}_{LR}$ in Eq. \eqref{ALRe} can be recognized, ensuring a non-zero asymmetry at very high energies. In this regime, the main contributions arise from the $h_3^H$ quadratic terms. Thus, as observed in Fig. \ref{PlotPol}, the $CP$ violating form factor is the main responsible for the deviations from the SM predictions since they are more relevant in the $q^2\rightarrow \infty$ limit.    

In summary, the window to detect  effects of anomalous couplings,  in particular  of the $CP$-violating one, increases for high energy polarized $Z$ gauge bosons. Therefore, the recent measurement of the production of two $Z$ gauge bosons from an off-shell Higgs boson hints the possibility of detecting  anomalous $HZZ$ couplings in the near future with the advent of LHC upgrades. Moreover, the implications of these contributions could also be observed via the four-lepton final state  $H^\ast\rightarrow ZZ\rightarrow 4\ell$ \cite{Maina:2021xpe}, where any deviation from the SM  would be a clear evidence of new physics. To our knowledge, only the case $H\rightarrow ZZ^\ast\rightarrow 4\ell$ has been explored up to now \cite{Soni:1993jc,Gao:2010qx,Bolognesi:2012mm,Buchalla:2013mpa,Berge:2015jra,He:2019kgh,Maina:2020rgd,Maina:2021xpe}, thus we expect that our calculation could be helpful for new lines of research. We also would like to point out that we refrain from presenting  the study of pair-production of longitudinally polarized  $Z$ gauge bosons as it yields no significant deviation from the SM.

\section{Conclusions and outlook}
\label{Conc}
Motivated by the recent measurement of $Z$ gauge boson pair production from an off-shell Higgs boson at the LHC, we have presented a calculation of the one-loop SM contributions to the $HZZ$ coupling, which can be written in terms of two $CP$-conserving $h^V_{1,2}$ and one $CP$-violating $h_3^V$ form factors, where $V$ stands for the off-shell Higgs boson.   Since measurements at future LHC Runs could be sensitive to the potential effects of any anomalous contributions to this Higgs boson coupling, an up-to-date calculation is necessary to disentangle any anomalous effects from its SM contributions. We first obtain general results for the contributions to the  $H^\ast Z^\ast Z^\ast$ coupling via the BFM in terms of Passarino-Veltman scalar functions, which are too lengthy to be presented here, but are available for the interested reader at our GitLab repository \cite{urlcode}. From such general results,  the corresponding contributions to the $H^*ZZ$ and $HZZ^*$ couplings can be straightforwardly obtained and are reported in Appendix \ref{PassVel}.

We then present an analysis of the behavior of the   $H^\ast ZZ$ and $HZZ^\ast$ couplings, with focus on 
the effects of their absorptive parts, which have been overlooked in the past.  For the numerical analysis, we consider the most popular parametrizations used in the literature, and the relationships necessary to map from one to another parametrization are obtained. We choose to work in the parametrization of Eq. \eqref{Lag2} as it allows one to express the $h_2^V$ form factor in terms of only one anomalous coupling. It is found that the main one-loop SM contribution to the $h_2^V$  ($V=H$, $Z$) form factor arises from Feynman diagrams with a virtual $W$ gauge boson.  Although $h_2^H$ can be as large as $10^{-2}$ at low energies, in general, both $h_2^H$ and $h_2^Z$  can be of the order of $10^{-3}-10^{-4}$ at high energies. By means of our analytical results and the current constraints on the $HZZ$ coupling ratios reported by the CMS collaboration, we obtain  new bounds on the anomalous  $HZZ$ couplings, which are of the order of $10^{-2}-10^{-4}$ depending on the value of the transfer momentum of the off-shell boson.  It is worth emphasizing that while the limits on the real parts of the anomalous  $HZZ$ couplings are up to two orders of magnitude tighter than the previous ones, the bounds on the absorptive parts are a novel contribution as no previous limits of this kind have been reported before. 

To assess the potential effects of the absorptive parts of the $HZZ$ anomalous couplings, we estimate their contributions to the partial decay width  $\Gamma_{H^\ast\rightarrow ZZ}$, which are compared with the SM tree-level contribution. It is found that at low energies the anomalous contributions can yield a deviation on $\Gamma_{H^\ast\rightarrow ZZ}$  up to 3\% from the SM tree-level  result, whereas at high energies there is a negligibly  deviation. Along this line,  polarized $Z$ gauge bosons could be useful for the study of non-SM contributions via a  new left-right asymmetry $\mathcal{A}_{LR}$, which is sensitive to  $CP$-violating  complex form factors. Such asymmetry can be as large as the unity for large values of the $CP$-violating form factor, but it is of the order of $10^{-3}-10^{-4}$ in a more conservative scenario. Such values are indeed still larger by five or four orders of magnitude than the SM prediction for $\mathcal{A}_{LR}$, which has a non-vanishing contribution up to the three-loop level. 

The partial decay widths  $\Gamma_{H^\ast\rightarrow Z_LZ_L}$ and $\Gamma_{H^\ast\rightarrow Z_RZ_R}$ are also studied in several scenarios of new physics and a comparison is made with the tree and one-loop level SM contributions. Once again, significant deviations from the SM results are due to the $CP$-violating $HZZ$ form factor.  It is observed that such deviations can be large at high energies, and become even larger as the energy increases. Thus, for polarized $Z$ gauge bosons, there is a broad energy interval where the measurements could be sensitive to anomalous contributions, in contrast with the case of unpolarized $Z$ gauge bosons, where there is only sensitivity to anomalous contributions at low energies. Thus,  new physics effects may be searched for via the four-lepton final state of the process $H^\ast\rightarrow ZZ\rightarrow 4\ell$.

\begin{acknowledgements}
We acknowledge support from Consejo Nacional de Ciencia y Tecnolog\'ia and Sistema Nacional de Investigadores (Mexico). Partial support from Vicerrector\'ia de Investigaci\'on y Estudios de Posgrado de la Ben\'emerita Universidad Aut\'onoma de Puebla is also acknowledged.
\end{acknowledgements}

\appendix

\section{Parametrization mappings}\label{ParRel}
In this appendix se present explicit expressions for the relations between the coupling constants appearing in the parametrization of Eq. \eqref{Lag2} and those of the LHC framework and the SMEFT.   
\subsection{LHC Framework}
Using the kinematics for $H^*\rightarrow ZZ$, the interaction of Eq. \eqref{vertex2}, which is used at the LHC analyses, can be rewritten in terms  of the $h_i^V$ form factors of Eq. \eqref{vertex} as follows

\begin{align}
&h_1^H(q^2,p_1^2,m^2_Z)=\frac{a_1^{ZZ}}{2}+\frac{\kappa_1^{ZZ} p_1^2+\kappa_2^{ZZ} p_2^2}{2\big(\Lambda^{ZZ}_1\big)^2}+\frac{a_2^{ZZ}}{2} \frac{q^2-p_1^2-p_2^2}{m_Z^2}, \label{h111}\\
&h_2^H(q^2,p_1^2,m^2_Z)=- a_2^{ZZ},\label{h222}\\
&h_3^H(q^2,p_1^2,m^2_Z)=- a_3^{ZZ}.\label{h333}
\end{align}
By matching with Eqs. \eqref{h11} , \eqref{h22}  and \eqref{h33} from the Hagiwara basis, we  obtain
\begin{align}
1+a_Z&=\frac{a_1^{ZZ}}{2}, \label{az}\\
-   b_Z \frac{q^2-p_1^2-p_2^2}{m_Z^2}+ c_Z \frac{q^2}{m_Z^2}&=\frac{\kappa_1^{ZZ} p_1^2+\kappa_2^{ZZ} p_2^2}{2\big(\Lambda^{ZZ}_1\big)^2}+\frac{a_2^{ZZ}}{2} \frac{q^2-p_1^2-p_2^2}{m_Z^2}\label{eqaZZbZ2},\\
b_Z-c_Z&=-\frac{a_2^{ZZ}}{2} \label{eqaZZbZ},\\
\widetilde{b}_Z&=-\frac{a_3^{ZZ}}{2}\label{cpbZ},
\end{align}
which  yields the following relationship
\begin{equation}
c_Z \frac{p_1^2+p_2^2}{m_Z^2}= \frac{\kappa_1^{ZZ} p_1^2+\kappa_2^{ZZ} p_2^2}{2\big(\Lambda^{ZZ}_1\big)^2}.
\end{equation}
In the case of $\kappa_1^{ZZ}=\kappa_2^{ZZ}$ and $\Lambda^{ZZ}_1\equiv m_Z$  we obtain
\begin{equation}
c_Z=\frac{\kappa_1^{ZZ}}{2}.
\end{equation}
On the other hand, from the basis of Lagrangian \eqref{Lag2} given by Eqs. \eqref{H11}, \eqref{H22} and \eqref{H33}, we obtain the relations
\begin{align}
 -   \hat{b}_Z \frac{q^2-p_1^2-p_2^2}{m_Z^2}+ \frac{\hat{c}_Z}{2} \frac{p_1^2+p_2^2}{m_Z^2}&=\frac{\kappa_1^{ZZ} p_1^2+\kappa_2^{ZZ} p_2^2}{2\big(\Lambda^{ZZ}_1\big)^2}+\frac{a_2^{ZZ}}{2} \frac{q^2-p_1^2-p_2^2}{m_Z^2}   \\
     \hat{b}_Z & =-\frac{a_2^{ZZ}}{2}\label{hatbZ},
\end{align}
whereas Eqs. \eqref{az} and \eqref{cpbZ} remain valid. For $\kappa_1^{ZZ}=\kappa_2^{ZZ}$ and $\Lambda^{ZZ}_1\equiv m_Z$, we obtain
\begin{equation}
\hat{c}_Z= \kappa_1^{ZZ}. \label{chat}
\end{equation} 
Therefore, the parametrization of the $H^*ZZ$ vertex given by Eq. \eqref{vertex2}  is redundant  since  only the form factor $k_1^{ZZ}$  is necessary in both bases. In this work we have considered the relations \eqref{az}, \eqref{cpbZ}, \eqref{hatbZ} and \eqref{chat}.

\subsection{SMEFT}

The Lagrangian of Eq. \eqref{LagSMEFT} can be straightforwardly written in a more familiar form 
\begin{align}
 \mathcal{L}  &=\frac{g }{ 2 c_W m_Z}H \Big[\big(1+\delta c_z\big)m_Z^2 Z_\mu Z^\mu +c_{zz}\frac{g_L^2+g_Y^2}{4}Z_{\mu\nu}Z^{\mu\nu}+c_{z\Box} g_L^2 Z_\mu\partial_\nu Z^{\mu\nu}+\widetilde{c}_{zz}\frac{g_L^2+g_Y^2}{4}Z_{\mu\nu}\widetilde{Z}^{\mu\nu} \Big],
\end{align}
which allows one to identify the following relations with the form factors of  Lagrangian \eqref{Lag2}  
\begin{align}
   \delta c_z &= a_Z   \\
  c_{zz}  &= \frac{4}{g_L^2+g_Y^2}\hat{b}_Z,\label{czzEq}\\
    c_{z\Box}  &= \frac{1}{g_L^2} \hat{c}_Z,\\
      \widetilde{c}_{zz}  &= \frac{4}{g_L^2+g_Y^2}\widetilde{b}_Z,
\end{align}
they agree with the relations reported in Ref. \cite{CMS:2021nnc}. The mappings with the form factors of the Hagiwara basis follows easily from the relations of Sec. \ref{theofram}.

\section{Analytical results}\label{PassVel}

In this Appendix we present the analytical expression for the one-loop contributions  to the $HZZ$ form factor $h_2^V$ of Eq.  \eqref{h2} in terms of Passarino-Veltman scalar functions, which are obtained from  our general results for the form factors of the $H^*Z^* Z^*$ coupling available as a Mathematica code at our gitlab site  \cite{urlcode}.

We first present the convention used for the  two- and three-point Passarino-Veltman scalar functions:
\begin{align}
\text{B}_0(r^2,m_1^2,m_2^2)&=\frac{1}{i\pi^2}\int \frac{d^Dk}{(k^2-m_1^2)((k+r)^2-m_2^2)},\\
\text{C}_0(r_1^2,(r_1+r_2)^2,r_2^2,m_1^2,m_2^2,m_3^2)&=\frac{1}{i\pi^2}\int \frac{d^Dk}{(k^2-m_1^2)((k+r_1)^2-m_2^2)((k+r_2)^2-m_3^2)},
\end{align}
and introduce the following shorthand notation 
\begin{align}
\text{B}_{0ij}(r^2)&=\text{B}_0\big(r^2,m_i^2,m_j^2 \big),\nonumber\\
\text{C}_{0ijk}\left( q^2\right)&=\text{C}_0\left(m_Z^2,m_Z^2,q^2,m_i^2,m_j^2,m_k^2 \right),\nonumber\\
\text{C}_{0ijk}\left( p_1^2\right)&=\text{C}_0\left(m_H^2,m_Z^2,p_1^2,m_i^2,m_j^2,m_k^2 \right).
\end{align}
It is useful to note that the following symmetry relations are obeyed:
\begin{align}
\text{B}_{ij}(r^2)&=\text{B}_{ji}(r^2),\nonumber\\
\text{C}_{ijk}\left( q^2\right)&=\text{C}_{kji}\left( q^2\right),\nonumber\\
%C_{iij}\left( q^2\right)&= C_{jji}\left( q^2\right),\nonumber\\
%C_{iji}\left( q^2\right)&=C_{jij}\left( q^2\right).
\end{align}

\subsection{Contributions to the $H^*ZZ$ coupling}
The fermion contributions  to the  $A^H_{Vf,Af}$ functions read as

\begin{align}
A^H_{Vf}(q^2,m_Z^2)= &\frac{1}{q^2\left(q^2-4  m_Z^2\right){}^2}\Big\{4 q^2 m_Z^2
   \big[\text{B}_{0ff}\left(q^2\right)-\text{B}_{0ff}\left(m_Z^2\right)
   -3\big]+8 m_Z^4
   \big[\text{B}_{0ff}\left(q^2\right)-\text{B}_{0ff}\left(m_Z^2\right)
   +2\big]\nonumber\\
   &-\left(q^2-2 m_Z^2\right) \left(-4 m_f^2 \left(q^2-4 m_Z^2\right)-6 q^2
   m_Z^2-4 m_Z^4+q^4\right) \text{C}_{0fff}\left(q^2\right)+2
   q^4\Big\},
   \end{align}

\begin{align}
A^H_{Af}(q^2,m_Z^2)= &\frac{1}{q^2\left(q^2-4  m_Z^2\right){}^2}\Big\{\left(q^2-2 m_Z^2\right) \left(4 \big(q^2-m_Z^2\right)
  \big[ \text{B}_{0ff}\left(q^2\right)-
   \text{B}_{0ff}\left(m_Z^2\right)\big]\nonumber\\
   &+\left(-2 m_Z^2 \left(8 m_f^2+q^2\right)+4
   q^2 m_f^2+4 m_Z^4+q^4\right)
   \text{C}_0\left(q^2,m_Z^2,m_Z^2,m_f^2,m_f^2,m_f^2\right)+2 \left(q^2-4
   m_Z^2\right)\big)\Big\}.
   \end{align}

As for the contributions from  $W$ gauge boson ($\mathcal{W}$), and $H-Z$ boson ($\mathcal{HZ}$) exchange, they are given by 
   
\begin{align}
   A^H_W(q^2,m_Z^2)=&\frac{1}{ 8 q^2 \big(q^2-4 m_Z^2\big){}^2} \Big\{ 2 \Big(\big(1-2 c_W^2\big)^2 m_H^2 m_Z^2 \big(2 m_Z^2+q^2\big)+2 m_W^2
   \big(\big(12 c_W^4-4 c_W^2-7\big) q^2 m_Z^2\nonumber\\
   &+\big(24 c_W^4-8
   c_W^2+2\big) m_Z^4+2 q^4\big)\Big) \big[ \text{B}_{0WW}\big(m_Z^2\big)-\text{B}_{0WW}\big(q^2\big)\big]-2 \Big(\big(1-2 c_W^2\big)^2 m_H^2 \big(2 m_Z^4 \big(q^2-m_Z^2\big)\nonumber\\
   &+m_W^2
   \big(-6 q^2 m_Z^2+8 m_Z^4+q^4\big)\big)+2 m_W^2 \big(4 c_W^2 \big(1-2
   c_W^2\big) q^6+2 \big(32 c_W^4-16 c_W^2+1\big) q^4 m_Z^2\nonumber\\
   &-2\big(52 c_W^4-28 c_W^2+3\big) q^2 m_Z^4+\big(12 c_W^4-4
   c_W^2+1\big) m_W^2 \big(-6 q^2 m_Z^2+8 m_Z^4+q^4\big)\nonumber\\
   &+\big(-24
   c_W^4+8 c_W^2-2\big) m_Z^6\big)\Big)  \text{C}_{0WWW}\big(q^2\big)+\big(-6 q^2 m_Z^2+8 m_Z^4+q^4\big) \big(-\big(1-2 c_W^2\big)^2 m_H^2\nonumber\\
   &-2
   \big(12 c_W^4-4 c_W^2+1\big) m_W^2\big)\Big\},
   \end{align}
and
\begin{align}
   A^H_{ZH}(q^2,m_Z^2)=&\frac{1}{8 q^2 \big(q^2-4 m_Z^2\big){}^2}\Big\{ -2 \big(m_Z^2-m_H^2\big) \big(q^2-4 m_Z^2\big) \Big[m_Z^2
   \text{B}_{0ZZ}\big(0\big)-m_H^2
   \text{B}_{0HH}\big(0\big)\Big]\nonumber\\
   &+\big(6 m_H^4 \big(q^2-m_Z^2\big)-9 q^2 m_H^2 m_Z^2\big)
   \text{B}_{0HH}\big(q^2\big)+2 \big(8 m_Z^4 \big(m_H^2-q^2\big)-q^2
   m_H^4\nonumber\\
   &+2 m_Z^2 \big(2 q^2 m_H^2-m_H^4+q^4\big)\big)
   \text{B}_{0HZ}\big(m_Z^2\big)-\big(2 q^2 m_H^4+m_Z^2 \big(3 q^2 m_H^2-2
   m_H^4+4 q^4\big)\nonumber\\
   &-18 q^2 m_Z^4+8 m_Z^6\big) \text{B}_{0ZZ}\big(q^2\big)+m_H^2 \Big(\big(-2 m_H^4 \big(q^2-m_Z^2\big)-m_H^2 \big(-2 q^2 m_Z^2+4
   m_Z^4+q^4\big)\nonumber\\
   &-6 q^4 m_Z^2+28 q^2 m_Z^4-16 m_Z^6\big)
   \text{C}_{0ZHZ}\big(q^2\big)-3 \big(2 m_H^4
   \big(q^2-m_Z^2\big)+m_H^2 \big(8 m_Z^4-8 q^2 m_Z^2\big)\nonumber\\
   &+q^2 m_Z^2 \big(2
   m_Z^2+q^2\big)\big)
   \text{C}_{0HZH}\big(q^2\big)\Big)+\big(4 m^2_Z-q^2\big)  \big(2 m_H^2 \big(q^2-4 m_Z^2\big)+2
   m_H^4+q^2 m_Z^2\big) \Big\}.
   \end{align}
   
\subsection{Contributions to the $HZZ^*$ coupling}
The corresponding contributions to the $HZZ^*$ coupling can be written as

\begin{align}
A^Z_{Vf}(p_1^2,m_Z^2,m_H^2)= &\frac{1}{\left(-2 m_H^2 \left(m_Z^2+p_1^2\right)+m_H^4+\left(m_Z^2-p_1^2\right){}^2\right){}^2}\Big\{ \Big( 2 \big[m_H^4 \left(m_Z^2+p_1^2\right)-2 m_H^2 \left(-4 p_1^2
   m_Z^2+m_Z^4+p_1^4\right)\nonumber\\
   &+\left(m_Z^2-p_1^2\right){}^2 \left(m_Z^2+p_1^2\right)\big]\Big)\text{B}_{0ff}\big(m_H^2\big)+\Big(-2 m_Z^2 \big[4 p_1^2 \left(m_H^2+m_Z^2\right)\nonumber\\
   &+\left(m_H^2-m_Z^2\right){}^2-5
   p_1^4\big]\Big)\text{B}_{0ff}\big(m_Z^2\big)+\Big(-2 p_1^2 \big[m_H^2 \left(4 m_Z^2-2 p_1^2\right)+m_H^4+4 p_1^2 m_Z^2\nonumber\\
   &-5
   m_Z^4+p_1^4\big]\Big)\text{B}_{0ff}\big(p_1^2\big)+\Big(\left(m_H^2-m_Z^2-p_1^2\right) \big[-p_1^4 \left(-4 m_f^2+3 m_H^2+m_Z^2\right)\nonumber\\
   &-p_1^2
   \left(8 m_f^2 \left(m_H^2+m_Z^2\right)-10 m_H^2 m_Z^2-3
   m_H^4+m_Z^4\right)+\left(m_H^2-m_Z^2\right){}^2 \left(4
   m_f^2-m_H^2+m_Z^2\right)\nonumber\\
   &+p_1^6\big]\Big)\text{C}_{0fff}\big(p_1^2\big)+ 2 \big[-3 m_H^4 \left(m_Z^2+p_1^2\right)+m_H^2 \left(2 p_1^2 m_Z^2+3 m_Z^4+3
   p_1^4\right)+m_H^6\nonumber\\
   &-\left(m_Z^2-p_1^2\right){}^2 \left(m_Z^2+p_1^2\right)\big] \Big\},
   \end{align}

\begin{align}
A^Z_{Af}(p_1^2,m_Z^2,m_H^2)= &\frac{1}{\left(-2 m_H^2 \left(m_Z^2+p_1^2\right)+m_H^4+\left(m_Z^2-p_1^2\right){}^2\right){}^2}\Big\{ \Big(\left(m_H^2-m_Z^2-p_1^2\right) \big[-2 m_H^2 \left(m_Z^2+p_1^2\right)+4 m_H^4\nonumber\\
&-2
   \left(m_Z^2-p_1^2\right){}^2\big]\Big)\text{B}_{0ff}\big(m_H^2\big)+\Big(-2 \left(m_H^2-m_Z^2-p_1^2\right) \big[m_H^2 \left(m_Z^2-2 p_1^2\right)+m_H^4+p_1^2
   m_Z^2\nonumber\\
   &-2 m_Z^4+p_1^4\big]\Big)\text{B}_{0ff}\big(m_Z^2\big)+\Big(-2 \left(m_H^2-m_Z^2-p_1^2\right) \big[p_1^2
   \left(m_H^2+m_Z^2\right)+\left(m_H^2-m_Z^2\right){}^2\nonumber\\
   &-2 p_1^4\big]\Big)\text{B}_{0ff}\big(p_1^2\big)+\Big(\left(m_H^2-m_Z^2-p_1^2\right) \big[-p_1^4 \left(-4 m_f^2+m_H^2+m_Z^2\right)\nonumber\\
   &-p_1^2
   \left(8 m_f^2 \left(m_H^2+m_Z^2\right)-6 m_H^2
   m_Z^2+m_H^4+m_Z^4\right)+\left(m_H^2-m_Z^2\right){}^2 \left(4
   m_f^2+m_H^2+m_Z^2\right)\nonumber\\
   &+p_1^6\big]\Big)\text{C}_{0fff}\big(p_1^2\big)+ 2 \big[-3 m_H^4 \left(m_Z^2+p_1^2\right)+m_H^2 \left(2 p_1^2 m_Z^2+3 m_Z^4+3
   p_1^4\right)+m_H^6\nonumber\\
   &-\left(m_Z^2-p_1^2\right){}^2 \left(m_Z^2+p_1^2\right)\big]\Big\},
   \end{align}
   
   \begin{align}
   A^Z_W(p_1^2,m_Z^2,m_H^2)=&\frac{1}{ 8 \big(-2 m_H^2 \big(m_Z^2+p_1^2\big)+m_H^4+\big(m_Z^2-p_1^2\big){}^2\big){}^2} \Big\{\Big[ -m_H^6 \big(\big(1-2 c_W^2\big){}^2 \big(m_Z^2+p_1^2\big)+8
   m_W^2\big)\nonumber\\
   &-2 m_H^4 \big(\big(12 c_W^4-4 c_W^2-7\big) m_W^2
   \big(m_Z^2+p_1^2\big)-\big(1-2 c_W^2\big){}^2 \big(-4 p_1^2
   m_Z^2+m_Z^4+p_1^4\big)\big)\nonumber\\
   &-m_H^2 \big(4 m_W^2 \big(16 p_1^2 c_W^2 \big(3
   c_W^2-1\big) m_Z^2+\big(-12 c_W^4+4 c_W^2+1\big) m_Z^4+p_1^4 \big(-12 c_W^4+4
   c_W^2+1\big)\big)\nonumber\\
   &+\big(1-2 c_W^2\big){}^2 \big(m_Z^2-p_1^2\big){}^2
   \big(m_Z^2+p_1^2\big)\big)-2 \big(12 c_W^4-4 c_W^2+1\big) m_W^2
   \big(m_Z^2-p_1^2\big){}^2 \big(m_Z^2+p_1^2\big) \Big]\nonumber\\
   &\times \text{B}_{0WW}\big(m_H^2\big)+\Big[ 2 m_H^4 \big(m_W^2 \big(\big(12 c_W^4-4 c_W^2-1\big) m_Z^2-6 p_1^2\big)-\big(1-2
   c_W^2\big){}^2 m_Z^2 \big(m_Z^2-2 p_1^2\big)\big)\nonumber\\
   &+m_H^2 \big(4 m_W^2 \big(8
   p_1^2 c_W^2 \big(3 c_W^2-1\big) m_Z^2-2 \big(6 c_W^4-2 c_W^2+1\big) m_Z^4+3
   p_1^4\big)\nonumber\\
   &+\big(1-2 c_W^2\big){}^2 m_Z^2 \big(4 p_1^2 m_Z^2+m_Z^4-5
   p_1^4\big)\big)+m_H^6 \big(\big(1-2 c_W^2\big){}^2 m_Z^2+4 m_W^2\big)\nonumber\\
   &+2
   m_W^2 \big(m_Z^2-p_1^2\big) \big(p_1^2 \big(60 c_W^4-20 c_W^2+1\big)
   m_Z^2+\big(12 c_W^4-4 c_W^2+3\big) m_Z^4+2 p_1^4\big)\Big] \text{B}_{0WW}\big(m_Z^2\big)\nonumber\\
   &+\Big[ -2 m_H^4 \big(m_W^2 \big(p_1^2 \big(-12 c_W^4+4 c_W^2+1\big)+6 m_Z^2\big)+p_1^2
   \big(1-2 c_W^2\big){}^2 \big(p_1^2-2 m_Z^2\big)\big)\nonumber\\
   &+m_H^2 \big(4 m_W^2
   \big(8 p_1^2 c_W^2 \big(3 c_W^2-1\big) m_Z^2-2 p_1^4 \big(6 c_W^4-2
   c_W^2+1\big)+3 m_Z^4\big)\nonumber\\
   &+p_1^2 \big(1-2 c_W^2\big){}^2 \big(4 p_1^2 m_Z^2-5
   m_Z^4+p_1^4\big)\big)+m_H^6 \big(p_1^2 \big(1-2 c_W^2\big){}^2+4
   m_W^2\big)\nonumber\\
   &-2 m_W^2 \big(m_Z^2-p_1^2\big) \big(p_1^2 \big(60 c_W^4-20
   c_W^2+1\big) m_Z^2+p_1^4 \big(12 c_W^4-4 c_W^2+3\big)+2 m_Z^4\big)]\text{B}_{0WW}\big(p_1^2\big)\nonumber\\
   &+2 \Big[m_H^6 \big(-\big(52 c_W^4-20 c_W^2-1\big) m_W^2 \big(m_Z^2+p_1^2\big)-2
   p_1^2 \big(1-2 c_W^2\big){}^2 m_Z^2\nonumber\\
   &+\big(-24 c_W^4+8 c_W^2-2\big)
   m_W^4\big)+m_H^4 \big(6 \big(12 c_W^4-4 c_W^2+1\big) m_W^4
   \big(m_Z^2+p_1^2\big)\nonumber\\
   &+m_W^2 \big(2 p_1^2 \big(4 c_W^4-4 c_W^2-1\big)
   m_Z^2+\big(84 c_W^4-36 c_W^2+3\big) m_Z^4+3 p_1^4 \big(28 c_W^4-12
   c_W^2+1\big)\big)\nonumber\\
   &+p_1^2 \big(1-2 c_W^2\big){}^2 m_Z^2
   \big(m_Z^2+p_1^2\big)\big)+m_H^2 \big(-2 \big(12 c_W^4-4 c_W^2+1\big) m_W^4
   \big(2 p_1^2 m_Z^2+3 m_Z^4+3 p_1^4\big)\nonumber\\
   &-m_W^2 \big(m_Z^2+p_1^2\big) \big(-4
   p_1^2 \big(36 c_W^4-16 c_W^2+3\big) m_Z^2+\big(60 c_W^4-28 c_W^2+5\big)
   m_Z^4\nonumber\\
   &+p_1^4 \big(60 c_W^4-28 c_W^2+5\big)\big)+p_1^2 \big(1-2 c_W^2\big){}^2
   m_Z^2 \big(m_Z^2-p_1^2\big){}^2\big)+\big(12 c_W^4-4 c_W^2-1\big) m_H^8
   m_W^2\nonumber\\
   &+2 m_W^2 \big(m_Z^2-p_1^2\big){}^2 \big(-p_1^2 \big(1-2 c_W^2\big){}^2
   m_Z^2+\big(12 c_W^4-4 c_W^2+1\big) m_W^2 \big(m_Z^2+p_1^2\big)\nonumber\\
   &+\big(8 c_W^4-4
   c_W^2+1\big) m_Z^4+p_1^4 \big(8 c_W^4-4 c_W^2+1\big)\big)\Big]\text{C}_{0WWW}\big(p_1^2\big)-\big(\big(1-2 c_W^2\big){}^2 m_H^2\nonumber\\
   &+2 \big(12 c_W^4-4 c_W^2+1\big) m_W^2\big)
   \big(-3 m_H^4 \big(m_Z^2+p_1^2\big)+m_H^2 \big(2 p_1^2 m_Z^2+3 m_Z^4+3
   p_1^4\big)+m_H^6\nonumber\\
   &-\big(m_Z^2-p_1^2\big){}^2 \big(m_Z^2+p_1^2\big)\big)\Big\},
   \end{align}
and
\begin{align}
   A^Z_{ZH}(q^2,m_Z^2)=&\frac{1}{16 \big(-2 m_H^2 \big(m_Z^2+p_1^2\big)+m_H^4+\big(m_Z^2-p_1^2\big){}^2\big){}^2}\Big\{-4 \big(m_H^2-m_Z^2\big) \big(-2 m_H^2
   \big(m_Z^2+p_1^2\big)+m_H^4\nonumber\\
   &+\big(m_Z^2-p_1^2\big){}^2\big) \Big[m_H^2
   \text{B}_{0HH}\big(0\big)-m_Z^2
   \text{B}_{0ZZ}\big(0\big)\Big] + 3 m_H^2 \Big[-m_H^4 \big(7 m_Z^2+3 p_1^2\big)+2 m_H^2 \big(m_Z^4-p_1^2
   m_Z^2\big)\nonumber\\
   &+4 m_H^6+\big(m_Z^2-p_1^2\big){}^3\Big] \text{B}_{0HH}\big(m_H^2\big)+\Big[m_H^6 \big(p_1^2-11 m_Z^2\big)+4 m_H^4 \big(p_1^2 m_Z^2+7 m_Z^4+p_1^4\big)\nonumber\\
   &-m_H^2
   \big(7 p_1^2 m_Z^4+p_1^4 m_Z^2+7 m_Z^6+p_1^6\big)-4 m_H^8-2 \big(3
   m_Z^2+p_1^2\big) \big(m_Z^3-p_1^2 m_Z\big){}^2\Big]\text{B}_{0ZZ}\big(m_H^2\big)\nonumber\\
   &-2 \Big[-2 m_H^6 \big(m_Z^2+p_1^2\big)+m_H^4 \big(6 p_1^2 m_Z^2-7
   m_Z^4+p_1^4\big)+2 m_H^2 m_Z^2 \big(-9 p_1^2 m_Z^2+8 m_Z^4+p_1^4\big)+m_H^8\nonumber\\
   &+2
   p_1^6 m_Z^2+6 p_1^2 m_Z^6-8 m_Z^8\Big]\text{B}_{0HZ}\big(m_Z^2\big)-2 \Big[m_H^6 \big(2 p_1^2-6 m_Z^2\big)+m_H^4 \big(-11 p_1^2 m_Z^2+12
   m_Z^4-p_1^4\big)\nonumber\\
   &-2 m_H^2 \big(-3 p_1^2 m_Z^4-3 p_1^4 m_Z^2+5
   m_Z^6+p_1^6\big)+m_H^8+3 m_Z^2 \big(m_Z^2-p_1^2\big)
   \big(m_Z^2+p_1^2\big){}^2\Big]\text{B}_{0HZ}\big(p_1^2\big)\nonumber\\
   &-6 m_H^4 \Big[-3 m_H^4 \big(2 m_Z^2+p_1^2\big)+m_H^2 \big(p_1^2 m_Z^2+6 m_Z^4+3
   p_1^4\big)+2 m_H^6-2 \big(m_Z^2-p_1^2\big){}^2 \big(m_Z^2+p_1^2\big)\Big]\nonumber\\
   &\times\text{C}_{0HHZ}\big(p_1^2\big)-2 \Big[m_H^8 \big(4 m_Z^2-2 p_1^2\big)-2 m_H^6 \big(11 m_Z^4+p_1^4\big)+m_H^4
   \big(5 p_1^2 m_Z^4-2 p_1^4 m_Z^2+12 m_Z^6+p_1^6\big)\nonumber\\
   &+m_H^2 \big(11 m_Z^2+5
   p_1^2\big) \big(m_Z^3-p_1^2 m_Z\big){}^2+3 m_H^{10}-2 \big(m_Z^3-p_1^2
   m_Z\big){}^2 \big(p_1^2 m_Z^2+4 m_Z^4+p_1^4\big)\Big]\text{C}_{0ZZH}\big(p_1^2\big)\nonumber\\
   &-2 \Big[-m_H^6 \big(13 m_Z^2+10 p_1^2\big)+m_H^4 \big(5 p_1^2 m_Z^2+15 m_Z^4+8
   p_1^4\big)+m_H^2 \big(8 p_1^2 m_Z^4+p_1^4 m_Z^2-7 m_Z^6-2 p_1^6\big)\nonumber\\
   &+4
   m_H^8+m_Z^2 \big(m_Z^2-p_1^2\big){}^3\Big]\Big\}.
   \end{align}

\bibliography{BiblioH}

\end{document}